\pdfoutput=1
\documentclass[11pt,a4paper]{article}
\usepackage{jheppub,bm,booktabs,multirow}

\usepackage[mathletters]{ucs}
\usepackage[utf8x]{inputenc}

\makeatletter
\def\@fpheader{~}
\makeatother

\usepackage[Q=yes,pverb-linebreak=no]{examplep}

\newcommand{\TT}{T_{\perp}}
\newcommand{\tT}{\tau_{\perp}}
\newcommand{\ts}{\tau_{s\perp}}
\newcommand{\slsh}{\!\!\!\slash}

\newcommand{\perptop}{\perp\hspace{-0.11cm}\top}

\title{Factorization and resummation for transverse thrust}
\author[a,b]{Thomas Becher,}
\author[a]{and Xavier Garcia i Tormo}
\affiliation[a]{Albert Einstein Center for Fundamental Physics, Institut f\"ur Theoretische Physik, Universit\"at Bern,
  Sidlerstrasse 5, CH-3012 Bern, Switzerland}
\affiliation[b]{Center for the Fundamental Laws of Nature, Harvard University,
Cambridge, MA 02138, USA}

\emailAdd{becher@itp.unibe.ch}
\emailAdd{garcia@itp.unibe.ch}

\date{\today}

\abstract{
We analyze transverse thrust in the framework of Soft Collinear Effective Theory and obtain a factorized expression for the cross section that permits resummation of terms enhanced in the dijet limit to arbitrary accuracy. The factorization theorem for this hadron-collider event-shape variable involves collinear emissions at different virtualities and suffers from a collinear anomaly. We compute all its ingredients at the one-loop order, and show that the two-loop input for next-to-next-to-leading logarithmic accuracy can be extracted numerically, from existing fixed-order codes.}

\begin{document}

\maketitle

\section{Introduction}

Event-shape variables are designed to measure geometrical properties of energy flow in  collider events. They were among the first observables proposed to test quantum chromodynamics (QCD), and can also be used to discriminate beyond-the-Standard-Model (BSM) physics against the QCD background. Numerous event-shape studies have appeared over the years, notably including extractions of the strong coupling, $\alpha_s$. The majority of the existing work has focused on leptonic collisions, and on deep-inelastic scattering (DIS). Nevertheless, event shapes are also of great interest in the much richer environment of hadronic collisions. Indeed, a lot of recent work is using event shapes as a tool to study jet substructure. Event shapes could also be instrumental to get a better handle on some poorly understood aspects of hadronic collisions, such as the underlying event.
 
 Several event-shape variables for hadronic collisions were  studied in refs.~\cite{Banfi:2004nk,Banfi:2010xy}. They were defined in analogy to the ones in leptonic collisions, but in terms of the components of the three-momenta transverse to the beam direction. In this paper we focus on the archetypal event shape, thrust. Transverse thrust, which we denote by $\TT$, is defined, in analogy to ordinary thrust, as
\begin{equation}\label{eq:def_trans_thrust}
\TT=\max_{\vec{n}_{\perp}}\frac{\sum_i\left|\vec{p}_{i\perp}\cdot\vec{n}_{\perp}\right|}{\sum_i |\vec{p}_{i\perp}|},
\end{equation}
where the sum is over all the particles in the final state, with
momenta $\vec{p}_i$. Throughout the paper, the
subindex $\perp$ denotes the (two) momentum components transverse to
the beam direction. The vector $\vec{n}_{\perp}$ which maximizes the ratio on the right-hand side of eq.~(\ref{eq:def_trans_thrust}) is called the transverse-thrust axis. Transverse thrust has been measured at the Large Hadron Collider (LHC)~\cite{Khachatryan:2011dx,Aad:2012np,Aad:2012fza,Khachatryan:2014ika} and previously also at the Tevatron~\cite{Aaltonen:2011et}. Here, we will study this quantity in the dijet limit, where $\TT\to 1$, and obtain a factorization formula that allows us to resum the enhanced terms arising in this limit. Resummation for hadron-collider event shapes in the dijet limit at next-to-leading-logarithmic (NLL) accuracy was studied in refs.~\cite{Banfi:2004nk,Banfi:2010xy} within the automated resummation framework CAESAR~\cite{Banfi:2004yd}. In the present paper we analyze transverse thrust using Soft Collinear Effective Theory (SCET)~\cite{Bauer:2000yr,Bauer:2001yt,Beneke:2002ph} (see \cite{Becher:2014oda} for an introduction), and obtain an all-order factorization formula that allows for resummation at any desired accuracy. 

\begin{figure}[t!]
\centering
\includegraphics[width=0.82\textwidth]{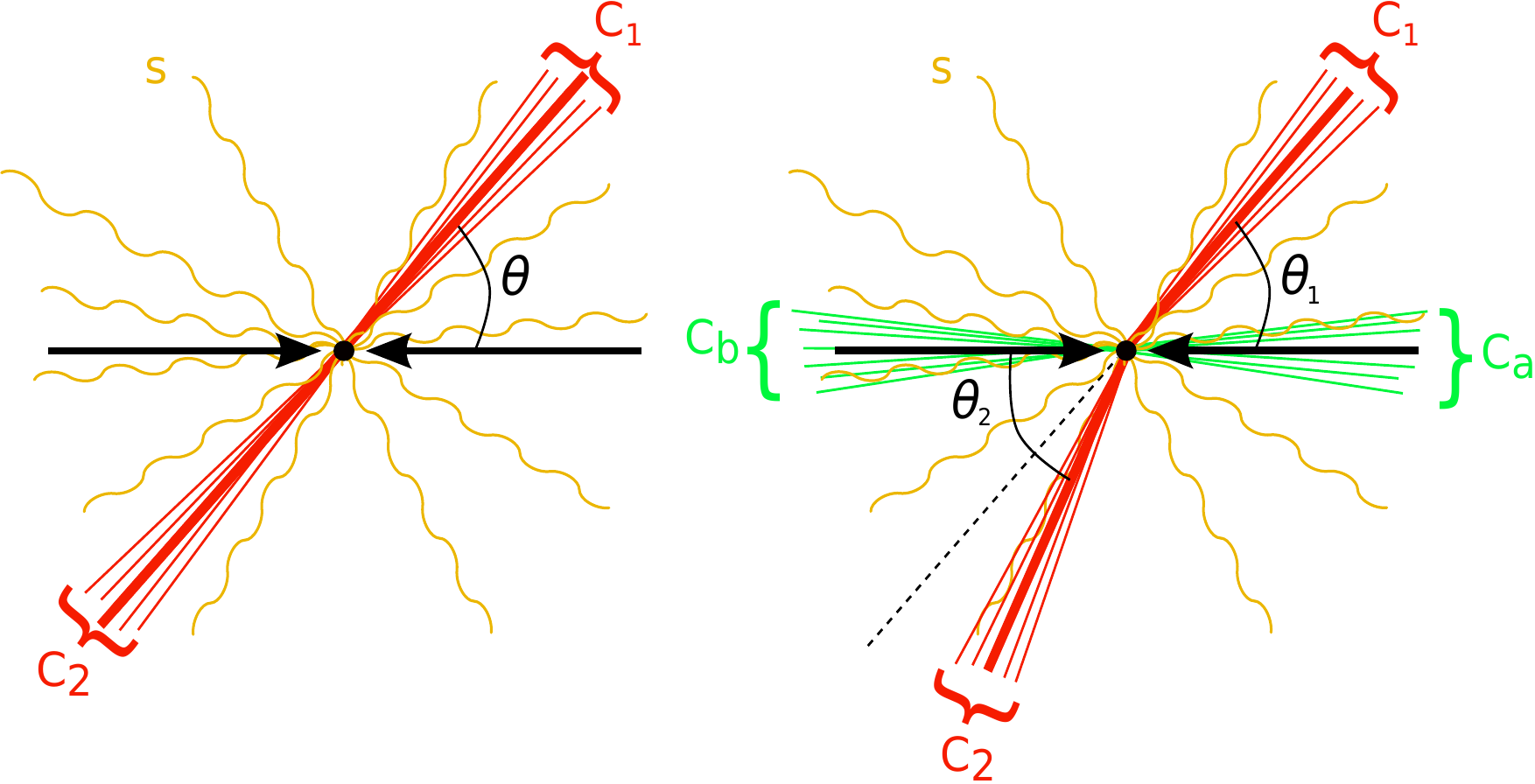}
\caption{Schematic representation of a dijet event in the $\TT\to 1$ limit for leptonic (left panel) and hadronic (right panel) collisions. The soft radiation $s$ and the collinear emissions $c_1$, $c_2$, $c_a$, $c_b$ are represented by different fields in the effective theory. The typical virtuality of the fields $c_a$, $c_b$ and $s$ is the same, and is lower than the virtuality of $c_1$ and $c_2$.}\label{fig:pic_coll}
\end{figure}

To derive the factorization theorem, we start with the lepton-collider case. There is no need to restrict oneself to the plane transverse to the beam in this case, but doing so provides us with a simpler environment to analyze the factorization of transverse observables. In contrast to standard thrust, $\TT\to 1$ does not imply that the event consists of two low-mass jets. Nevertheless, the terms which are enhanced in this limit do arise from two-jet configurations, such as the one shown on the left-hand side of figure~\ref{fig:pic_coll}, in which all the radiation is soft or collinear to the two low-mass jets. The resulting factorization formula has the same structure as the one for thrust. It involves a hard function which collects the virtual corrections to the hard scattering process, two jet functions describing the collinear emissions and a soft function. The energy of the soft emissions is parametrically lower than the typical mass of the jets.

In the hadron-collider case also the incoming partons carry color charge and the effective theory involves additional collinear fields which describe the initial-state radiation. The proton matrix element of these fields defines beam functions, which can be factorized into a perturbative kernel, describing the emissions, convolved with the standard parton distribution functions (PDFs). Interestingly, the virtuality of the initial-state collinear fields is parametrically of the same order as the one of the soft fields, and is lower than the virtuality of the collinear fields of the final-state jets. 
As is typical for problems which involve soft and collinear fields of the same virtuality, transverse thrust suffers from a collinear anomaly: the soft and beam functions are not well defined individually and their product involves large logarithms associated with the large rapidity difference between the emissions from the two incoming particles \cite{Becher:2010tm}. To compute the beam and soft functions individually, one needs to introduce an additional regulator, which can be removed after combining the functions. Traditionally, this regularization was achieved by taking the Wilson lines describing soft and collinear emissions in these functions off the light-cone, see e.g.\ \cite{Collins:1984kg}. However, in an effective theory context, it is more convenient to use an analytic regulator which does not introduce additional scales into the problem. 
 The cancellation of the divergences in the additional regulator imposes constraints on the form of the large logarithms generated by the collinear anomaly. These constraints are particularly interesting in our case due to the nontrivial color structure and angular dependence of the soft function for transverse thrust. The fact that the problem involves nontrivial color structure, collinear fields at different virtualities and a collinear anomaly illustrates that factorization for transverse thrust is quite nontrivial.

The resummation of large logarithms is achieved by solving the renormalization group (RG) equations of the ingredients in Laplace space. Transforming back to momentum space, we provide an analytic form of the resummed partonic cross section. Towards the goal of achieving next-to-next-to-leading logarithmic (N$^2$LL) accuracy we evaluate all the constituents of the theorem at one-loop accuracy. The other ingredients for N$^2$LL resummation are the two-loop anomalous dimensions and the two-loop anomaly coefficient. Using factorization constraints, we show that the only unknown quantities are three two-loop coefficients. We determine one of these coefficients numerically by comparing to the next-to-next-to-leading order (N$^2$LO) fixed-order result for transverse thrust in leptonic collisions. This also provides a numerical check on our factorization formula. We then show that also the remaining coefficients can be extracted numerically, by considering transverse thrust in Drell-Yan and Higgs production events, and using existing N$^2$LO fixed-order codes for Higgs and Drell-Yan production. This determination, together with a numerical implementation and phenomenological analysis of the resummed cross section, will performed in a future publication.

Our paper is organized as follows: We derive the factorization formula for the transverse-thrust differential distribution in section~\ref{sec:facform}  and compute its ingredients at the one-loop level in section~\ref{sec:onelin}. Solving the associated RG equations, we derive in section~\ref{sec:resum} resummed expressions for the cross section in the dijet limit.  In section~\ref{sec:N2LLlep}, we first compare our resummed results for the lepton-collider case with the fixed-order computation. This allows us to determine the two-loop anomalous dimensions, so that we have all the ingredients for N$^2$LL resummation in the lepton-collider case. We then show that a similar procedure can be used in the hadron-collider case. Our conclusions and an outlook on future work are presented in section~\ref{sec:concl}. The appendices collect anomalous dimensions, and provide details on the one-loop computations of the jet, soft and beam functions.

\section{Factorization formula}\label{sec:facform}

\begin{figure}
\centering
\begin{tabular}{ccc}
\includegraphics[width=0.45\textwidth]{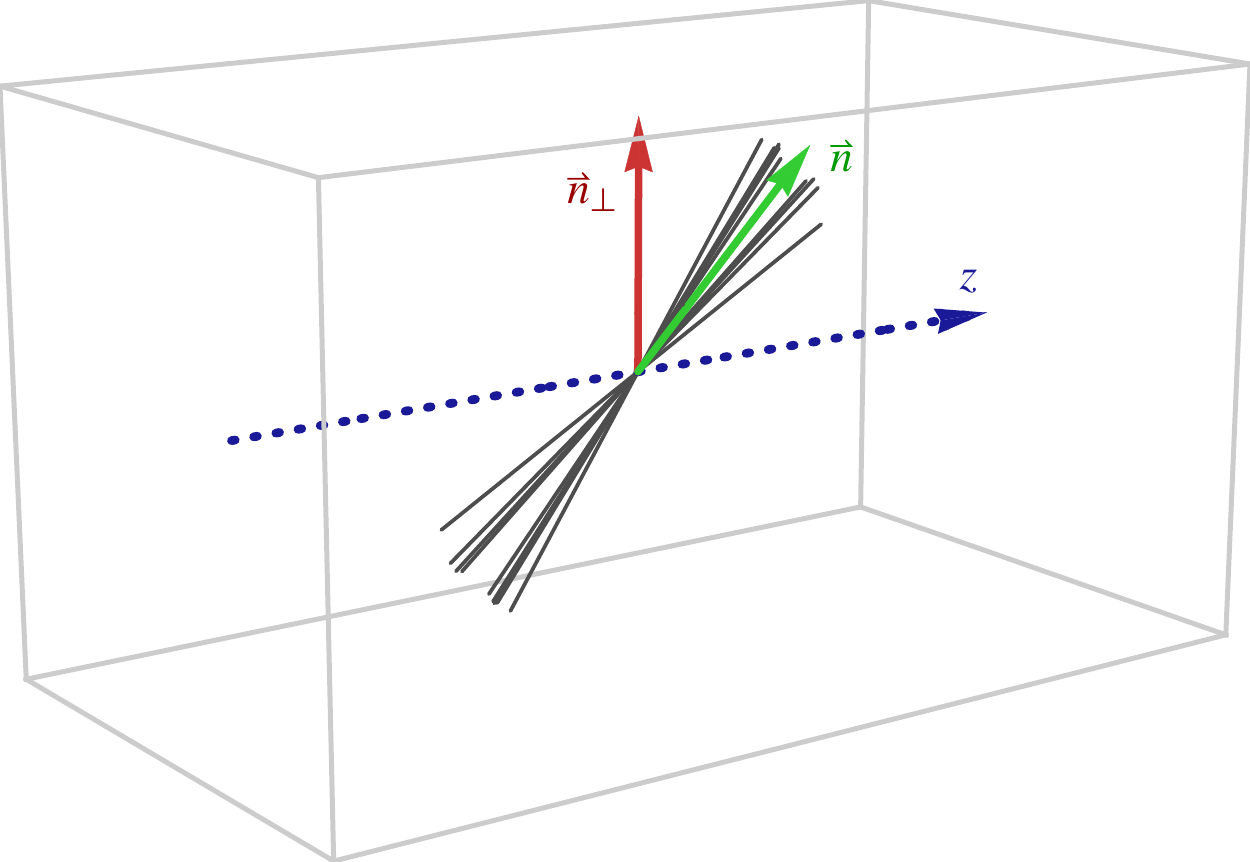} &\hspace{0.2\textwidth}& \includegraphics[width=0.45\textwidth]{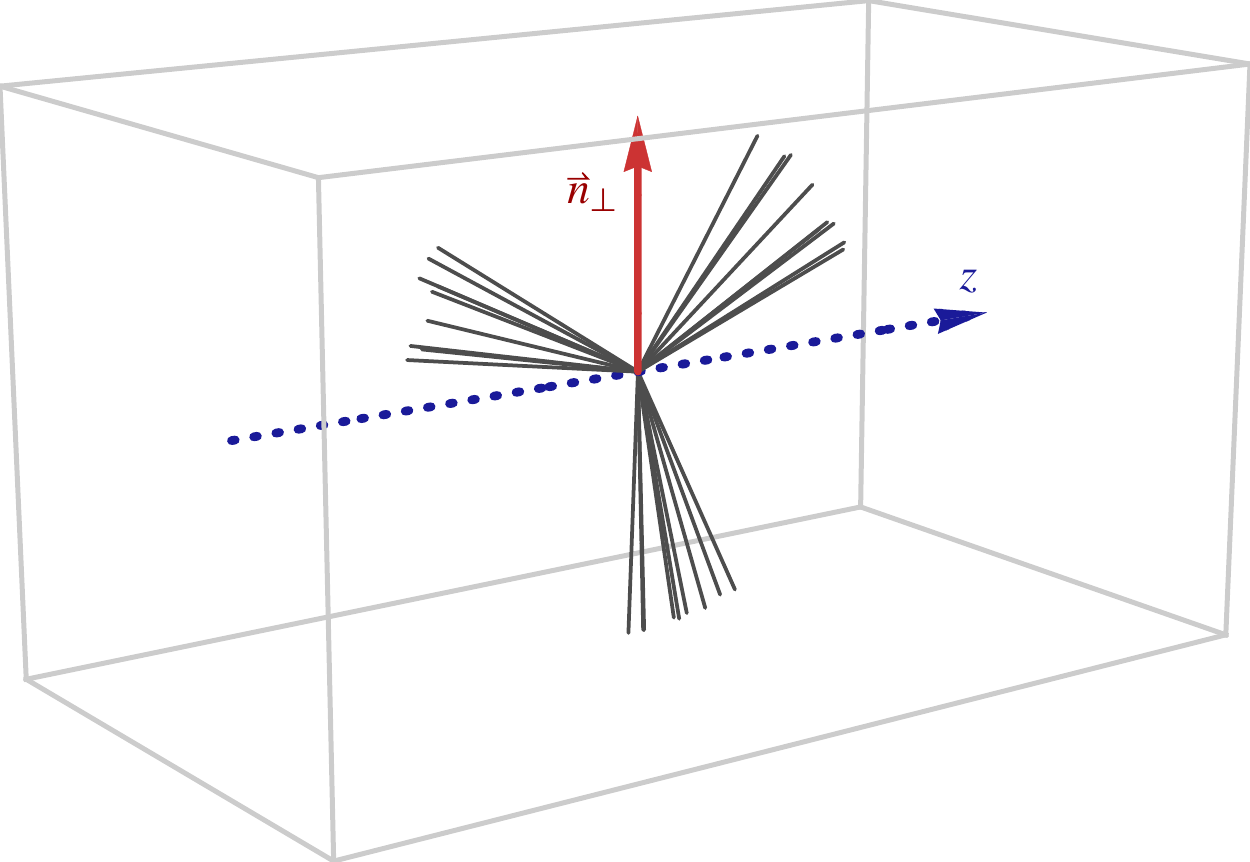}
\end{tabular}
\caption{Left: A two-jet configuration with low $\tT$. The figure shows the thrust axis $\vec{n}$ (green), the transverse thrust axis $\vec{n}_{\perp}$ (red), and the beam as a dashed line in the $z$-direction (blue). Right: A planar three-jet configuration with  $\tT=0$. Any distribution of particles that is restricted to a plane which contains the beam has $\tT=0$.}\label{fig:nonsingconf}
\end{figure}

Our goal is to derive a factorization formula for the transverse-thrust differential distribution that is valid in the dijet limit. If we define, as usual,
\begin{equation}
\tT:=1-\TT,
\end{equation}
the dijet limit, where the event contains two low-mass jets, corresponds to $\tT\to0$. 
However unlike the usual thrust case, the limit $\tT\to0$ contains not only dijet configurations but also configurations where all the particles lie in a plane which contains the beam, see figure~\ref{fig:nonsingconf}. The dijet configurations give singular perturbative contributions to the cross section at low $\tT$
\begin{equation}
\frac{d\sigma}{d\tT} \sim \frac{\alpha_s}{\tT}\,,
\end{equation}
while the multi-jet configurations are regular and thus power-suppressed at low $\tT$. In the following, we will study the singular terms in the limit $\tT\to0$ in detail, and resum their contribution to all orders in perturbation theory. The power-suppressed terms can be added by matching to fixed-order results. We will find that the matching corrections are larger for transverse than for regular thrust, the reason could be the presence of multi-jet configurations even at low $\tT$. 

In order to understand how one should treat event-shape variables that involve only momenta transverse to the beam direction in SCET, we will first consider leptonic collisions. At lepton colliders, there is not much experimental motivation to restrict event shapes to the transverse plane, but doing so provides us with a simplified environment to analyze factorization for such observables.  A typical dijet configuration in a leptonic collision is depicted in the left panel of figure~\ref{fig:pic_coll}. The incoming electron and positron are taken in the $z$ direction, and are represented by the black arrows in the figure. The jets are emitted at an angle $\theta$ with respect to the beam. As for the usual lepton-collider event shapes, the effective theory needs two collinear modes, which we call $c_1$ and $c_2$, to account for the energetic particles in the final state (represented by the red lines in figure~\ref{fig:pic_coll}), and a soft mode $s$ which describes final-state soft radiation (represented by the orange wavy lines in figure~\ref{fig:pic_coll}). We will perform the detailed factorization analysis below, but the astute reader will have guessed that the result will be a factorization formula of the form 
\begin{equation}
\frac{dσ}{dτ_{\perp}d\theta} \propto H(Q)\, \cdot (J_{c_2\perp}\otimes J_{c_1\perp} \otimes S_{\perp} )(\tau_\perp)\,.
\end{equation}
The hard function $H(Q)$ collects the virtual corrections to the hard-scattering process and is the same as for regular thrust. The jet and soft functions, on the other hand, differ from the standard case, because the phase-space constraints associated with $\tT$ only act in the transverse plane.

The lepton-collider case can be contrasted with the hadronic case depicted in the right panel of figure \ref{fig:pic_coll}. A complication on the kinematic level is that the partonic collisions, which produce the jets at hadron colliders, are not taking place in the hadronic center-of-mass frame and do not have fixed energy. Therefore the jets are not back-to-back and their energies are not fixed. To be able treat the process perturbatively, one needs to ensure that each event involves an underlying hard collision. One way to do this is to impose a minimum transverse momentum.

A complication for the theoretical description is that there is also initial-state radiation and the effective theory  includes two additional collinear modes $c_a$ and $c_b$ along the beam directions. The initial-state radiation is described by beam functions, which are proton matrix elements of these collinear fields. For perturbative values of $\tT$, these can be factorized into PDFs convolved with perturbative kernels describing the emissions. We will show below that the collinear modes $c_a$ and $c_b$ have a low virtuality, of the same size as the one of the soft fields, and that the convolution of the beam functions and the soft function suffers from a collinear anomaly \cite{Becher:2010tm}, i.e. the functions are not individually well-defined and their product has a logarithmic dependence on the hard scale. 

\subsection{Leptonic collisions\label{sec:lept}}

Let us now analyze the leptonic case in detail. In the dijet limit the final state $X$ consists of particles $X_{c_1}$ and $X_{c_2}$ with large energies flying along the jet direction, defined by the thrust axis $\vec{n}$, and soft particles $X_s$ which are radiated at arbitrary angles. Up to power suppressed terms, we can thus simplify the denominator in the transverse-thrust expression as
\begin{equation}\label{eq:denttlep}
\sum_{i\in c_1, c_2,s}
|\vec{p}_{i\perp}|= \sum_{i\in c_1, c_2,s}E_i|\sin\theta_i|
 \simeq |\sin\theta|\sum_{i\in c_1,c_2}E_i+\sum_{i\in s}|\sin\theta_i|E_i\simeq Q |\sin\theta|,
\end{equation}
where $E_i$ is the energy of the particle with 3-momentum $\vec{p}_i$, and $Q$ is the total center-of-mass energy. The sum over $i$ in eq.~(\ref{eq:denttlep}) runs over the three sectors in the effective theory. In the second step we used the approximation that the collinear particles fly approximately along the jet direction $\theta_{i}\simeq\theta$, and we have neglected the small contribution of the soft particles in the final step. In order for this approximation to be valid, the angle $\theta$ cannot be too small (the scaling of the fields derived below will yield the condition $\theta > \tT$, which is always satisfied in the dijet $\tT\to0$ limit).

For a final state $X$ in the dijet limit, we thus have
\begin{equation}\label{eq:tT}
\tT(X)=\frac{1}{Q|\sin\theta|}\sum_{i\in X} \left(|\vec{p}_{i\perp}|-|\vec{n}_{\perp}\cdot\vec{p}_{i\perp}|\right)=\tau_{\perp}(X_{c_1})+\tau_{\perp}(X_{c_2})+ \tau_{\perp}(X_s)\,.
\end{equation}
Eq.~(\ref{eq:tT}) is in a suitable form to derive a factorization theorem in SCET, because the sum over particles naturally separates into the different sectors of the effective theory, as indicated by the right-hand side of eq.~(\ref{eq:tT}). From eq.~(\ref{eq:tT}) one can also read off the relevant scaling of the different modes. To do so, let us first introduce the light-like vectors $n_i:=(1,\vec{n}_i)$ and $\bar{n}_i:=(1,-\vec{n}_i)$, where $\vec{n}_i$ is the direction of jet $i$. Therefore, in the lepton-collider case we are considering in this section, we have $n_1=\bar{n}_2$, $\bar{n}_1=n_2$, and the vector $\vec{n}_1=:\vec{n}$ is the thrust axis. Throughout the paper, we denote momentum components transverse to the thrust vector $\vec{n}$ with the subindex $\top$. Table~\ref{tab:notation} summarizes the notations we employ for the different axis and relevant directions.
\begin{table}
\centering
\begin{tabular}{cclc}
& direction & \phantom{1235}vector & perp.\ dir.\ \\
\toprule
\multirow{ 2}{*}{beams}&$a$ & $n_a^\mu=(1, \,0, \,0,\phantom{-}1)$ & \multirow{ 2}{*}{$\perp$} \\
 & $b$ & $n_b^\mu=(1, \,0, \,0,-1)$& \\ \midrule
\multirow{ 2}{*}{jets} & 1 & $n_1^\mu=(1,\phantom{-}\sin\theta_1,\, 0,\phantom{-}\cos\theta_1)$ & \multirow{ 2}{*}{$\top$}\\
& 2 & $n_2^\mu=(1,-\sin\theta_2,\,0,-\cos\theta_2)$ & \\ \midrule
thrust  & & $\vec{n}_{\phantom{\perp}}=(\sin\theta,\, 0, \cos\theta)$ & \multirow{ 1}{*}{$\top$} \\
transverse thrust  & & $\vec{n}_{\perp}=(\phantom{sin}1,\, 0, \phantom{cos} 0)$ & \\
\bottomrule
\end{tabular}
\caption{Summary of the notation used in the text for the relevant axes and directions. The last column shows the symbol we use to denote the momentum components perpendicular to the respective directions. We  use the notation $p^\mu_{\perptop}$ for the component of $p^\mu$ which is transverse to both axes. At $e^+e^-$ colliders one has $\theta_1=\theta_2=\theta$. At hadron colliders, the same is true in the center-of-mass frame of the underlying hard collision.\label{tab:notation}}
\end{table}
To analyze the scaling, we split a generic momentum $p^{\mu}$ into components along the jet and a remainder which is transverse to the jet:
\begin{equation}
p^{\mu}=(\bar{n}_1\cdot p)\frac{n_1^{\mu}}{2}+(n_1\cdot p)\frac{\bar{n}_1^{\mu}}{2}+p_{\top}^{\mu}=:p_{+}\frac{n_1^{\mu}}{2}+p_{-}\frac{\bar{n}_1^{\mu}}{2}+p_{\top}^{\mu} \,.
\end{equation}
Please note that we use the notation $p^\mu_{\top}$ and $\vec{p}_{\top}$ to indicate quantities transverse to the thrust axis. The notation $p^\mu_{\perp}$ is instead used to indicate quantities transverse to the beam axis and we use the notation $p^\mu_{\perptop}$ for quantities which are transverse to both axes. 
The contribution of a collinear particle to regular thrust is driven by the small light-cone component of its momentum, i.e.\ by $p_{-}= \vec{p}_{\top}^{\,2}/p_+$ for the $c_1$ particles. For transverse thrust, we will see that the relevant quantity is $\vec{p}_{\perptop}^{\,2}/p_+$ which scales in the same way. Eq.~(\ref{eq:tT}) then tells us that the components of the soft modes scale like the small components of the collinear modes. The components $(p_{+},p_{-},\vec{p}_{\top})$ of the different momenta therefore scale as
\begin{equation}
c_1: (1,\tT,\sqrt{\tT})Q\,,\quad c_2: (\tT,1,\sqrt{\tT})Q\,,\quad s:(\tT,\tT,\tT)Q\,,
\end{equation}
which is the same scaling that is relevant for ordinary thrust. The associated effective theory is usually called SCET$_{\mathrm I}$.  Given this scaling, one can further expand the contribution of the collinear particles to transverse thrust by using the fact that $ |\vec{p}_{i\perptop}| \ll |\vec{n}_{\perp}\cdot\vec{p}_{i\perp}|$, and write
\begin{align}\label{eq:collthr}
\tau_\perp(X_{c_{1,2}}) Q_{\perp}&=\sum_{i\in c_{1,2}}\left(|\vec{p}_{i\perp}|-|\vec{p}_i\cdot
  \vec{n}_{\perp}|\right)=\sum_{i\in c_{1,2}}\left(\sqrt{(\vec{p}_i\cdot \vec{n}_{\perp})^2+p_{i\perptop}^2}-|\vec{p}_i\cdot
  \vec{n}_{\perp}|\right) \nonumber\\
&\simeq \sum_{i\in c_{1,2}}\frac{1}{2}\frac{p_{i\perptop}^2}{|\vec{p}_i\cdot
  \vec{n}_{\perp}|}\simeq\sum_{i\in c_{1,2}}\frac{1}{2}\frac{p_{i\perptop}^2}{|\vec{p}_{i\perp}|} \simeq \frac{1}{|\sin\theta|}\sum_{i\in c_{1,2}}\frac{p^2_{i\perptop}}{2E_i}\, ,
\end{align}
where $Q_{\perp}:=Q|\sin\theta|$.

To put forward a factorization formula in SCET, we start from the
expression for the QCD cross section differential in $\tT$
\begin{equation}\label{eq:sigQCD}
\frac{dσ}{d\tT} =\frac{1}{2Q^2}\sum_X|\mathcal{M}(e^+e^-\to
X)|^2(2π)^4{\delta}^{(4)}(q-p_X){\delta}(\tT-\tT(X)),
\end{equation}
where $\tT(X)$ is given by eq.~(\ref{eq:tT}). At leading order in
the electroweak couplings, the matrix element squared can be written
as 
\begin{equation}
|\mathcal{M}(e^+e^-\to
X)|^2=\sum_{i=V,A}L^i_{μν}\left<0\right|j_i^{μ\dagger}(0)\left|X\right>\left<X\right|j_i^ν(0)\left|0\right>,
\end{equation}
where
\begin{equation}
j_i^μ(x)=\sum_{a,f}\bar{q}_f^a(x)\hat{Γ}_i^μq_f^a(x),
\end{equation}
are the vector ($V$) and axial ($A$) currents, with $\hat{Γ}_V^μ=γ^μ$, and
$\hat{Γ}_A^μ=γ^μγ^5$; $a$ is a color index, and $f$ denotes flavor. In
the following we will leave the sum over color and flavor implicit. The photonic contribution to the lepton tensor $L^i_{μν}$ is given by
\begin{equation}
L_{μν}^V  = 
-\frac{e^4}{2Q^2}\left(g_{μν}-2\frac{p_{1μ}p_{2ν}+p_{2μ}p_{1ν}}{Q^2}\right)Q_f^2\quad;\quad L_{μν}^A =  0,
\end{equation}
with $p_1$ and $p_2$ the electron and positron momenta,
respectively, $e$ the charge of the electron, and $Q_f$ the electric charge of fermion $f$.

The steps needed to obtain a factorized form of the cross section in eq.~(\ref{eq:sigQCD}) above are analogous to other SCET derivations that exist in the literature, the main difference being only that we want to leave the angle $\theta$ unintegrated.\footnote{For regular $e^+e^-$ event-shape variables, the dependence on $\theta$ has been considered in ref.~\cite{Mateu:2013gya}.} For this reason, we will just go through the main points of the derivation here; we refer to ref.~\cite{MScSilvan} for further details. The first step is to match the currents onto SCET operators; then we perform field redefinitions on the SCET fields to decouple the soft gluons. After the decoupling, the different sectors no longer interact and we can write the final state as
\begin{equation}
\left|X\right>=\left|X_s\right>\left|X_{c_1}\right>\left|X_{c_2}\right>\,
\end{equation}
and the transverse thrust constraint in the form
\begin{align}
δ(\tT-\tT (X))&=δ\left(\tT-τ_\perp(X_{s})-τ_\perp(X_{c_{1}})-τ_\perp(X_{c_{2}})\right) \nonumber\\ 
&=\int
dτ_{c_{1}\perp} dτ_{c_{2}\perp} dτ_{s\perp} \,δ(\tT-τ_{s\perp}-τ_{c_{1}\perp}-τ_{c_{2}\perp}) \nonumber \\
&\quad\quad \times δ(τ_{c_{1}\perp}-τ_\perp(X_{c_{1}}))
δ(τ_{c_{2}\perp}-τ_\perp(X_{c_{2}})) δ(τ_{s\perp}-τ_{\perp}(X_s))\,.
\end{align}
We then obtain the cross section as a convolution of matrix elements in the different sectors of the effective theory
\begin{align} 
\frac{dσ}{d\tT} =&\frac{1}{2Q^2}
\sum_{i=V,A}L^i_{μν} |\widetilde{C}(Q^2)|^2\,  \int
dτ_{c_1\perp} dτ_{c_2\perp} dτ_{s\perp} \,δ(\tT-τ_{s\perp}-τ_{c_1\perp}-τ_{c_2\perp}) \nonumber \\
& \times\left(\sum_{X_{c_2}}\left<0\right|\bar{χ}_{c_2,α}^{j}(0)\left|X_{c_2}\right>\left<X_{c_2}\right|χ_{c_2,β'}^{k'}(0)\left|0\right>δ(τ_{c_2\perp}-τ_\perp(X_{c_{2}}))\right) \nonumber \\
&\times \left(\sum_{X_{c_1}}\left<0\right|χ_{c_1,β}^{j'}(0)\left|X_{c_1}\right>\left<X_{c_1}\right|\bar{χ}_{c_1,α'}^{k}(0)\left|0\right>δ(τ_{c_1\perp}-τ_\perp(X_{c_{1}})) \right)  \nonumber \\
&
\times\left(\sum_{X_{s}}\Big<0\Big|\left[Y^{\dagger}_{c_2}(0)Y_{c_1} (0)\right]_{jj'}\Big|X_{s}\Big>\Big<X_{s}\Big|\left[Y^{\dagger}_{c_1}(0)Y_{c_2}(0)\right]_{kk'}\Big|0\Big>δ(τ_{s\perp}-τ_\perp(X_s)) \, \right) \nonumber \\
&\times  Γ_{i,αβ}^μΓ_{i,α'β'}^ν \,  (2π)^4 δ^{(4)}(q-p_{X_{c_1}}-p_{X_{c_2}}-p_{X_{s}})\,,
\label{eq:dsdtintrdelt}
\end{align}
where $\widetilde{C}$ is the Fourier transform of the matching coefficient from the QCD current to the SCET operators. We use latin indices to denote color and greek indices for the Dirac structure. The $\chi$'s are quark-jet fields in SCET, which include the quark coming out of the hard collision and its interactions with collinear particles. The $Y$'s are soft Wilson lines, which encode soft interactions at leading power. As a final step, we want to expand away small components in the momentum conservation $\delta$-function and make the dependence on the angle $\theta$ explicit. Up to power suppressed terms, we have
\begin{equation}\label{eq:deltarew}
δ^{(4)}(q-p_{X_{c_1}}-p_{X_{c_2}}-p_{X_{s}}) = 2\, δ(\bar{n}_1\cdot p_{X_{c_1}}-Q) δ(\bar{n}_2\cdot p_{X_{c_2}}-Q) δ^{(2)}(p_{X_{c_1}}^\top+p_{X_{c_2}}^\top)\,.
\end{equation}
To make the dependence on the angle $\theta$ explicit, we now explicitly distinguish the reference vector $\vec{n}$ in SCET from the thrust axis $\vec{n}_T$, which is derived from the particles in a given event. We then introduce
\begin{equation}
1 = \int \! d^3 \vec{n}\, δ^{(3)}(\vec{n} - \vec{n}_T),
\end{equation} 
into eq.~\eqref{eq:dsdtintrdelt}. In the effective theory, the thrust axis is given by $\vec{n}_T = \vec{p}_{X_{c_1}}/|\vec{p}_{X_{c_1}}|$ up to power corrections. Inserting this into the above equation and using the fact that  momentum conservation fixes $|\vec{p}_{X_{c_1}}|=Q/2$, we can rewrite it in the form
\begin{equation}
 \int \! d^3 \vec{n} \,δ^{(3)}(\vec{n} - \vec{n}_T) = (2\pi) \int d\cos\theta \left(\frac{Q}{2}\right)^2 
 δ^{(2)}(p_{X_{c_1}}^\top)\, .
\end{equation} 
After these manipulations, the momentum conservation δ functions only act on a single sector and the cross section factorizes into separate collinear and soft matrix elements. The collinear matrix elements define jet functions and can be written in the form
\begin{align}\label{eq:defjetf1}
\frac{δ^{jk}}{2(2π)^3}\left[\frac{n\slsh_{1}}{2}\right]_{βα}J_{c_1\perp}(τ_{c_1\perp})
&:=\sum_{X_{c_1}}\left<0\right|χ_{c_1,β}^{j}(0)\left|X_{c_1}\right>\left<X_{c_1}\right|\bar{χ}_{c_1,α}^{k}(0)\left|0\right> \nonumber \\
&\quad\quad\timesδ(τ_{c_1\perp}-τ_{\perp}(X_{c_1}))δ(Q-\bar{n}_{1}\cdot
p_{X_{c_1}})δ^{(2)}(p_{X_{c_1}\top})\,,  \\
\frac{δ^{jk}}{2(2π)^3}\left[\frac{n\slsh_{2}}{2}\right]_{βα}
J_{c_2\perp}(τ_{c_2\perp})
&:=\sum_{X_{c_2}}\left<0\right|\bar{χ}_{c_2,α}^{j}(0)\left|X_{c_2}\right>\left<X_{c_2}\right|χ_{c_2,β}^{k}(0)\left|0\right> \nonumber\\
&\quad\quad\times δ(τ_{c_2\perp}-τ_{\perp}(X_{c_2}))δ(Q-\bar{n}_{c_2}\cdot
p_{X_{c_2}})δ^{(2)}(p_{X_{c_2}\top})\,, \label{eq:defjetf2}
\end{align}
where the collinear-sector transverse-thrust constraint has been expanded according to eq.~\eqref{eq:collthr}. The soft matrix element has the form 
\begin{equation}\label{eq:defSoft}
S_{\perp}(τ_{s\perp}):=\frac{1}{N_c}\sum_{X_{s}}\Big<0\Big|\left[Y^{\dagger}_{c_2}(0)Y_{c_1} (0)\right]_{jk}\Big|X_{s}\Big>\Big<X_{s}\Big|\left[Y^{\dagger}_{c_1}(0)Y_{c_2}(0)\right]_{kj}\Big|0\Big>δ(τ_{s\perp}- τ_{\perp}(X_{s})).
\end{equation}
Expanding away power suppressed terms in the $\delta$-functions in eq.~(\ref{eq:dsdtintrdelt}), inserting the definitions of the jet and soft functions, and contracting the Dirac structure, we obtain the desired factorized expression for the cross section
\begin{align}\label{eq:factttSCET}
\frac{dσ}{dτ_{\perp}d\cos\theta}&=\frac{\pi N_c Q_f^2\alpha^2}{2Q^2}(1+\cos^2\theta)H(Q^2)\int
dτ_{c_{1}\perp}dτ_{c_{2}\perp}dτ_{s\perp} 
 \nonumber\\
&\hspace{1cm}  δ(τ_{\perp}-τ_{s\perp}-τ_{c_2\perp}-τ_{c_1\perp}) J_{c_2\perp}(τ_{c_{2}\perp})J_{c_1\perp}(τ_{c_1\perp})S_{\perp}(τ_{s\perp}),\end{align}
%% \[
%% =N(\theta)\int
%% d\bar{τ}_{c_1\perp}d\bar{τ}_{c_2\perp}d\bar{τ}_{s\perp}\bar{J}_{c_2\perp}(\bar{τ}_{c_2\perp})\bar{J}_{c_1\perp}(\bar{τ}_{c_1\perp})\bar{S}_{\perp}(\bar{τ}_{s\perp})
%% \]
%% \begin{equation}\label{eq:factttSCET}
%% \times\sin^2\theta\,δ\left(\sin^2\theta\,\tT-\frac{\bar{τ}_{c_2\perp}}{Q^2}-\frac{\bar{τ}_{c_1\perp}}{Q^2}-\frac{\bar{τ}_{s\perp}}{Q}\right),
%% \end{equation}
with $H(Q^2):=|\widetilde{C}(Q^2)|^2$, and where $\alpha=e^2/(4\pi)$ is the fine structure constant and $N_c$ the number of colors. To include also the contributions from $Z$ exchange, one should substitute $Q_f^2e^4/(2Q^2)\to (L^V+L^A)$, where $L^{V,A}$ are defined in appendix~\ref{app:lepttens}.

We already stressed above that the jet functions that appear in eq.~(\ref{eq:factttSCET}) are not the inclusive jet functions that one needs, for instance, in ordinary thrust. At lowest order, we have
\begin{equation}
S_{\perp}(τ_{s\perp})=δ(τ_{s\perp})\quad;\quad
J_{c\perp}(τ_{c\perp})=δ(τ_{c\perp})\quad ;\quad H(Q^2)=1.
\end{equation}
We compute all these functions at one loop in section~\ref{sec:onelin}.

\subsection{Hadronic collisions}\label{subsec:hadrcoll}
Having obtained the factorization formula in the lepton-collider case in the previous section, we now move to hadronic collisions, which is the actual situation of interest. The right panel of figure~\ref{fig:pic_coll} depicts a typical dijet configuration in a hadronic collision. The final-state jets do not need to be back-to-back in the lab frame, since the momentum fractions of the partons entering the hard interaction can be very different. To define the jet axes, one can use a jet algorithm. Since our treatment concerns events with two narrow energetic jets, any choice of the algorithm will lead to the same jet directions in the limit $\tT \to 0$, up to terms which are  power-suppressed in this limit. The jet algorithm provides us with the two angles $\theta_1$ and $\theta_2$ of the two energetic jets, as indicated in Figure \ref{fig:pic_coll}, and with their energies $E_{J_1}$ and $E_{J_2}$. Since the jets are massless, these angles are in one-to-one correspondence to the rapidities of the two jets. To ensure that we indeed deal with hard collisions, one needs to require that the two jets are hard. This can, for example, be achieved by imposing that
\begin{equation}
Q_\perp := |\sin\theta_1| E_{J_1} + |\sin\theta_2| E_{J_2} > Q_0\,,
\end{equation}
note that the definition of $Q_\perp$ now involves $\theta_{1,2}$ and $E_{J_{1,2}}$, as adequate for hadronic collisions. The scale $Q_0$ must be large enough so that the soft scale is still perturbative, $Q_\perp \tT \gg\Lambda_{\rm QCD}$. In the limit $\tT\to 0$, the two contributions to $Q_\perp$ correspond to the transverse momenta of the two, approximately massless jets and momentum conservation in the transverse plane requires that the two contributions to $Q_\perp$ must be equal. Below, we will perform a boost along the beam axis to the frame in which the jets are back-to-back and their energies are equal. In the limit $\tT\to 0$ this is simply the center-of-mass frame of the hard scattering after initial-state radiation.

The modes that we need in the effective theory include the collinear modes $c_{1,2}$ for the final-state jets and the soft mode $s$ that were already present in the lepton-collider case. On top of this, we need two additional collinear modes in the beam directions, which we denote by $c_{a,b}$ and describe the initial-state radiation. They are represented by the green lines in figure~\ref{fig:pic_coll}. To derive the desired factorization theorem, we start by writing the expression for the QCD cross section differential in $\tT$, in a proton-proton collision,
\begin{equation}\label{eq:QCDcshad}
\frac{dσ}{d\tT}=\frac{1}{2E_{CM}^2}\sum_X\left|\mathcal{M}\left(pp\to X\right)\right|^2(2π)^4δ^{(4)}\left(P_a+P_b-p_X\right)δ(\tT-\tT(X)) \theta(Q_\perp-Q_0),
\end{equation}
where $P_a$ and $P_b$ are the momenta of the protons, $p_X$ is the total final-state momentum, and $E_{CM}$ is the hadronic center-of-mass energy. 

Before moving to the factorization analysis, it is useful to set up the kinematics. The momenta of the protons are given by
\begin{equation}
P_a^\mu=E_{CM}\frac{n_a^μ}{2}\quad;\quad P_b^\mu=E_{CM}\frac{n_b^μ}{2},
\end{equation}
and we assume that the partons which produce the two jets carry fractions $x_a$ and $x_b$ of the proton momenta.\footnote{The quantities $x_a$ and $x_b$  are the momentum fractions after initial-state radiation, they are not equal to the momentum fractions inside the PDFs.}
Following ref.~\cite{Stewart:2009yx}, we write the total final-state collinear momenta in the beam directions as
\begin{eqnarray}
p_{X_{c_a}}^μ & = & (1-x_a)E_{CM}\frac{n_a^μ}{2}+b_{a\perp}^μ+b_{a-}\frac{\bar{n}_a^μ}{2}\,,\\
p_{X_{c_b}}^μ & = & (1-x_b)E_{CM}\frac{n_b^μ}{2}+b_{b\perp}^μ+b_{b-}\frac{\bar{n}_b^μ}{2}\,.
\end{eqnarray}
The first term is the proton remnant, the remainder arises because the leading parton radiates into the final state. The momenta of the partons that enter the hard interaction are
\begin{eqnarray}
p_a^μ & = & x_aE_{CM}\frac{n_a^μ}{2}-b_{a\perp}^μ-b_{a-}\frac{\bar{n}_a^μ}{2},\\
p_b^μ & = & x_bE_{CM}\frac{n_b^μ}{2}-b_{b\perp}^μ-b_{b-}\frac{\bar{n}_b^μ}{2}\,.
\end{eqnarray}
The total final-state momentum $p_X$ is given by
\begin{equation}
p_X=p_{X_s}+p_{X_{c_1}}+p_{X_{c_2}}+p_{X_{c_a}}+p_{X_{c_b}} \,,
\end{equation}
and momentum conservation $P_a+P_b = p_X$ then implies the partonic relation 
\begin{equation}\label{eq:momcons}
 p_a+p_b=p_{X_s}+p_{X_{c_1}}+p_{X_{c_2}}.
\end{equation}

\begin{figure}[t!]
\centering
\includegraphics[width=2.5cm]{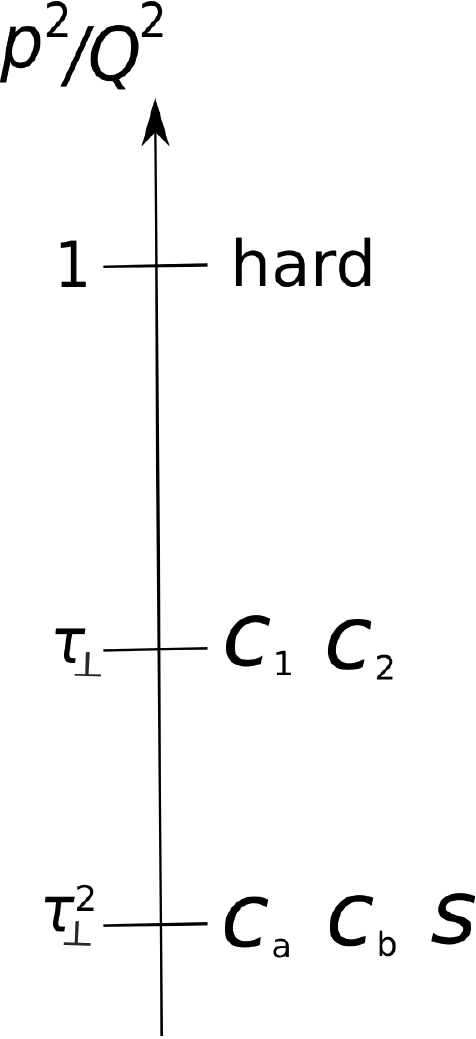}
\caption{Virtualities of the different modes present in the hadron-collider case.}\label{fig:virtmodes}
\end{figure}

As in the lepton-collider case, we can simplify the denominator in the expression for transverse thrust by dropping power-suppressed contributions
\begin{align}
\sum_i
|\vec{p}_{i\perp}| &=\sum_iE_i|\sin\theta_i| 
%\nonumber\\ & 
\simeq |\sin\theta_1|\sum_{i=c_1}E_i+|\sin\theta_2|\sum_{i=c_2}E_i+\sum_{i=s,c_a,c_b}E_i\sin\theta_i\,, \nonumber\\
& \simeq |\sin\theta_1|\, E_{J_1}+|\sin\theta_2|\, E_{J_2}  = Q_\perp.
\label{eq:dentthad}
\end{align}
We see that the denominator reduces to $Q_\perp$ in the dijet limit. To obtain this result, we have used that in the $c_{1,2}$ sectors the angles between each particle and the beam are approximately equal to the jet direction. The contribution from the soft sector is negligible, as in the leptonic case, and also the $c_{a,b}$ sectors do not contribute to eq.~(\ref{eq:dentthad}) at leading power, since $\sin\theta_{i}\simeq0$ for particles collinear to the beam. As in the leptonic case, we thus have
\begin{equation}\label{eq:tThad}
Q_\perp \tT(X)=\sum_i\left(|\vec{p}_{i\perp}|-|\vec{n}_{\perp}\cdot\vec{p}_{i\perp}|\right)\,,\end{equation}
and the particle sum separates into sums in the different sectors of the theory
\begin{equation}\label{eq:tThadsep}
\tT(X)= \tT(X_{s})+\tT(X_{c_1})+\tT(X_{c_2})+\tT(X_{c_a})+\tT(X_{c_b})\, .
\end{equation}
From this, we can read off the relevant scaling of the different modes. Obviously, the $c_{1,2}$ and $s$ modes have the same scaling as in the lepton-collider case. In order to contribute, the transverse components $b_{a,b\perp}^μ$ of the $c_{a,b}$ modes need to scale like $\tT$, and therefore the small components of these modes scale like $\tT^2$. We are thus in a situation where the collinear modes $c_{a,b}$ and the soft mode $s$ have the same virtuality, which is usually called a SCET$_{\mathrm{II}}$  problem. The virtualities of all the different modes are summarised in figure~\ref{fig:virtmodes}. Since the collinear fields in the jets have a large virtuality, the study of transverse thrust in hadronic collisions thus involves, quite uniquely, SCET$_{\rm I}$ and SCET$_{\rm II}$ together in the same problem. Since we have collinear and soft modes with virtualities that are parametrically of the same order, we should expect a collinear factorization anomaly in the effective theory~\cite{Becher:2010tm}. Below, we will find that such an anomaly is indeed present.

\begin{figure}[t!]
\centering
\includegraphics[width=3cm]{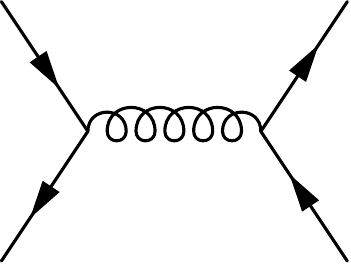}
\caption{Feynman diagram for the Born level $q\bar{q}\to q'\bar{q}'$ channel.}\label{fig:qqtoqpqp}
\end{figure}

There are many different partonic channels that contribute to the cross section since all four partons involved in the hard collision can be quarks or gluons. The different hard-scattering channels correspond to different leading-power operators in the effective theory. These operators are built from collinear quark fields $\chi_i$ and collinear gluon fields ${\cal A}^{\top}_i$, and involve one field for each direction. The corresponding operators and their one-loop Wilson coefficients were given in~\cite{Kelley:2010fn}, and recently these results were extended to two-loop order in \cite{Broggio:2014hoa}. To keep the following discussion simple, we focus on the hard process $q\bar{q}\to q'\bar{q}'$ where the outgoing quarks have a different flavor than the incoming ones. For this case a basis of the relevant SCET operators is given by
\begin{align}\label{eq:basis}
O_1(x; s,t,u,v) & =  \bar{\chi}_b(x+ t \bar{n}_b) \gamma_{\mu}t^A \chi_a(x+s \bar{n}_a)\, \bar{\chi}_2(x+v \bar{n}_2) \gamma^\mu t^A\chi_1(x+u \bar{n}_1) \,, \nonumber\\
O_2(x; s,t,u,v) & =  \bar{\chi}_b(x+ t \bar{n}_b) \gamma_{\mu} \chi_a(x+s \bar{n}_a)\, \bar{\chi}_2(x+v \bar{n}_2) \gamma^\mu \chi_1(x+u \bar{n}_1)\,.
\end{align}
As usual, the operators are smeared over the light-cone direction conjugate to the large momentum flow and their contribution to the effective Lagrangian is obtained after convolution with the Wilson coefficients:
\begin{equation}\label{eq:matchophad}
\Delta {\cal L}_{\rm SCET} = \int \!d^4x \int \! ds\, dt \,du \,dv\, C_I (s,t,u,v)\,O_I(x; s,t,u,v) \,,\end{equation}
where the sum over the different operators $I$ is implied.
The Fourier transforms of the Wilson coefficients $\tilde C_I (\bar{n}_a\cdot p_a, \bar{n}_b\cdot p_b, \bar{n}_1\cdot p_1,\bar{n}_2\cdot p_2)$ depend on the large momentum components of the collinear fields and are directly related to the relevant scattering amplitudes~\cite{Becher:2009qa}; for brevity we will not write the arguments explicitly in the following. In our channel only the single diagram shown in Fig.~\ref{fig:qqtoqpqp} contributes to the coefficients at leading order. 

Due to its color structure, it only generates a nonzero Wilson coefficient $C_{1}$, but at higher orders also the second operator will be present. Under renormalization the two operators mix, so that the RG equation becomes matrix-valued. As usual, one can perform the necessary field redefinitions in SCET to achieve the decoupling of the soft gluons and finds that the resulting soft function is a matrix in color space. After the decoupling, the final state is decomposed as
\begin{equation}
\left|X\right>=\left|X_s\right>\left|X_{c_1}\right>\left|X_{c_2}\right>\left|X_{c_a}\right>\left|X_{c_b}\right>,
\end{equation}
and using manipulations analogous to the lepton-collider case one obtains
\begin{align}
\frac{dσ}{d\tT}=\frac{1}{2E_{CM}^2}&\int
dτ_{c_1\perp} dτ_{c_2\perp}dτ_{c_a\perp}dτ_{c_b\perp} dτ_{s\perp} δ(\tT-τ_{s\perp}-τ_{c_1\perp}-τ_{c_2\perp}-τ_{c_a\perp}-τ_{c_b\perp}) \nonumber\\
&\times\left(\sum_{X_{c_1}}\left<0\right|χ_{c_1,δ}\left|X_{c_1}\right>\left<X_{c_1}\right|\bar{χ}_{c_1,δ'} \left|0\right>δ(τ_{c_1\perp}-τ_{\perp}(X_{c_1}))\right) \nonumber\\
&\times\left(\sum_{X_{c_2}}\left<0\right|\bar{χ}_{c_2,γ} \left|X_{c_2}\right>\left<X_{c_2}\right|χ_{c_2,γ'} \left|0\right>δ(τ_{c_2\perp}-τ_{\perp}(X_{c_2}))\right) \nonumber\\
&\times\left(\sum_{X_{c_a}}\left<P_a\right|\bar{χ}_{c_a,α} \left|X_{c_a}\right>\left<X_{c_a}\right|χ_{c_a,α'} \left|P_a\right>δ(τ_{c_a\perp}-τ_{\perp}(X_{c_a}))\right) \nonumber\\
&\times\left(\sum_{X_{c_b}}\left<P_b\right|χ_{c_b,β} \left|X_{c_b}\right>\left<X_{c_b}\right|\bar{χ}_{c_b,β'} \left|P_b\right>δ(τ_{c_b\perp}-τ_{\perp}(X_{c_b}))\right) \widetilde{C}_I\widetilde{C}_J^* \nonumber\\
& \times  \frac{1}{N_c^3}\mathcal{S}_{JI}(τ_{s\perp})  γ^μ_{αβ}γ^μ_{γδ}γ^ν_{δ'γ'}γ^ν_{β'α'} (2π)^4 δ^{(4)}(P_a+P_b-p_X) \, \theta(Q_\perp-Q_0). \label{eq:cshadintstep}
\end{align}
In writing the above formula, we have made use of the fact that all four collinear matrix elements are color-diagonal and have replaced
\begin{equation}
\bar{χ}_{c_m,α}^i χ_{c_m,α'}^j \to \frac{1}{N_c} \delta^{ij} \bar{χ}_{c_m,α} χ_{c_m,α'}\,,
\end{equation}
where it is understood that the color indices of the fields on the right-hand side are contracted. The resulting color contractions then act on the soft Wilson lines and produce the soft function 
\[
\mathcal{S}_{IJ}(τ_{s\perp}):=\frac{1}{N_c} \sum_{X_s}\left<0\right|\left[Y^{\dagger}_{c_a}T_{I}Y_{c_b}\right]_{ij}\left[Y^{\dagger}_{c_2}T_{I}Y_{c_1}\right]_{kl}\left|X_s\right>\left<X_s\right|\left[Y^{\dagger}_{c_1}T_{J}Y_{c_2}\right]_{lk}\left[Y^{\dagger}_{c_b}T_{J}Y_{c_a}\right]_{ji}\left|0\right>
\]
\begin{equation}\label{eq:softhadqq}
\times δ(τ_{s\perp}-\tau_{\perp}(X_s)),
\end{equation}
which, as anticipated, is a matrix in the color indices $I$ and $J$ . For the $q\bar{q}$-channel we are considering here, there are two possible color structures $\bm{T}_1=t^A$ and $\bm{T}_2=\mathbb{I}$, see eq.~\eqref{eq:basis}, and the soft function is therefore a $2\times2$ matrix.

In a next step, we again expand out the power-suppressed components in the momentum conservation $\delta$-function. After this step, only the large components of the four different collinear fields remain in the $\delta$-function. We already anticipated this step in our definition of the soft function in eq.~\eqref{eq:softhadqq}, which was defined without a constraint on the soft momenta other than their thrust. Next, we take the matrix elements of the collinear fields along the beam which then define quark beam functions $\mathcal{B}_{q/c_i\perp}(τ_{c_i\perp},x_i)$, according to
\begin{align}
\frac{1}{2}\left[\frac{n\slsh_{a}}{2}\right]_{α'α}\mathcal{B}_{q/c_a\perp}(τ_{c_a\perp},x_a) &:= \sum_{X_{c_a}}  δ(τ_{c_a\perp}-τ_{\perp}(X_{c_a}))  δ\left(\bar{n}_a \cdot P_a \,(1-x_a)-\bar{n}_a\cdot p_{X_{c_a}}\right)  \nonumber \\ 
&\hspace{2cm} \times \left<P_a\right|\bar{χ}_{c_a,α}(0)\left|X_{c_a}\right>\left<X_{c_a}\right|χ_{c_a,α'}(0)\left|P_a\right> \,, \nonumber \\
\frac{1}{2}\left[\frac{n\slsh_{b}}{2}\right]_{ββ'}\mathcal{B}_{\bar{q}/c_b\perp}(τ_{c_b\perp},x_b) & := 
\sum_{X_{c_b}}δ(τ_{c_b\perp}-τ_{\perp}(X_{c_b})) δ\left(\bar{n}_b \cdot P_b \,(1-x_b)-\bar{n}_b\cdot p_{X_{c_b}}\right) \nonumber \\
&\hspace{2cm} \times \left<P_b\right|χ_{c_b,β}(0)\left|X_{c_b}\right>\left<X_{c_b}\right|\bar{χ}_{c_b,β'}(0)\left|P_b\right>\, , \label{eq:beamquark}
\end{align}
where $\bar{n}_a \cdot P_a = \bar{n}_b \cdot P_b = E_{CM}$.  We can introduce the integrations over the momentum fractions, $x_{a,b}$, in the expression for the cross section by writing
\begin{equation}
1=(\bar{n}_a\cdot P_a)(\bar{n}_b\cdot P_b)\int \! dx_adx_b \,\delta(\bar{n}_a\cdot P_a(1-x_a)-\bar{n}_a\cdot p_{X_{c_a}})\delta(\bar{n}_b\cdot P_b(1-x_b)-\bar{n}_b\cdot p_{X_{c_b}}).
\end{equation}
We then have
\begin{align}
\frac{dσ}{d\tT}=\frac{1}{4N_c^3}&\int\!
dτ_{c_1\perp} dτ_{c_2\perp}dτ_{c_a\perp}dτ_{c_b\perp} dτ_{s\perp} δ(\tT-τ_{s\perp}-τ_{c_1\perp}-τ_{c_2\perp}-τ_{c_a\perp}-τ_{c_b\perp}) \nonumber\\
&\times \int \!dx_a \int \! dx_b\, \mathcal{B}_{q/c_a\perp}(τ_{c_a\perp},x_a) \mathcal{B}_{\bar{q}/c_b\perp}(τ_{c_b\perp},x_b) \left(-g_{\mu\nu}^\perp\right) γ^μ_{γδ}γ^ν_{δ'γ'}  \nonumber\\
&\times\left(\sum_{X_{c_1}}\left<0\right|χ_{c_1,δ}\left|X_{c_1}\right>\left<X_{c_1}\right|\bar{χ}_{c_1,δ'} \left|0\right>δ(τ_{c_1\perp}-τ_{\perp}(X_{c_1}))\right) \nonumber\\
&\times\left(\sum_{X_{c_2}}\left<0\right|\bar{χ}_{c_2,γ} \left|X_{c_2}\right>\left<X_{c_2}\right|χ_{c_2,γ'} \left|0\right>δ(τ_{c_2\perp}-τ_{\perp}(X_{c_2}))\right) \widetilde{C}_I\widetilde{C}_J^*  \,\mathcal{S}_{JI}(τ_{s\perp})\nonumber\\
& \times  (2π)^4 δ^{(4)}(x_a P_a+x_b P_b-p_{X_s}-p_{X_{c_1}}-p_{X_{c_2}}) \, \theta(Q_\perp-Q_0),
\end{align}
where $g^{\perp}_{μν}=g_{μν}-\frac{n_a^μn_b^ν+n_b^μn_a^ν}{2}$. At this point, the cross section has the form which is usually obtained in perturbative QCD, except that the initial state is described by beam functions instead of PDFs. To make the following discussion similar to the lepton collider case, it is convenient to now perform a boost to the frame where the jets are back-to-back, i.e. the frame where $\theta_1=\theta_2$. Up to terms suppressed by powers of $\tT$, the total momentum of the partons that enter the hard interaction in the original lab frame is given by
\begin{equation}
q^μ=x_a P_a^μ+x_b P_b^μ =x_aE_{CM}\frac{n_a^μ}{2}+x_bE_{CM}\frac{\bar{n}_a^μ}{2},
\end{equation}
where we used that $n_b=\bar{n}_a$. At leading power a boost along the $z$-direction is thus sufficient to make the jets back-to-back. We need to perform a boost such that the total momentum of the partons that enter the hard interaction is given by
\begin{equation}
q^μ\rightarrow\hat{q}^μ=Q\frac{n_a^μ}{2}+Q\frac{\bar{n}_a^μ}{2}+\mathcal{O}(Q\tT),
\end{equation}
where we denote quantities in the boosted frame with a hat. The plus and minus components of the $c_a$ particles transform under a boost in the $z$ direction according to $\bar{n}_a\cdot p\to e^ξ\bar{n}_a\cdot p$, $n_a\cdot p\to e^{-ξ}n_a\cdot p$, therefore the boost we need is given by
\begin{equation}\label{eq:boostpar}
ξ=\frac{1}{2}\ln\frac{x_b}{x_a}\quad;\quad Q=E_{CM}\sqrt{x_ax_b}.
\end{equation}
Since the constraint for transverse thrust only involves momenta transverse to the beam direction, the expression for the cross section in eq.~(\ref{eq:QCDcshad}) is invariant under the boost. The beam functions are invariant and can be evaluated in the lab frame, while we will adopt the boosted frame for the jet and soft functions. From now on we will exclusively work in the boosted frame, and for simplicity drop the hats on the momenta. We then rewrite the momentum conservation $\delta$-function as in eq.~\eqref{eq:deltarew} and, also as in the lepton case, introduce an integration over the angle $\theta$ of the jet in the boosted frame by writing
\begin{equation}
1=  \int \! d^3 \vec{n} \,δ^{(3)}(\vec{n} - \vec{n}_1) = (2\pi) \int d\cos\theta \left(\frac{Q}{2}\right)^2 
 δ^{(2)}(p_{X_{c_1}}^\top)\, .
\end{equation} 
Finally, we take the collinear matrix elements, which are the jet functions defined in eqs.~(\ref{eq:defjetf1})-(\ref{eq:defjetf2}), where $Q$ is now given by eq.~(\ref{eq:boostpar}). Putting all the ingredients together we obtain
\begin{align}
\frac{d\sigma}{d\tT d \cos\theta}=&\int dτ_{c_{1}\perp}dτ_{c_{2}\perp}dτ_{s\perp}dτ_{c_{a}\perp}dτ_{c_{b}\perp}
δ(τ_{\perp}-τ_{s\perp}-τ_{c_2\perp}-τ_{c_1\perp}-τ_{c_b\perp}-τ_{c_a\perp})\nonumber \\
&\times \int \!dx_a \int \! dx_b\, \mathcal{B}_{q/c_a\perp}(τ_{c_a\perp},x_a)\mathcal{B}_{\bar{q}/c_b\perp}(τ_{c_b\perp},x_b) J_{c_2\perp}(τ_{c_{2}\perp})J_{c_1\perp}(τ_{c_1\perp})\nonumber \\
&\times H_{IJ}(Q,\theta)  \, 
 \mathcal{S}_{JI\perp}(τ_{s\perp})\, \theta(Q_\perp-Q_0)\,,
\label{eq:naivefactqq}
\end{align}
where we defined the hard function as
\begin{equation}\label{eq:hardfunHad}
H_{IJ}(Q,\theta) = \frac{Q^2}{32\pi N_c}(1+\cos^2\theta)\widetilde{C}_I\widetilde{C}_J^*\, .
\end{equation}
Note that, unlike in the lepton-collider case above, we have not normalized it such that the entries of the matrix are only ones and zeros at tree level. The angle $\theta$, which corresponds to the angle between jet 1 and the beam in the partonic center-of-mass frame, is typically not measured experimentally, but it is convenient to keep the expressions differential in $\cos\theta$ for the present discussion. 

We can verify that we reproduce the lowest-order cross section for two-jet production by evaluating the ingredients at leading order. The LO hard coefficient is $\widetilde{C}_I = g_s^2\,(1, 0)/Q^2$. Its denominator $Q^2$ is from the hard gluon propagator which is integrated out when matching onto SCET. The relevant entry of the soft-function matrix is $\mathcal{S}_{11\perp}(τ_{s\perp}) = \frac{C_F}{2} δ(τ_{s\perp})$. The LO jet functions are equal to $δ(\tau_{c_i\perp})$ and the beam functions are given by the standard PDFs multiplied by the same  $δ$-function. Putting the ingredients together, and integrating over $\tT$, one reproduces the standard LO partonic cross section for $q\bar{q} \to q'\bar{q}'$.

\subsection{Collinear anomaly}\label{subsec:collan}

For leptonic collisions, the  theorem eq.~\eqref{eq:factttSCET} achieves the complete factorization of scales we are aiming for, but the same is not true for the hadron collider formula eq.~(\ref{eq:naivefactqq}). The reason is that the beam and soft functions, as given above, are not individually well defined. Beyond leading order, one encounters phase-space integrals that are ill-defined within dimensional regularization. To properly define them, one needs to introduce an additional regulator. The convolution of the regularized beam and soft functions is then regulator independent, as it needs to be since the cross section is a physical observable, but contains an additional dependence on the hard scale, on top of the one encoded in the hard function. This effect, i.e.\ the appearance of a hidden additional dependence on the hard scale, is called the collinear anomaly~\cite{Becher:2010tm}. One encounters it when the effective theory contains collinear and soft modes which have virtualities that are parametrically of the same order. Consistency conditions restrict the form of this additional hard-scale dependence to be a pure power to all orders in perturbation theory~\cite{Chiu:2007dg,Becher:2010tm}. The additional dependence can be factorized and in this way one achieves the desired separation of scales. One can show that for transverse-momentum dependent quantities, such as the one we consider here, only real-emission diagrams need additional regularization and one can use a simple analytic regulator to render them well defined~\cite{Becher:2011dz}. In this way, gauge invariance and the eikonal structure of the soft and collinear emissions in the effective theory are explicitly maintained.  Alternatively, one can regularize the soft and collinear Wilson lines \cite{Chiu:2011qc}.

To discuss the form of the collinear anomaly it is convenient to perform a Laplace transform of the cross section, since the Laplace-transformed expression is a simple product, rather than a convolution, of the jet, beam, and soft functions. We write
\begin{align}
\widetilde{t}(\kappa)&=\int d\tT e^{-\tT z}\left(\frac{d\sigma}{d\tT d(\cos\theta)dx_adx_b}\right)
\nonumber\\
& = H_{IJ}(Q,\theta)  \,\widetilde{\mathcal{S}}_{JI\perp}(\kappa,\theta)\,\widetilde{j}_{c_2\perp}(\kappa)\,\widetilde{j}_{c_1\perp}(\kappa)\,\widetilde{\mathcal{B}}_{q/c_a\perp}(\kappa,x_a)\,\widetilde{\mathcal{B}}_{\bar{q}/c_b\perp}(\kappa,x_b), \label{eq:LTcs}
\end{align}
where $z=1/(e^{\gamma_E}\kappa)$. The factor $e^{\gamma_E}$ is included to avoid a proliferation of such factors in the Laplace transforms of the jet, soft, and beam functions, which are indicated by a tilde. For example, for the jet function we define
\begin{equation}\label{eq:Lapltrj}
\widetilde{j}_{c_1\perp}(\kappa)=\int d\tau_{\perp} e^{-\tau_{\perp} z }J_{c_1\perp}(\tau_{\perp}),
\end{equation}
and analogously for the other ingredients. For later convenience we also define the dimensionful variable $\bar{\kappa}=2\kappa\, Q\,\sin\theta$, which is of the order of the soft energy scale. As mentioned before, the soft and beam functions in eq.~(\ref{eq:LTcs}) depend on the analytic regulator, and contain hidden dependence on the hard scale. Their product is regulator independent and, to all orders in perturbation theory, has the form
\begin{align}\label{eq:SBBtoFW}
\widetilde{\mathcal{S}}_{JI\perp}(\kappa,\theta)\,\widetilde{\mathcal{B}}_{q/c_a\perp}(\kappa,x_a)\,\widetilde{\mathcal{B}}_{\bar{q}/c_b\perp}(\kappa,x_b)& =\left(\frac{Q^2}{c_0^2\bar{\kappa}^2}\right)^{-F_{\perp}^{q\bar{q}}(\kappa)} \widetilde{W}_{JI}(\kappa,\theta,x_a,x_b)\,, 
\end{align}
where
\begin{align}
\widetilde{W}_{JI}(\kappa,\theta,x_a,x_b)&=\widetilde{S}_{JI\perp}(\kappa,\theta)\,\widetilde{B}_{q/c_a\perp}(\kappa,x_a)\,\widetilde{B}_{\bar{q}/c_b\perp}(\kappa,x_b). \label{eq:SBB}
\end{align}
All the dependence on the hard scale $Q^2$ is now manifest. The structure we find here closely resembles the one encountered in previous applications involving the collinear anomaly, see e.g. refs.~\cite{Becher:2011pf,Becher:2012qa}, and the derivation of eq.~(\ref{eq:SBBtoFW}) parallels the steps presented there. We call the function $F_{\perp}^{q\bar{q}}(\kappa)$ the anomaly exponent, and $W_{JI}(\kappa)$ the remainder function. We split the remainder function into three parts in eq.~(\ref{eq:SBB}). This decomposition is not unique, but it is useful because it shows that the dependencies on $x_a$, $x_b$ and $\theta$ factorize. The constant $c_0$ is conventional and will be given in section~\ref{sec:onelin}. We can now write the final factorized form for the Laplace transformed  cross section. For a general partonic channel, $ab\to12$, the corresponding factorization formula reads
\begin{multline}\label{eq:factformgenchann}
\widetilde{t}^{ab\to12}(\kappa)=
H^{ab\to12}_{IJ}(Q,\theta)\;\left(\frac{Q^2}{c_0^2\bar{\kappa}^2}\right)^{-F_{\perp}^{ab}(\kappa)}\widetilde{S}^{ab\to12}_{JI\perp}(\kappa,\theta)\,\\
\widetilde{B}_{a/c_a\perp}(\kappa,x_a)\,\widetilde{B}_{b/c_b\perp}(\kappa,x_b)\widetilde{j}_{c_1\perp}(\kappa)\widetilde{j}_{c_2\perp}(\kappa)\,,
\end{multline}
where we explicitly indicated the partons upon which each term depends. Equation~(\ref{eq:factformgenchann}) is the main result of this work. In section~\ref{sec:onelin}, we compute all the ingredients that enter in this formula at the one-loop level. 

As discussed above, the cross section must be independent of the analytic regulator, while the individual pieces have divergences as the regulator goes to zero. This cancellation of divergences is non-trivial in our case because the soft function is a color matrix, while the beam functions are color-diagonal. It is interesting to look at the structure of the analytic divergences in the soft function in detail to check how they cancel the divergences in the beam functions. In our computations in the next section, we will use the standard form of the phase-space regulator \cite{Becher:2011dz}
\begin{equation}%\label{eq:regulator}
\int \!d^dk \,  \delta(k^2) \,\theta(k^0)\;\; \to\;\; \int \!d^dk \,  \delta(k^2) \,\theta(k^0) \,  \left(\frac{\nu}{n_b\cdot k}\right)^{\alpha}\,,
\end{equation}
with $n_b\cdot k = k_0 + k_z$, but for the discussion in this paragraph, it is convenient to introduce the analytic regulator as follows \cite{Becher:2013xia}
\begin{equation}\label{splitreg}
\int d^4k\, \delta(k^2)\,\theta(k^0) \left[    \left( \frac{\nu}{ n_b \cdot k } \right)^\alpha \theta(n_b \cdot k - n_a \cdot k)
    + \left( \frac{\nu}{n_a \cdot k} \right)^\beta \theta(n_a \cdot k - n_b \cdot k) \right] \,,
 \end{equation}
where $\bar{n}_a=n_b$ and $\bar{n}_b=n_a$. This second form distinguishes the divergences from left- and right-moving particles and is symmetric if one chooses $\alpha=\beta$. The implications of having separate regulators for left and right sectors were first discussed in ref.~\cite{Echevarria:2012js}.  Since the components $k_+$ and $k_-$ scale differently in the collinear regions, the regulator \eqref{splitreg} must be expanded in the small components. In the collinear region $c_a$ we have $n_b \cdot k  \gg n_a \cdot k$ and the term involving the regulator $\beta$ vanishes upon expanding the $\theta$-functions, while the $\alpha$ term vanishes in the collinear region $c_b$. We conclude that the divergences in the $c_a$ beam function only involve the regulator $\alpha$ and vice versa for the $c_b$ beam function. The question is then how the cancellation can be possible, given that the soft function depends on the color charges of all four particles participating in the hard scattering, while the beam functions only involve the color generators of the incoming partons.

To see how the cancellation arises, it is instructive to consider the one-loop case, where the soft function is given by a sum of diagrams in which a gluon is emitted from leg $i$ and absorbed at leg $j$ (see figure \ref{fig:softf} below). The amplitude squared associated with this process has the form
\begin{equation}
|{\cal M}(k)|^2  \propto \bm{T}_i\cdot \bm{T}_j  \,\frac{n_i\cdot n_j}{(n_i\cdot k) \, (n_j \cdot k)}\,.
\end{equation}
We use the notation of the color-space formalism~\cite{Catani:1996jh}, in which $\bm{T}_i$ denotes the color generator associated with parton $i$. 
Only the contributions which involve leg $a$  ($i=a$ or $j=a$), can give rise to divergences in the regulator $\alpha$. Computing the corresponding diagrams explicitly, we find that they have the structure
\begin{equation}\label{eq:nadivS}
\frac{1}{α}\left(\bm{T}_a\cdot \bm{T}_b+\bm{T}_a\cdot \bm{T}_1+\bm{T}_a\cdot \bm{T}_2\right)=\frac{1}{α}\left(-\bm{T}_a^2\right)\, ,
\end{equation}
where we have used color-conservation $\sum_i \bm{T}_i = 0$. The quantity $\bm{T}_a^2=C_a$ is the quadratic Casimir operator of the representation of parton $a$. Similarly, we find
\begin{equation}\label{eq:nbdivS}
\frac{1}{β}\left(\bm{T}_a\cdot \bm{T}_b+\bm{T}_b\cdot \bm{T}_2+\bm{T}_b\cdot \bm{T}_1\right)=\frac{1}{β}\left(-\bm{T}_b^2\right).
\end{equation}
From eqs.~(\ref{eq:nadivS})-(\ref{eq:nbdivS}) one clearly sees that the divergences in each hemisphere can cancel with the corresponding analytic-regulator divergences in the beam functions, which are proportional to the corresponding Casimir. The same structure must also arise at higher orders. Factorization thus imposes nontrivial constraints on the structure of the divergences in the analytic regulator, similar to the constraints it imposes on the infrared structure of the scattering amplitudes \cite{Becher:2009cu,Gardi:2009qi,Becher:2009qa,Dixon:2009ur}. The above structure implies that the collinear anomaly has the form
\begin{equation}
F_{\perp}^{ab}(\kappa) = \frac{(C_a + C_b)}{2} F_{\perp}(\kappa)\,.
\end{equation}
The anomaly is thus given by a universal function $F_{\perp}(\kappa)$. Casimir scaling of the anomaly was observed earlier in the simpler case of transverse-momentum resummation \cite{Becher:2010tm}.

\section{One-loop ingredients}\label{sec:onelin}
In this section we compute all the different ingredients that appear in the factorization formula at the one-loop level. We work in dimensional regularization with $d=4-2\varepsilon$ dimensions, and $α_s=g_s^2/(4\pi)$ is always understood to be the coupling constant in the $\overline{\rm MS}$ scheme at scale $\mu$. The relation to the bare coupling is $α_s^0=Z_αα_s\tilde{μ}^{2ε}$, with $\tilde{μ}^2=μ^2e^{γ_E}(4π)^{-1}$, and $Z_α=1$ at the order we are working. We expand the anomalous dimensions appearing in our expressions according to 
\begin{equation}
γ_{\perp}=\sum_{n=0}^{\infty}γ_{n\perp}\left(\frac{α_s}{4π}\right)^{n+1}\quad ; \quad
γ_{\rm cusp}=\sum_{n=0}^{\infty}Γ_n\left(\frac{α_s}{4π}\right)^{n+1},
\end{equation}
where the coefficients $\Gamma_n$ of the cusp anomalous dimension are collected in appendix \ref{app:anom}.

\subsection{Jet functions}
For simplicity we denote $\bar{n}_{c_i}\to \bar{n}$ and $c_i\to c$ in this section, since the diagrams only involve one collinear direction, and represent the quark jet function by $J_{q\perp}(τ_{c\perp})$ and the gluon jet function by $J_{g\perp}(τ_{c\perp})$. The definition of the quark jet functions was given in eqs.~(\ref{eq:defjetf1}) and (\ref{eq:defjetf2}). For hadronic collisions, we also need the gluon jet function whose definition reads
\begin{align}
\frac{\delta^{ab}}{2Q(2π)^{d-1}} \,(-g_{\mu\nu}^{\top})\, J_{g\perp}(τ_{c\perp})&:=\sum_{X_{c}}\left<0\right|\mathcal{A}_{c\top\mu}^{a}(0)\left|X_{c}\right>\left<X_{c}\right|\mathcal{A}_{c\top\nu}^{b}(0)\left|0\right> \nonumber \\
&\hspace{1cm}\times δ(τ_{c\perp}(X_c)-τ_{c\perp})δ(Q-\bar{n}\cdot
p_{X_c})δ^{(d-2)}(p_{X_c,\top}),
\end{align}
where $\mathcal{A}_{\top\mu}$ is the SCET gluon jet field (normalized such that $\mathcal{A}_{\top\mu}=A_{\top\mu}$ at lowest order in the coupling constant), and $g_{\mu\nu}^{\top}=\left(g_{\mu\nu}-\frac{\bar{n}_{\mu}n_{\nu}+n_{\mu}\bar{n}_{\nu}}{2}\right)$. The normalization is chosen such that $J_{g\perp}(τ_{c\perp}) = \delta(τ_{c\perp})$ at lowest order. To ensure that the same is true for the quark jet function in $d$ dimensions one has to replace $(2π)^3 \to (2π)^{d-1}$ in the definitions in eqs.~(\ref{eq:defjetf1}) and (\ref{eq:defjetf2}) and work with a $(d-2)$-dimensional transverse-momentum $\delta$-function.

\begin{figure}
\centering
\includegraphics[width=15cm]{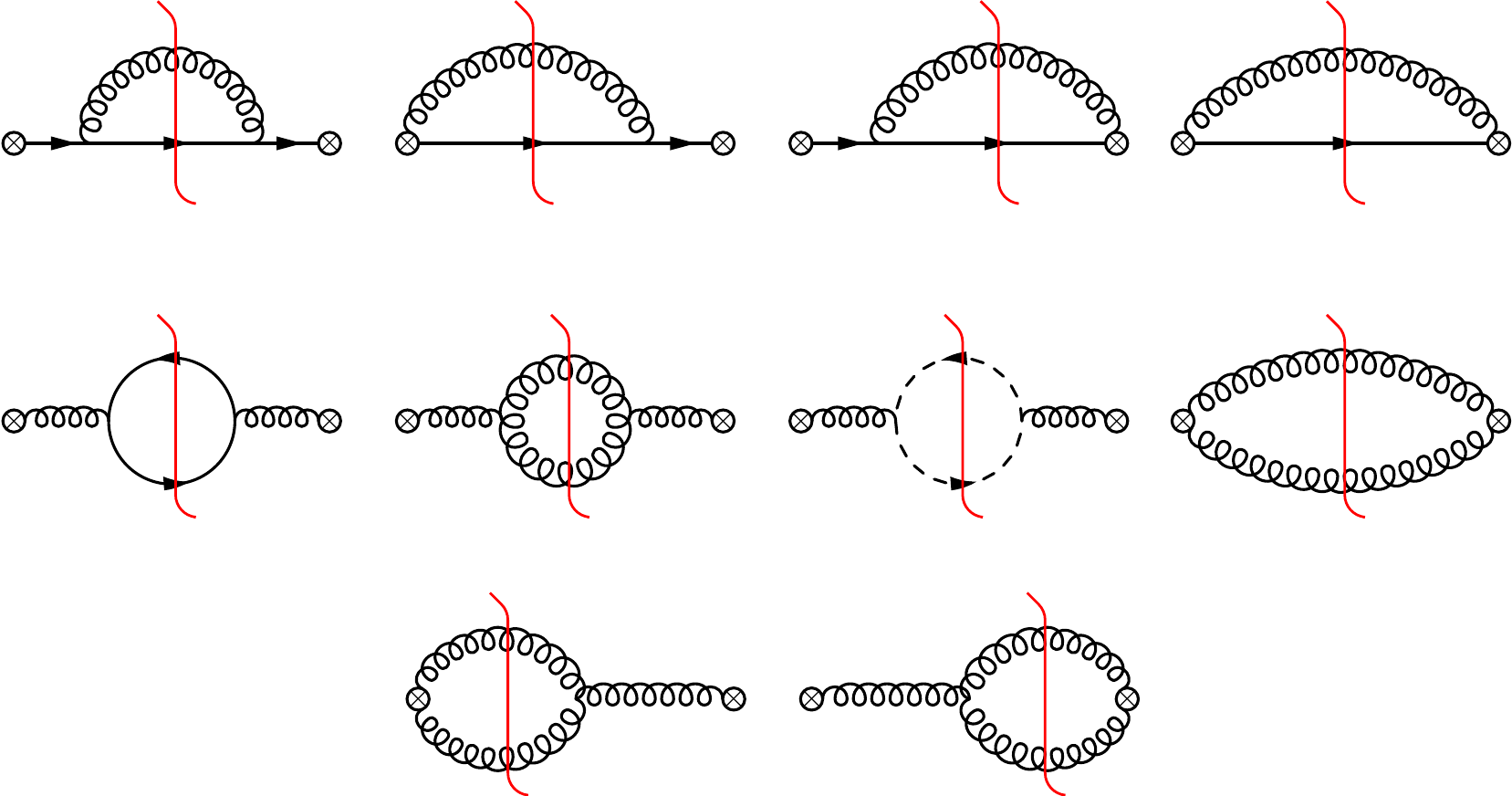}
\caption{Next-to-leading order real-emission diagrams for the quark jet function (first line), and for the gluon jet function (second and third lines). The red vertical lines represent the final-state cut. The crosses indicate the collinear Wilson lines.}\label{fig:jetf}
\end{figure}

The diagrams that contribute to the jet functions at the one-loop level are shown in figure~\ref{fig:jetf}. We only show real-emission diagrams since the one-loop virtual diagrams are scaleless and vanish in dimensional regularization. It is convenient to follow ref.~\cite{Becher:2010pd} and perform the calculation of the jet functions in light-cone gauge $n\cdot A(x)=0$. In this gauge, the collinear Wilson lines which multiply the fields $\chi_c$ and ${\cal A}_c^\mu$ become trivial, therefore only the first diagram contributes to the quark jet function and only the first two diagrams in the second line to the gluon jet function. The relevant phase-space integrals can easily be performed in $d$-dimensions and we then obtain the bare quark jet function at one loop as
\begin{equation}\label{quarkjet}
J_{q\perp}^{\rm bare}(τ_{c\perp})=δ(τ_{c\perp})-α_s\,C_F \frac{%
(4-\varepsilon) 2^{-3+2 \varepsilon } e^{\gamma_E  \varepsilon } \Gamma (2-\varepsilon)
}{\varepsilon \, \Gamma\left(\frac{1}{2}-\varepsilon \right) \Gamma \left(\frac{3}{2}-\varepsilon \right)} \,\frac{1}{τ_{c\perp}}\left(\frac{τ_{c\perp}Q^2 \sin^2\theta}{μ^2}\right)^{-ε}.
\end{equation}
The explicit expressions for the two diagrams contributing to the gluon jet function are given in appendix \ref{app:oneloop}. 

The bare jet function has divergences for $ε\to 0$. Since the jet  functions are distributions in $τ_{c\perp}$,  it is convenient to perform a Laplace transform
\begin{equation}
\widetilde{j}_{i\perp}(L,μ)=\int_0^{\infty}dτ_{c\perp}\, e^{-τ_{c\perp}/(\kappa e^{γ_E})}J_{i\perp}(τ_{c\perp},μ),
\end{equation}
where, for later convenience, we have written the Laplace transform as a function of the logarithm $L=\ln(\frac{4\kappa Q^2\sin^2\theta}{μ^2})$ of the Laplace variable $\kappa$. The
renormalized Laplace-transformed jet function is related to the bare
one according to $\widetilde{j}_{i\perp}=Z_{j_i\perp}\,\widetilde{j}_{i\perp}^{\rm
  bare}$. Since the bare function is $\mu$-independent, the renormalized jet function, $\widetilde{j}_{i\perp}$, and $Z_{j_i\perp}$ fulfill both the same RG equation
\begin{equation}
\frac{d}{d\lnμ}\widetilde{j}_{i\perp}(L,μ)=\left[-2C_i \gamma_{\rm
    cusp} \,L -2γ^{J_i}_{\perp}\right]\widetilde{j}_{i\perp}(L,μ),
\end{equation}
where $\gamma_{\rm cusp}$ is the cusp anomalous dimension, which gets multiplied by the relevant Casimir operator $C_q = C_F$ and $C_g = C_A$.
Solving the RG equation for $Z_{j_i\perp}$ one gets
\begin{equation}
\ln Z_{j_i\perp}=\frac{α_s}{4π}\left[-\frac{C_i Γ_0}{ε^2}+\frac{1}{ε}(C_i Γ_0 L+γ_{0\perp}^{J_i})\right].
\end{equation}
Expanding $\widetilde{j}^{\rm bare}_{i\perp}$ and $Z_{j_i\perp}$ in $ε$ and
$α_s$, we find that the divergences in $\widetilde{j}_{i\perp}$ cancel
with
\begin{align}
γ_{0\perp}^{J_q}&=γ_0^q=-3C_F\, , & γ_{0\perp}^{J_g}&=γ_0^g=-\beta_0\,.
\end{align}
The one-loop anomalous dimensions are thus identical to the inclusive jet-function with our choice of the scale of the logarithm $L$. The renormalized function $\widetilde{j}_{i\perp}$ is given by
\begin{equation}
\widetilde{j}_{i\perp}(L,μ)=1+\frac{α_s}{4π}\left(C_i Γ_0 \frac{L^2}{2}+γ_{0\perp}^{J_i}L+c_{1\perp}^{J_i}\right),
\end{equation}
with
\begin{align}
c_{1\perp}^{J_q} &=C_F\left(7-\frac{4 π^2}{3}\right)\,, & c_{1\perp}^{J_g}&=C_A\left(\frac{67}{9}-\frac{4π^2}{3}\right)-\frac{20}{9}T_Fn_f .
\end{align}

\subsection{Beam functions}
\begin{figure}[!t]
\includegraphics[width=15cm]{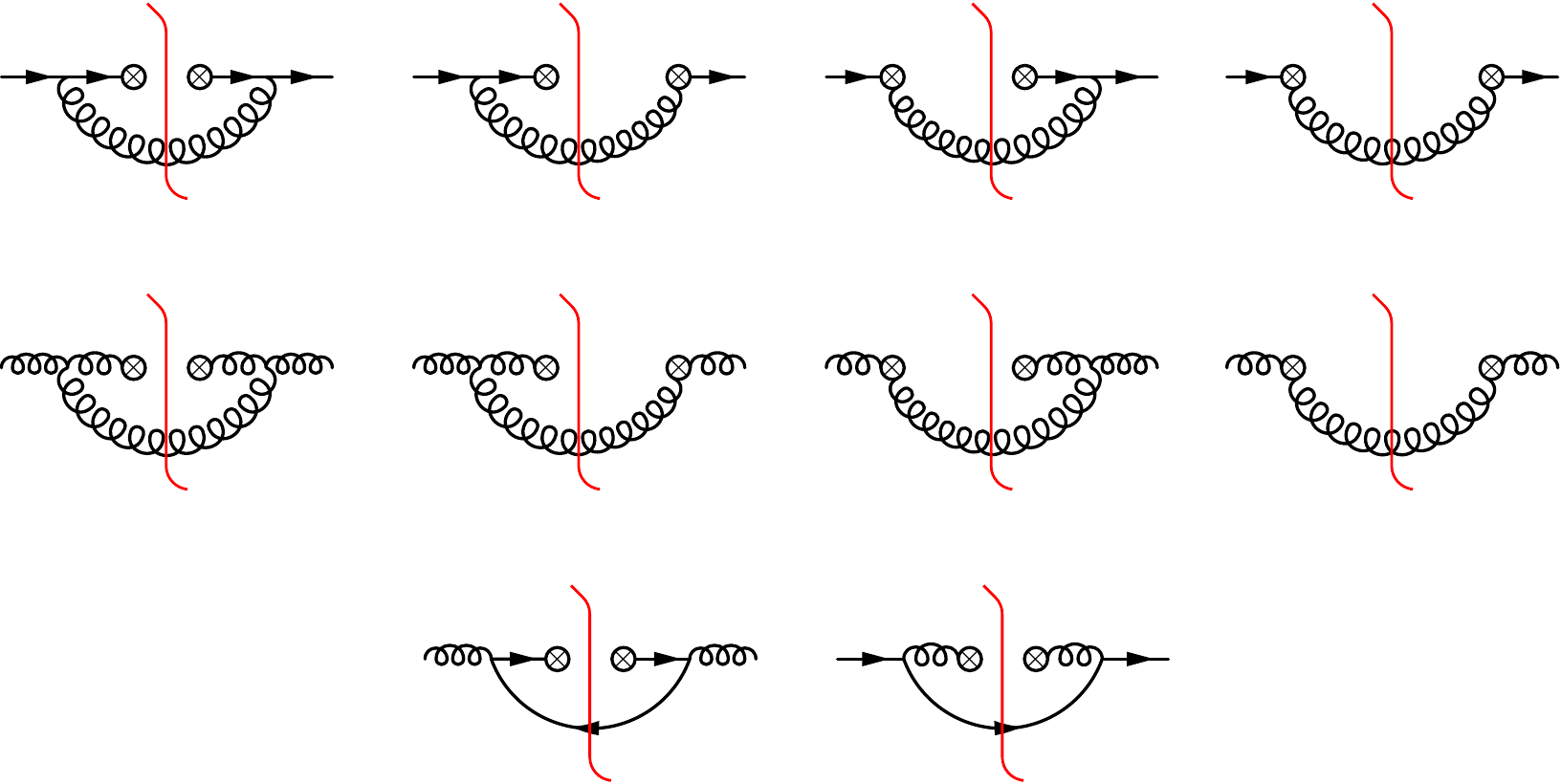}
\caption{\label{fig:beamgraphs}
One-loop diagrams contributing to the matching coefficients ${\cal I}_{q\leftarrow q}$ (top row), ${\cal I}_{g\leftarrow g}$ (middle row), and ${\cal I}_{q\leftarrow g}$ (bottom row left) and ${\cal I}_{g\leftarrow q}$ (bottom row right) of the beam functions. The red vertical lines represent the final-state cut.}
\end{figure}
The definition of the quark beam function was given in eq.~(\ref{eq:beamquark}) in section~\ref{subsec:hadrcoll}. In addition, we also need the gluon beam function which is defined as
\begin{align}\label{gluonbeam}
\mathcal{B}_{g/c_a\perp}(τ_{c_a\perp},x_a)
&:=-x_a(\bar{n}_a\cdot P_a)\sum_{X_{c_a}}\left<P_a\right|{\cal A}_{c_a\top}^{\mu,a}(0)\left|X_{c_a}\right>\left<X_{c_a}\right|{\cal A}_{c_a\top\mu}^a(0)\left|P_a\right>
\nonumber \\
& \hspace{1cm}\times δ(τ_{c_a\perp}(X_{c_a})-τ_{c_a\perp})δ\left((\bar{n}_a\cdot P_a)(1-x_a)-\bar{n}_a\cdot p_{X_{c_a}}\right).
\end{align}
For perturbative values of $\tau_{\perp}$, we can compute the final-state radiation in perturbation theory and match the beam functions onto standard PDFs, $f_{j/N}(x)$ for the hadron $N$, convolved with a perturbative coefficient, ${\cal I}_{i\leftarrow j}(x,\tT)$. The corresponding matching relation takes the form \cite{Collins:1981uw,Stewart:2009yx,Becher:2010tm}
\begin{equation}\label{OPE}
   {\cal B}_{i/c_k\perp}(\tT,\xi) 
   = \sum_j \int_\xi^1\!\frac{dx}{x}\, {\cal I}_{i\leftarrow j}(\xi/x,\tT)\,
    f_{j/N_k}(x).     
\end{equation}
At lowest order in perturbation theory the matching coefficients take the form
\begin{align}
{\cal I}_{i\leftarrow j}(x,\tT)  & = \delta_{ij} \delta(1-x) \delta(\tT)\,, 
\end{align}
so that the beam function reduces to the PDF times $\delta(\tT)$. When computing the matching coefficients, it is convenient to evaluate the beam-function matrix elements eqs.~\eqref{eq:beamquark} and \eqref{gluonbeam} with quark and gluon states instead of hadrons. In this case the PDFs are trivial  $f_{j/k}(x) = \delta_{jk} \delta(1-x)$ and the partonic computation directly yields the matching coefficients ${\cal I}_{i\leftarrow k}(\xi/x,\tT)$.

As explained in section~\ref{subsec:hadrcoll}, the transverse-thrust beam functions are not well defined within dimensional regularization and require an additional regulator. We regularize them using the analytic regulator of ref.~\cite{Becher:2011dz} which modifies the phase-space integration measure as follows
\begin{equation}\label{eq:regulator}
\int \!d^dk \,  \delta(k^2) \,\theta(k^0)\;\; \to\;\; \int \!d^dk \,  \delta(k^2) \,\theta(k^0) \,  \left(\frac{\nu}{k_0+k_z}\right)^{\alpha}\,.
\end{equation}
Note that the beam functions in both $c_a$ and $c_b$ directions are regularized with the same regulator. To discuss the renormalization of the beam functions, and the cancellation of the dependencies on the analytic regulator, we again perform a Laplace transform
\begin{equation}
\widetilde{\cal B}_{q/c_a\perp}(\kappa,x_a) = \int d\tT   e^{-\tT/(\kappa e^{γ_E})} {\cal B}_{q/c_a\perp}(\tT,x_a).
\end{equation} 
We have discussed the cancellation of the divergences in the analytic regulator in section~\ref{subsec:collan}. The divergences of the soft function with the form of the analytic regulator given in eq.~\eqref{eq:regulator} can be obtained from those in eqs.~\eqref{eq:nadivS}-\eqref{eq:nbdivS} by replacing $\beta \to -\alpha$. We see that the analytic divergences  cancel in the soft function for channels involving the same color representation of the incoming partons, such as $q\bar{q}$ and $gg$. In these cases, the remaining divergences must cancel out in the product of the Laplace transformed beam functions which takes the form
\begin{equation}\label{eq:refac}
\widetilde{\cal B}_{q/c_a\perp}(\kappa,x_a)\, \widetilde{\cal B}_{\bar{q}/c_b\perp}(\kappa, x_b) = \left(\frac{Q^2}{c_0^2 \bar{\kappa}^2}\right)^{-C_F F_\perp(\kappa,\mu)} \widetilde{B}_{q/c_a\perp}(\kappa,x_a,\mu)\, \widetilde{B}_{\bar{q}/c_b\perp}(\kappa,x_b, \mu)\,,
\end{equation}
and analogously for the $gg$ channel. Computing the product of the Laplace transforms of the quark beam functions, we find that this is indeed the case, and obtain the corresponding anomaly exponent and refactorized matching coefficients. The diagrams that contribute to it at one loop are shown in figure~\ref{fig:beamgraphs}. The diagrams need to be evaluated both in the $c_a$ and $c_b$ sectors because the analytic regularization is not symmetric, as we stressed above. The explicit expressions for each of the individual diagrams are listed in appendix \ref{app:oneloop}. 
Proceeding as discussed, we get
\begin{equation}\label{eq:Fperpqq}
   F_{\perp}(\kappa,\mu)
   = \frac{\alpha_s}{4\pi} \,\Gamma_0\,L_\perp\,,
\end{equation}
with $L_\perp =\ln\frac{\mu^2}{c_0^2\bar{\kappa}^2}$, with $\ln c_0 = \frac{4 G}{\pi}$, and where $G\approx 0.915966$ is the Catalan constant. The characteristic scale of the logarithm is $\bar{\kappa}=2 \kappa Q \sin \theta$. The refactorized matching coefficients are defined as 
\begin{equation}
B_{i/c_k\perp}(\tT,\xi,\mu)=\sum_j \int_\xi^1\!\frac{dz}{z}\, I_{i\leftarrow j}(\xi/z,\tT,\mu)\,
    f_{j}(z,\mu)\,,
\end{equation}
in analogy with the corresponding matching coefficients and are obtained by evaluating the beam functions with partonic states. At one-loop order, they have the general form
\begin{equation}\label{eq:Ires}
   \widetilde{I}_{i\leftarrow j}(x,\kappa,\mu) 
   = \delta(1-x)\,\delta_{ij} \left[ 1 + \frac{\alpha_s}{4\pi} \left( C_i \Gamma_0\,\frac{L_\perp^2}{4}
    - \gamma_0^i\,L_\perp \right) \right]
    + \frac{\alpha_s}{4\pi} \left[ - {\cal P}_{i\leftarrow j}^{(1)}(x)\,\frac{L_\perp}{2}
    + {\cal R}_{i\leftarrow j}(x) \right] \,,
\end{equation}
and contain the Altarelli-Parisi kernels
\begin{equation}
\begin{aligned}
   {\cal P}_{q\leftarrow q}^{(1)}(x) 
   &= 4C_F  \left(\frac{1+x^2}{1-x}\right)_+ \,, \\
   {\cal P}_{q\leftarrow g}^{(1)}(x) 
   &= 4 T_F\,\left(x^2+(1-x)^2\right)\,,\\
    {\cal P}_{g\leftarrow g}^{(1)}(x) 
   &= 8C_A \left[ \frac{x}{\left(1-x\right)_+} + \frac{1-x}{x} + x(1-x) \right]
    + 2\beta_0\,\delta(1-x) \,, \\
   {\cal P}_{g\leftarrow q}^{(1)}(x) 
   &= 4C_F\,\frac{1+(1-x)^2}{x},
\end{aligned}
\end{equation}
which involve plus-distributions, and remainder functions
\begin{align}
   {\cal R}_{q\leftarrow q}(x) &= - C_F\left(\frac{\pi^2}{6} + \frac{64 G^2}{\pi^2} + 2 F \right)\delta(1-x)+ 2 C_F(1-x) \,, \nonumber \\
   {\cal R}_{q\leftarrow g}(x) &= 4 T_F x(1-x) \,,\nonumber\\
    {\cal R}_{g\leftarrow g}(x) &= - C_A\left(\frac{\pi^2}{6} + \frac{64 G^2}{\pi^2} + 2 F \right)\delta(1-x), \nonumber \\
      {\cal R}_{g\leftarrow q}(x) &= 2 C_F  x \,.
\end{align}
The numerical value of the constant $F$ in the equation above is $F\approx 8.20629$, an analytic expression in terms of Lerch's $\Phi$ function is given in the appendix. 

The anomalous dimensions of the anomaly exponent and the refactorized beam functions are given by
\begin{equation}\label{eq:Bevol}
\begin{aligned}
   \frac{d}{d\ln\mu} F_{\perp}
   &= 2\gamma_{\rm cusp} \,, \\
   \frac{d}{d\ln\mu}\widetilde{B}_{i/c_k\perp}
   &= \left[-c_{B_i}\gamma_{\rm cusp}\ln\frac{c_0^2 \bar{\kappa}^2}{\mu^2}
    - 2\gamma^{B_i}_{\perp} \right] \widetilde{B}_{i/c_k\perp} \,,
\end{aligned}
\end{equation}
with
\begin{equation}
c_{B_i}=C_i\quad \text{ and }\quad \gamma^{B_i}_{0\perp}=γ_0^i,
\end{equation}
according to our one-loop results above.

\subsection{Soft functions}

\begin{figure}
\centering
\includegraphics[width=15cm]{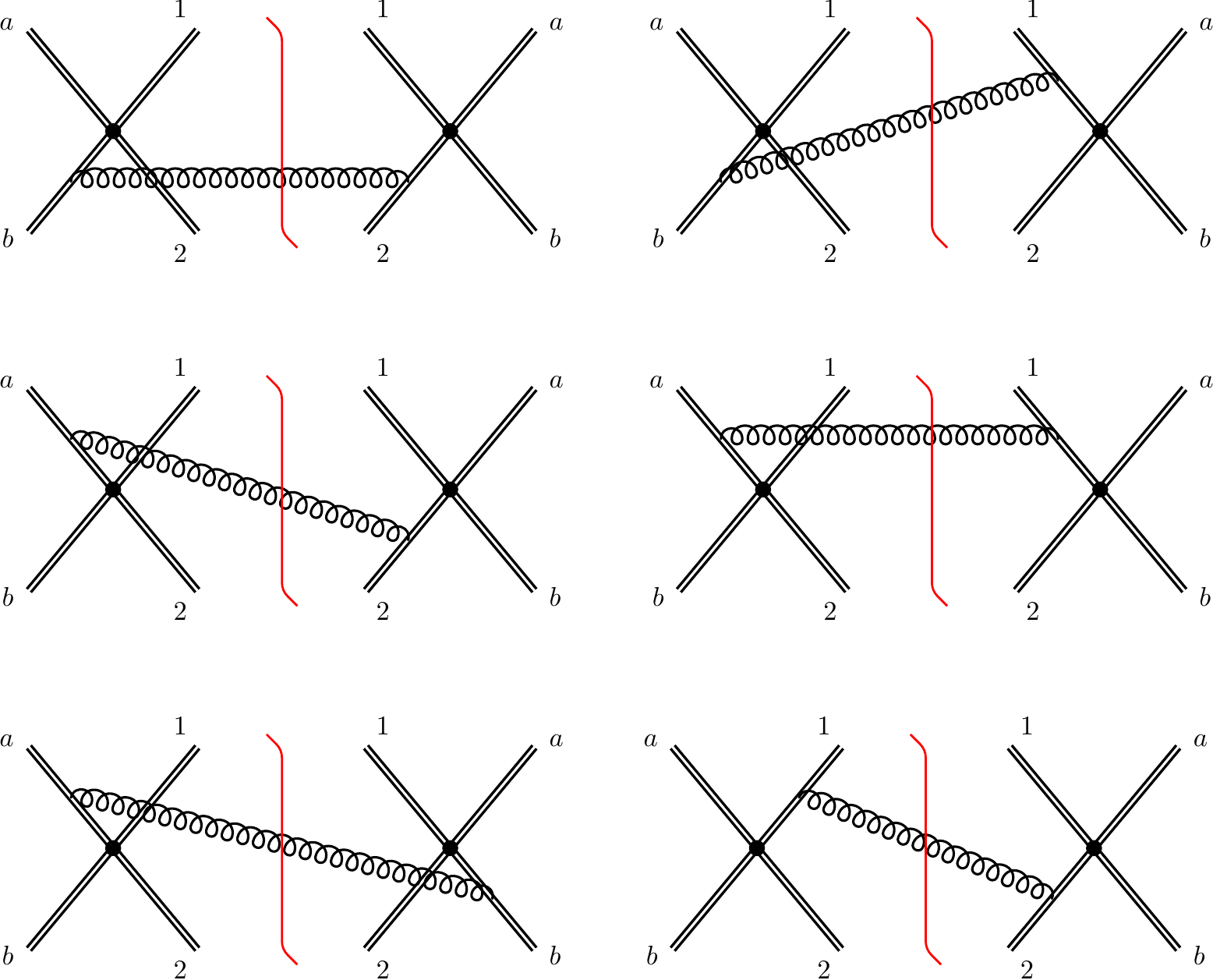}
\caption{Next-to-leading order real-emission diagrams for the soft function. The red vertical lines represent the final-state cut. Mirror diagrams and those that vanish because the gluon attach to two Wilson lines in the same direction are not shown. In the lepton-collider case the $a$ and $b$ Wilson lines are absent, and only the last diagram is possible.}\label{fig:softf}
\end{figure}

The soft function is, in general, a matrix in color space defined in the color basis adopted for the hard function. To discuss the color structure independently of the basis choice, it is most convenient to use the color-space formalism \cite{Catani:1996jh,Catani:1996vz}, where the hard function is written as
\begin{equation}
| \tilde C \rangle = \sum_I \tilde{C}_I | I \rangle\,,
\end{equation}
with $| I \rangle$ a basis of color states. For the $q\bar{q} \to q'\bar{q}'$ channel discussed above, the basis contains the color-singlet and the color-octet operators defined in eq.~\eqref{eq:basis}. In the chosen basis, the soft function acts as follows
\begin{equation}
\langle \tilde C|  \bm{S}(τ_{s\perp}) | \tilde C \rangle =   \tilde{C}_I^* \,  \langle I |  \bm{S}(τ_{s\perp}) | J \rangle\, \tilde{C}_J =   \tilde{C}_I^* \, S_{IJ}(τ_{s\perp}) \, \tilde{C}_J = H_{JI} \,S_{IJ}(τ_{s\perp}) \,.
\end{equation}
We can write it as 
\begin{equation}
\mathcal{S}_{IJ}(\tau_{s\perp})=\frac{1}{N_c} \sum_{X_s}\left<0\right|\mathcal{W}^{\dagger}_I\left|X_s\right>\left<X_s\right|\mathcal{W}_J\left|0\right>δ(τ_{s\perp}-\tau_{\perp}(X_s)),
\end{equation}
where $\mathcal{W}_I$ are combinations of Wilson lines, in the adequate representation for the partonic channel we are studying. The explicit expression relevant for the $q\bar{q}\to q'\bar{q}'$ channel was given in Eq.~(\ref{eq:softhadqq}). 

The diagrams that contribute at the one-loop level are shown in figure~\ref{fig:softf}. Once again only real-emission diagrams can give non-vanishing contributions in dimensional regularization. The soft function up to one-loop order is therefore given as 
\begin{equation}
\mathcal{S}_{IJ}(\tau)=\delta(\tau)D_{IJ}^{\rm tree}+\sum_{ij}I_{ij}D_{IJ}(i,j),
\end{equation}
where $i,j=a,b,1,2$, with $a$ and $b$ the partons in the initial state, and $1$ and $2$ the final-state jets. The color structures at tree level and one-loop order are
\begin{align}\label{eq:colfact}
D_{IJ}^{\rm tree} &= \frac{1}{N_c} \langle I | J \rangle\,, & &\text{ and }  & D_{IJ} &= \frac{1}{N_c} \langle I | (-\bm{T}_i\cdot \bm{T}_j) | J \rangle\,.
\end{align}
The $D_{IJ}$ encode the color factors, are matrices in color space, and depend on the partonic channel we are considering. For the $qq$-type channels they are $2\times2$ matrices, for the $qg$-type channels $3\times3$ matrices, and for the $gg$ channels $9\times9$ matrices. Explicit expressions for all of them were given in ref.~\cite{Kelley:2010qs}.\footnote{Note that the matrices in ref.~\cite{Kelley:2010qs} are defined without the $1/N_c$ prefactor in eq.~\eqref{eq:colfact}. Some of the entries in the matrices for the $gg$ channel in this reference are interchanged: the entries 4 and 19 (20 and 65) need to be swapped in the tree-level (one-loop) matrix. We thank Jan Piclum for pointing this out.} For leptonic collisions the color factors are numbers because the process only involves two color-charged legs which fulfill $-\bm{T}_1\cdot \bm{T}_2 = \bm{T}_1^2=\bm{T}_2^2 = C_F$. The soft integrals $I_{ij}$ only depend on the directions to which the soft gluon in figure~\ref{fig:softf} attaches, and not on the representation of each Wilson line. They are given by
\begin{equation}\label{eq:Iij}
I_{ij}=g_s^2\,\tilde{μ}^{2ε}\int\frac{d^dk}{(2π)^{d-1}}δ(k^2)\theta(k^0)\frac{n_i\cdot
n_{j}}{(n_i\cdot k) (n_{j}\cdot
k)}δ\left(τ_{s\perp}-\frac{1}{Q|\sin\theta|}\left(|\vec{k}_{\perp}|-|\vec{n}_{\perp}\cdot\vec{k}_{\perp}|\right)\right),
\end{equation}
with
\begin{align*}
n_a^\mu&=(1,0,0,1)\,, & n_b^\mu&=(1,0,0,-1)\, ,& n_1^\mu&=(1,\vec{n})\, ,& n_2^\mu&=(1,-\vec{n}) \,, & \vec{n}&=(\sin\theta,0,\cos\theta).
\end{align*}
For leptonic collisions, only $I_{12}$ arises, since there are no colored particles in the initial state. The $I_{ij}$ integrals are in general not well defined within dimensional regularization, and an additional regulator is required. In order for the divergences to cancel, we need to use the same analytic phase-space regulator as for the beam functions. The regularized versions of the $I_{ij}$ integrals have an additional $(\nu/(k_0+k_z))^{α}$ factor in the integrand. 

To compute the integrals $I_{ij}$, we find it useful to perform a change of variables $(k_{\perp},k_z)\to(x,y)$ with
\begin{align}
k_{\perp}&=\frac{\ts Q|\sin\theta|}{x}\,, & k_z&=\frac{\ts Q|\sin\theta|}{x}y.
\end{align}
The variable $x\in [0,1]$, while $y\in(-\infty,+\infty)$. To extract the divergences in $α$, it is convenient to introduce the variable $u=1/(1-y+\sqrt{1+y^2})\in[0,1]$. Parameterizing the integrals in this way, it is more or less straightforward to extract the divergences and to compute the remaining finite integrals numerically, but we have not managed to evaluate the finite parts analytically. The nontrivial soft integrals can be written in the form
\begin{eqnarray}\label{eq:Iijints}
I_{12} & = & \frac{n_1\cdot n_2}{2}N\,\mathcal{I}_0 ,\nonumber\\
2I_{a1} & = & \frac{n_a\cdot n_1}{2}N\left(\mathcal{I}_++\mathcal{I}_-+\mathcal{I'}_+-\mathcal{I'}_-\right),\nonumber\\
2I_{b2} & = & \frac{n_a\cdot n_1}{2}N\left(\mathcal{I}_+-\mathcal{I}_-+\mathcal{I'}_++\mathcal{I'}_-\right).\end{eqnarray} 
Below, we will discuss the evaluation of the integral $\mathcal{I}_0$ in detail. The normalization factor $N$ and explicit expressions for the integrals $\mathcal{I}_\pm$, $\mathcal{I}_\pm'$ are given in appendix~\ref{app:oneloop}. The three remaining soft integrals are given by
\begin{align}
I_{ab} & =  0\, , & I_{a2} & =  \left.  I_{a1} \right |_{\theta \to \pi - \theta} \,, & I_{b1} & =  \left.  I_{b2} \right |_{\theta \to \pi - \theta} \,.
\end{align}
 The integral $I_{ab}$ is zero because it is proportional to the scaleless integration
\begin{equation}\label{eq:sclsint}
\int_{-\infty}^{\infty}dy\frac{\left(y+\sqrt{1+y^2}\right)^{-α}}{\sqrt{1+y^2}}=\int_0^1du\, u^{-1-\alpha}(1-u)^{-1+\alpha}=0\,.
\end{equation}

The $I_{12}$ integral is well defined without the additional analytic regulator, since it does not involve the directions $a$ and $b$ of the initial-state partons, which involve collinear modes of the same virtuality as the soft modes. The soft function for leptonic collisions is therefore well defined within dimensional regularization, and there is no collinear factorization anomaly, as expected from the arguments in section~\ref{subsec:collan}. To illustrate how the soft integrals are computed, we explain now how to obtain the result for $\mathcal{I}_0$, which is given by
\begin{equation}
\mathcal{I}_{0} =
\int_0^1\!\!\! dx(2-x)^{-\frac{1}{2}-ε}x^{-\frac{1}{2}+ε}\!\!\!\int_{-\infty}^{\infty}\!\!\!\!\!\! dy\frac{1}{\sqrt{1+y^2}}\frac{1}{1+y^2-(y\cos\theta+(1-x)\sin\theta)^2}\,.
\end{equation}
To arrive at the above form from eq.~(\ref{eq:Iij}) one writes the integration measure as
\begin{equation}
\int \!\!\! d^dk=\int \!\!\! dk_0\, dk_z\, dk_{\perp}\, k_{\perp}^{d-3}\, d(\cosφ)\, (\sinφ)^{d-5}\, Ω_{d-3}\, ,
\end{equation}
and uses $δ(k^2)$ to perform the $k^0$ integration and the transverse-thrust constraint to obtain the integral over the angle $\varphi$.  Finally, one performs the change of variables $(k_{\perp},k_z)\to(x,y)$  specified above. The next step is the $y$ integral, which does not depend on ε. To compute it, we write
\[
\int_{-\infty}^{\infty}dy\frac{1}{\sqrt{1+y^2}}\frac{1}{1+y^2-\left(y\cos\theta+(1-x)\sin\theta\right)^2}
\]
\[
=\int_{-\infty}^{\infty}dy\frac{1}{\sqrt{1+y^2}}\frac{1}{(y_1-y_2)\sin^2\theta}\left(\frac{1}{y-y_1}-\frac{1}{y-y_2}\right)
\]
\begin{equation}
=\int_{-\infty}^{\infty}dy\frac{1}{\sqrt{1+y^2}}\frac{2i}{(y_1-y_2)\sin^2\theta}{\rm Im}\left\{\frac{1}{y-y_1}\right\},
\end{equation}
where
\begin{equation}\label{eq:y1y2}
y_1=(1-x)\cot\theta+i \frac{(2-x)^{\frac{1}{2}}x^{\frac{1}{2}}}{\sin\theta} \quad;\quad y_2=(1-x)\cot\theta- i\frac{(2-x)^{\frac{1}{2}}x^{\frac{1}{2}}}{\sin\theta},
\end{equation}
are the roots of $1+y^2-\left(y\cos\theta+(1-x)\sin\theta\right)^2$; note that $y_1-y_2$ is purely imaginary for $0<x<1$. Using
\begin{equation}
\int_{-\infty}^{\infty}dy\frac{1}{\sqrt{1+y^2}}\frac{1}{y-A}=\frac{1}{\sqrt{1+A^2}}\log\left(\frac{A-\sqrt{1+A^2}}{A+\sqrt{1+A^2}}\right),
\end{equation}
we find
\begin{equation}
\mathcal{I}_0=(\sin\theta)^{-1}\int_0^1dx\,(2-x)^{-1-ε}x^{-1+ε}\,{\rm Im}\left\{\frac{1}{\sqrt{1+y_1^2}}\ln\left(\frac{y_1-\sqrt{1+y_1^2}}{y_1+\sqrt{1+y_1^2}}\right)\right\}.
\end{equation}
The divergences in the integral over $x$ only come from the $x^{-1+ε}$ term, and the result for $\mathcal{I}_0$ as a series in ε can be easily found by using the expansion
\begin{equation}
x^{-1+ε}=\frac{1}{ε}δ(x)+\left[\frac{1}{x}\right]_++ε\left[\frac{\ln x}{x}\right]_++\mathcal{O}(ε^2),
\end{equation}
where the plus distributions are defined by
\begin{equation}
\int_0^1\left[f(x)\right]_+g(x)dx=\int_0^1f(x)(g(x)-g(0))dx.
\end{equation}
The integrals $\mathcal{I}_+$ and $\mathcal{I}_+'$ are finite as $\alpha\to 0$ and can be evaluated in exactly the same way as $\mathcal{I}_0$. A divergence in the analytic regulator $\alpha$ arises only in the integrals $\mathcal{I}_-$ and $\mathcal{I}_-'$ in eq.~(\ref{eq:Iijints}), which are simpler than the plus-type integrals so that the $y$-integration can be performed analytically for arbitrary $\alpha$.
The divergent terms of the integrals in eq.~(\ref{eq:Iijints}) are given by
\begin{eqnarray}\label{eq:divergences}
I_{12} & = & \frac{\alpha_s}{\pi} \frac{1}{\tau_{s\perp}}\left(\frac{4Q\tau_{s\perp}\sin\theta}{\mu}\right)^{-2\varepsilon}\left(\frac{1}{ε}-2\ln\sin\theta+\mathcal{O}(\varepsilon)\right),\nonumber\\
2I_{a1} & = & \frac{\alpha_s}{\pi} \frac{1}{\tau_{s\perp}}\left(\frac{4Q\tau_{s\perp}\sin\theta}{\mu}\right)^{-2\varepsilon}\nonumber\\
&&\times\left(\frac{1}{\varepsilon}-2\ln\cot\frac{\theta}{2}+\left(\frac{2Q\tau_{s\perp}\sin\theta}{\nu}\right)^{-\alpha}2\left(\frac{1}{\alpha}+H\right)+\mathcal{O}(\varepsilon,\alpha)\right),\nonumber\\
2I_{b2} & = & \frac{\alpha_s}{\pi} \frac{1}{\tau_{s\perp}}\left(\frac{4Q\tau_{s\perp}\sin\theta}{\mu}\right)^{-2\varepsilon}\nonumber\\
&&\times\left(\frac{1}{\varepsilon}-2\ln \cot\frac{\theta}{2}-\left(\frac{Q\tau_{s\perp}\sin\theta}{2\nu}\right)^{-\alpha}2\left(\frac{1}{\alpha}+H\right)+\mathcal{O}(\varepsilon,\alpha)\right), 
\end{eqnarray}
where the numerical value of the constant $H$ is $H\approx -1.85939$. We refer to appendix~\ref{app:oneloop} for the full expressions, including the finite parts. 

 Due to the structure of the $D_{IJ}$ color matrices, see e.g. ref.~\cite{Kelley:2010qs}, the $I_{ij}$ integrals always appear in the combinations $I_{12}+I_{ab}$, $I_{a1}+I_{b2}$, and $I_{a2}+I_{b1}$ for the $qq$- and $gg$-channels. As we can see from eq.~(\ref{eq:Iijints}), and explicitly in eq.~(\ref{eq:divergences}), in these combinations the $1/\alpha$ divergent terms cancel, as expected from the discussions in section~\ref{subsec:collan}, and the explicit calculation of the beam functions in the previous section. For the $qg$-type channels, on the other hand, the structure of the $D_{IJ}$ matrices is such that some individual $I_{ij}$ integrals appear, and not only the above combinations. Therefore, in the $qg$-type channels, the divergences in $\alpha$ do not cancel within the soft function itself, but one needs to combine it with the beam functions to obtain a regulator independent result, again as discussed in section~\ref{subsec:collan}. 
 
The RG equation of the soft function is given by
\begin{equation}
\frac{d}{d\lnμ}\widetilde{S}_{IJ\perp}^{ab\to12}  =  -\widetilde{S}_{IL\perp}^{ab\to12}\left(Γ_S^{ab\to12}\right)_{LJ}-\left(Γ_S^{ab\to12\dagger}\right)_{IM}\widetilde{S}^{ab\to12}_{MJ\perp},
\end{equation}
with
\begin{equation}
\left(Γ_S^{ab\to12}\right)_{IJ}  =  \left(\gamma_{\rm cusp}c_S\ln\frac{4\kappa Q\sin^2\theta}{μ}+γ_{s\perp}\right)δ_{IJ}+\gamma_{\rm cusp}\,M_{IJ}(\theta).
\end{equation}
The coefficients $c_S$, $γ_{s\perp}$, and the remainder function $M_{IJ}(\theta)$ depend on the partonic channel. Up to two-loop order the remainder function is fixed by general constraints on the structure of soft anomalous dimensions, see section~\ref{sec:resum}.
From our results above we get
\begin{equation}
c_S=-(C_1+C_2)\quad;\quad γ_{0s\perp}=0.
\end{equation}

\subsection{Hard functions}
Like the soft function, the hard function $H_{IJ}$ in eq.~\eqref{eq:hardfunHad} is a matrix in color space. Up to the conventional normalization factor, it is given by the Fourier transform of the matching coefficients $\widetilde{C}_I$ of the QCD currents to the SCET operators, according to
\begin{equation}
H_{IJ}\, \propto\,\widetilde{C}_I\widetilde{C}_J^*\,.
\end{equation}
The hard function is extracted from the results for the amplitudes for $2\to2$ processes in QCD. This was done at the one-loop level in ref.~\cite{Kelley:2010fn}, where the RG evolution of the hard function was also considered. Recently, the two-loop result for the hard function has been presented in ref.~\cite{Broggio:2014hoa}, based on the results for the QCD amplitudes of refs.~\cite{Glover:2003cm,Glover:2004si,Bern:2003ck,DeFreitas:2004tk,Bern:2002tk}. The results of ref.~\cite{Broggio:2014hoa} are conveniently given in an electronic form, and can be readily used. We refer to refs.~\cite{Kelley:2010fn,Broggio:2014hoa} for the explicit results for the hard functions, and do not copy them here.

\section{Resummation}\label{sec:resum}

\subsection{Renormalization group equations and scale independence}\label{subsec:checkRG}
The cross section, or equivalently its Laplace transform, must be RG invariant, i.e. it must be independent of $\mu$. That this is indeed the case at order $\alpha_s$ can be explicitly verified with the one-loop results presented above. The Laplace transform of the cross section was given in eq.~(\ref{eq:factformgenchann}). For concreteness, we now consider the cross section in the $q\bar{q}$ channel whose Laplace transform is given by
\begin{equation}
\widetilde{t}
= H_{IJ}\;\left(\frac{Q^2}{c_0^2\bar{\kappa}^2}\right)^{-F_{\perp}^{qq}}\widetilde{S}_{JI\perp}\,\widetilde{B}_{q/c_a\perp}\,\widetilde{B}_{\bar{q}/c_b\perp}\widetilde{j}_{c_1\perp}\widetilde{j}_{c_2\perp}\,,
\end{equation}
where $H_{IJ}:=H_{IJ}^{q\bar{q} \to q\bar{q} }$ and $S_{IJ}:=S_{IJ}^{q\bar{q} \to q\bar{q} }$. The cross section must satisfy
\begin{equation}\label{eq:RGinvcs}
\frac{d\widetilde{t}}{d\ln\mu}=0\,,
\end{equation}
which implies relations among the anomalous dimensions of the different ingredients. 
Their RG equations can be written as
\begin{eqnarray}\label{eq:allRGs}
\frac{d}{d\lnμ}H_{IJ} & = & \left(Γ_H\right)_{IK}H_{KJ}+H_{IK'}\left(Γ_H^{\dagger}\right)_{K'J},\nonumber\\
\frac{d}{d\lnμ}\widetilde{S}_{IJ\perp} & = & -\widetilde{S}_{IL\perp}\left(Γ_S\right)_{LJ}-\left(Γ_S^{\dagger}\right)_{IM}\widetilde{S}_{MJ\perp},\nonumber\\
\frac{d}{d\lnμ}F^{qq}_{\perp} & = & 2C_F\gamma_{\rm cusp},\nonumber\\
\frac{d}{d\lnμ}\widetilde{B}_{q/c_k\perp} & = & \left[-c_{B_q}\gamma_{\rm cusp}\ln\frac{c_0^2\bar{\kappa}^2}{\mu^2}-2\gamma^{B_q}_{\perp}\right]\widetilde{B}_{q/c_k\perp},\nonumber\\
\frac{d}{d\lnμ}\widetilde{j}_{q\perp} & = & \left[-c_{j_q}\gamma_{\rm cusp}\ln\frac{4\kappa Q^2\sin^2\theta}{\mu^2}-2\gamma^{J_q}_{\perp}\right]\widetilde{j}_{q\perp}.
\end{eqnarray}
The general structure of the hard-function anomalous dimension was derived in \cite{Becher:2009cu,Gardi:2009qi,Becher:2009qa,Dixon:2009ur}. Following \cite{Kelley:2010fn}, we rewrite the anomalous dimensions as a diagonal contribution and a remainder $M_{IJ}(\cos\theta)$
\begin{equation}
\left(Γ_H\right)_{IJ}  =  \left(\gamma_{\rm cusp}\frac{c_H}{2}\ln\frac{Q^2}{μ^2}+γ_H\right)δ_{IJ}+\gamma_{\rm cusp}M_{IJ}(\cos\theta).
\end{equation}
The remainder depends on ratios of Mandelstam variables, which can be rewritten in terms of the scattering angle $\theta$. In order for the related angular dependence to cancel, the soft anomalous dimension must involve the same remainder $M_{IJ}(\cos\theta)$
\begin{eqnarray}
\left(Γ_S\right)_{IJ} & = & \left(\gamma_{\rm cusp}c_S\ln\frac{4\kappa Q\sin^2\theta}{μ}+γ_{s\perp}\right)δ_{IJ}+\gamma_{\rm cusp}M_{IJ}(\cos\theta).
\end{eqnarray}
For the case of the soft function, the angle dependence arises via scalar products of the light-like reference vectors defining the Wilson lines. In addition, eq.~(\ref{eq:RGinvcs}) also imposes some constraints on the coefficients and anomalous dimensions appearing in the diagonal parts of the equations above, which are 
\begin{equation}\label{eq:RGinv}
c_H-c_S-2c_{B_q}-2c_{j_q}=0\quad;\quad \gamma_H-\gamma_{s\perp}-2\gamma^{B_q}_{\perp}-2\gamma^{J_q}_{\perp}=0.
\end{equation}
These conditions for the diagonal and non-diagonal parts are verified by our one-loop results in the previous section, which provides a check of the computations. To verify the scale independence, one uses relations such as
\begin{align}
\frac{u}{t}&=\frac{n_a\cdot n_2}{n_a\cdot n_1}=\cot^2\frac{\theta}{2}\,,&  \frac{s(-t)}{u^2}=\frac{n_1\cdot n_2\,n_a\cdot n_1}{(n_a\cdot n_2)^2}&=2\frac{1-\cos\theta}{(1+\cos\theta)^2}\,,
\end{align}
where $s=(p_a+p_b)^2$, $t=(p_a-p_1)^2$, and $u=(p_b-p_1)^2$ are the Mandelstam variables.

\subsection{Resummation of large logarithms}\label{subsec:resum}

With the RG equations at hand, we can now derive general resummed expressions for the cross section. To do so, we solve the equations in Laplace space and then invert the results back to momentum space, using the technique of \cite{Becher:2006nr}. All RG-equations in eq.~\eqref{eq:allRGs} are of the form
\begin{equation}\label{eq:RGgeneric}
\frac{d}{d\lnμ}\,\widetilde{f}\!\left(\ln\frac{\Lambda_f}{\mu},\mu\right)  =  \left[- C_f\, \gamma_{\rm cusp} \ln\frac{\Lambda_f}{\mu} +\gamma_{f}\right] \widetilde{f}\!\left(\ln\frac{\Lambda_f}{\mu},\mu\right) \,,
\end{equation}
where $\Lambda_f$ is the characteristic scale of the given function,  $C_f$ the relevant combination of Casimir operators, and $\gamma_{f}$ its anomalous dimension. For the hard function one has $\Lambda_h=Q$, for the jet function  $\Lambda_j^2 = 4 \kappa Q^2 \sin^2 \theta$, the soft scale is $\Lambda_s = 4 \kappa Q \sin^2 \theta$, and the beam functions depend on $\Lambda_b = c_0 \bar{\kappa} = 2 c_0 \kappa Q \sin \theta$. 
The RG equations for the hadron-collider soft and hard functions are matrix valued; to bring them to the form shown in eq.~\eqref{eq:RGgeneric}, one first has to diagonalize the anomalous dimension. 

The solution of the template RG equation \eqref{eq:RGgeneric} reads 
\begin{equation}
\widetilde{f}\!\left(\ln\frac{\Lambda_f}{\mu},μ\right)=\exp\!\left[ -C_f \,S(\mu_f,\mu)- A_{\gamma_f}(\mu_f,\mu)\right]\left(\frac{\Lambda_f}{\mu_f} \right)^{\eta_{f}} \widetilde{f}\!\left(\frac{\Lambda_f}{\mu_f},μ_f\right)\,.
\end{equation}
The evolution factors $S(\mu_f,\mu)$ and $A_{f}(\mu_f,\mu)$ are given in appendix~\ref{app:anom} and the exponent $\eta_f =  C_f A_{\gamma_{\rm cusp}}(\mu_f,\mu)$.
To obtain the solution of the RG equation in momentum space, one makes use of the fact that at any order in perturbation theory $\widetilde{f}(\ln\frac{\Lambda_f}{\mu},\mu)$ is just a polynomial in the logarithm so that one can  rewrite
\begin{equation}
\widetilde{f}\!\left(\ln\frac{\Lambda_f}{\mu_i},μ_i\right)\, \left(\frac{\Lambda_f}{\mu_i} \right)^{\eta_{f}} = \widetilde{f}\!\left(\partial_{\eta_f},μ_i\right)\, \left(\frac{\Lambda_f}{\mu_i} \right)^{\eta_{f}}\,.
\end{equation}
After this, the $\Lambda_f$ dependence, and therefore also the $\kappa$ dependence, is a pure power for each ingredient of the factorization formula. Since the cross section factorizes into a product in Laplace space,  the entire cross section is proportional to a power of $\kappa$ and using the fact that
\begin{equation}
 \label{eq:LapInt}
\kappa^{a} = \int_0^\infty d \tT \, e^{-\tT/(\kappa e^{\gamma_E})}\,  \frac{\tT^{a-1} e^{-a \gamma_E}}{\Gamma(a)}  \,,
\end{equation}
we can then invert the Laplace transform and obtain the resummed result in momentum space.
Using the above template, we can solve the RG equation for all the ingredients in the factorization theorem. By evaluating each one at its characteristic scale and then combining them at a common scale $\mu$, one resums the large logarithms. Below, we choose $\mu=\mu_s$ for simplicity in order to avoid evolving the soft function.

The factorization formula in eq.~(\ref{eq:factttSCET}) for leptonic collisions resembles the cross section for ordinary thrust.  Its Laplace transform reads
\begin{align}
\widetilde{t}(\kappa)=&\int_0^{\infty}d\tT e^{-\tT /(\kappa e^{\gamma_E})}\left(\frac{dσ}{dτ_{\perp}d(\cos\theta)}\right)=\frac{\pi N_c Q_f^2\alpha^2}{2Q^2}(1+\cos^2\theta)H(Q^2,\mu)\, \nonumber\\
& \times\widetilde{j}_{c\perp}\left(\ln\frac{4\kappa Q^2\sin^2\theta}{μ^2},μ\right)\widetilde{j}_{c\perp}\left(\ln\frac{4\kappa Q^2\sin^2\theta}{μ^2},μ\right)\widetilde{s}_{\perp}\left(\ln\frac{4\kappa Q\sin^2\theta}{μ},μ\right)\,,
\end{align}
where we denote both jet functions by $\widetilde{j}_{c\perp}$. We wrote the Laplace-transformed jet and soft functions as functions of the logarithm of the arguments so that we can directly use the template eq.~\eqref{eq:RGgeneric} to solve the corresponding RG equations and to invert the Laplace transform. After a few simplifications, one then obtains the resummed result
\begin{align}
\frac{1}{σ_0}\frac{dσ}{d\tT d(\cos\theta)}=&\frac{3}{8}\left(1+\cos\theta^2\right) \exp\left[ 4 C_F S(\mu_h,\mu_j) -2A_{\gamma_H}(\mu_h,\mu_s)\right] H(Q^2,μ^2_h) \nonumber \\
&\times \exp\left[ 4 C_F S(\mu_s,\mu_j)+4A_{\gamma_{J_q}}(\mu_j,\mu_s)\right] \left(\frac{Q^2}{\mu_h^2}\right)^{-2 C_FA_{\gamma_{\rm cusp}}(\mu_h,\mu_j)} \nonumber\\
&\times \left[\widetilde{j}_{c\perp}\left(\ln\frac{μ_sQ}{μ_j^2}+\partial_{\eta},μ_j\right)\right]^2\widetilde{s}_{\perp}\left(\partial_{\eta},μ_s\right)(4\sin^2\theta)^{\eta}\frac{1}{\tT}\left(\frac{\tT Q}{μ_s}\right)^{\eta}\frac{e^{-γ_E\eta}}{Γ(\eta)}, \label{eq:dsdttres}
\end{align}
where $\mu_h$, $\mu_j$ and $\mu_s$ are the hard, jet, and soft matching scales, at which the respective functions are evaluated. Up to the additional angle-dependence, the result has the same form as the one for ordinary thrust derived in \cite{Becher:2008cf}. The  Born-level cross section is $\sigma_0=\frac{4\pi N_c Q_f^2\alpha^2}{3Q^2}$  and $\eta=4 C_F A_{\gamma_{\rm cusp}}(\mu_j,\mu_s)$.  The factor $3/8$ on the r.h.s. accounts for the integral over $\cos\theta$
\begin{equation}
\int_{-1}^1\!d\cos\theta\left(1+\cos\theta^2\right)=\frac{8}{3}.
\end{equation}
At NLL accuracy, one can perform the full angular integral in eq.~\eqref{eq:dsdttres} analytically using the identity
\begin{equation}
 \int_{-1}^1\!d\cos\theta \, (\sin^2\theta)^{\eta} = \frac{\sqrt{\pi } \,\Gamma (1+\eta )}{\Gamma \left(\frac{3}{2}+\eta\right)}\,.
 \end{equation}

To obtain the resummed result also for the hadron-collider case, we start from eq.~(\ref{eq:factformgenchann}) and evolve the hard and jet functions from the scales $\mu_h$ and $\mu_j$ down to the scale $\mu$ at which the soft and beam functions are evaluated. 
The resummed expression for the hard functions in the different channels were given in ref.~\cite{Kelley:2010fn} and we will not reproduce them here. The solution for the jet functions is the same as in the lepton collider case. What remains is the anomaly times the remainder function $W^{ab\to12}_{JI}$, given by the product of beam and soft functions, see eq.~\eqref{eq:SBB}. To be able to use eq.~\eqref{eq:LapInt} to invert the Laplace transform, we follow ref.~\cite{Becher:2012qc} and write 
\begin{align}\label{eq:anomRewrite}
W_{JI}^{ab\to12}(L_\perp,x_a,x_b,\mu) \left(\frac{Q^2}{c_0^2\bar{\kappa}^2}\right)^{-F^{ab}_{\perp}(L_\perp,\mu)}&=W_{JI}^{ab\to12}(L_\perp,x_a,x_b,\mu)  E^{ab}_\perp(L_\perp,\mu) \left(\frac{c_0 \bar{\kappa}}{\mu} \right)^{\bar{\eta}}\nonumber \\
& = W_{JI}^{ab\to12}(-2\partial_{\bar{\eta}},x_a,x_b,\mu) E^{ab}_\perp(-2\partial_{\bar{\eta}},\mu) \left(\frac{c_0 \bar{\kappa}}{\mu} \right)^{\bar{\eta}}\,.
\end{align}
We have rewritten the anomaly and the remainder $W_{JI}^{ab\to12}$ as functions of the logarithm $L_\perp =2 \ln\frac{\mu}{c_0\bar{\kappa}}$ and have introduced the exponent
\begin{equation}
\bar{\eta}=\frac{\alpha_s}{4\pi}(C_a+C_b)\Gamma_0\ln\frac{Q^2}{\mu^2}\,,
\end{equation}
 as well as the quantity $E_\perp(L_\perp,\mu)$ which contains the higher-log contributions to the anomaly
\begin{equation}
E_\perp(L_\perp,\mu)=\exp\!\left(-L_\perp F^{ab}_{\perp}(L_\perp)-\bar{\eta} f_{\perp}(L_\perp)\right),
\end{equation}
and involves the function $f_{\perp}(L_\perp)$ defined through
\begin{equation}
F^{ab}_{\perp}(L_\perp)=\frac{\alpha_s}{8\pi}(C_a+C_b)\,\Gamma_0\left[L_\perp+2 f_{\perp}(L_\perp)\right].
\end{equation}
The function $E_\perp(L_\perp,\mu)$ depends on $\bar{\eta}$, but the derivatives only act on the exponent in eq.~\eqref{eq:anomRewrite}. After combining eq.~\eqref{eq:anomRewrite} with the solution for the two jet functions, which involve the evolution factor 
\begin{equation}
U_j(\mu_j,\mu)= \exp\left[-4(C_1+C_2) S(\mu_j,\mu)+2A_{\gamma_{J_{2}}}(\mu_j,\mu)+2A_{\gamma_{J_{2}}}(\mu_j,\mu)\right]\, ,
\end{equation}
and the quantity $\eta_{j_i} = 2 C_i A_{\gamma_{\rm cusp}}(\mu_j,\mu)$,
we obtain the resummed cross section for the hadron collider case
\begin{align}\label{eq:hadresum}
&\frac{d\sigma}{d\tT d(\cos\theta)dx_a dx_b}  =\nonumber\\
&\hspace{1.5cm} H_{IJ}(Q,\theta, \mu)\, U_j(\mu_j,\mu) \left(\lambda_j \right)^{-\eta_{j_1}-\eta_{j_2}}
  W^{ab\to12}_{JI}(-2\partial_{\eta},\theta,\mu) E^{ab}(-2\partial_{\eta},\mu)  \nonumber\\
  & \hspace{1.5cm}\quad \times  \widetilde{j}_{c_1\perp}\left(\partial_{\eta}+\ln\lambda_j,μ_j\right)  \widetilde{j}_{c_2\perp}\left(\partial_{\eta}+\ln\lambda_j,μ_j\right)  \frac{1}{\tT}\left(\frac{2c_0 Q \sin\theta\, \tT}{\mu}\right)^{\eta}\frac{e^{-\gamma_E\eta}}{\Gamma(\eta)},
\end{align}
where
\begin{align}
\lambda_j &:=\frac{2 \mu Q \sin\theta}{c_0 \mu_j^2}\,, & \eta&:=\bar{\eta}+\eta_{j_1}+\eta_{j_2}.
\end{align}
To resum all logarithms, one combines eq.~\eqref{eq:hadresum} with the RG evolved hard function from  \cite{Kelley:2010fn}. 

\section{Two-loop anomalous dimensions for N$^2$LL resummation}\label{sec:N2LLlep}

\subsection{Leptonic collisions}

In this section we compare our results for the lepton-collider case with the fixed-order expression for the transverse-thrust spectrum, obtained numerically with the fixed-order Monte-Carlo program \verb|EVENT2|~\cite{Catani:1996vz}. If we expand the resummed expression in SCET in fixed-order perturbation theory, it has to reproduce the $\tT \to 0$ singularities of the full fixed-order result. The \verb|EVENT2| code allows us to verify this agreement numerically. 
More importantly, the comparison allows us to numerically determine the two-loop anomalous dimensions of the jet and soft functions, the only ingredients that are missing in order to achieve N$^2$LL accuracy.

The \verb|EVENT2| code provides the $\mathcal{O}(α_s)$ and $\mathcal{O}(α_s^2)$ corrections to differential event-shape spectra, i.e. the coefficients $A(\tT)$ and $B(\tT)$ in the expression
\begin{equation}\label{eq:coeffsdiffcs}
\frac{1}{σ_0}\frac{dσ}{d\tT}=δ(\tT)+\left(\frac{α_s}{2π}\right)A(\tT)+\left(\frac{α_s}{2π}\right)^2B(\tT)+\cdots,
\end{equation}
where $σ_0$ is the Born-level cross section. To obtain the transverse-thrust spectrum with \verb|EVENT2| we modified the usual thrust computation in \verb|EVENT2|. This can be easily done by evaluating the thrust after dropping the longitudinal components of the momenta.

\begin{figure}
\centering
\includegraphics[width=7.5cm]{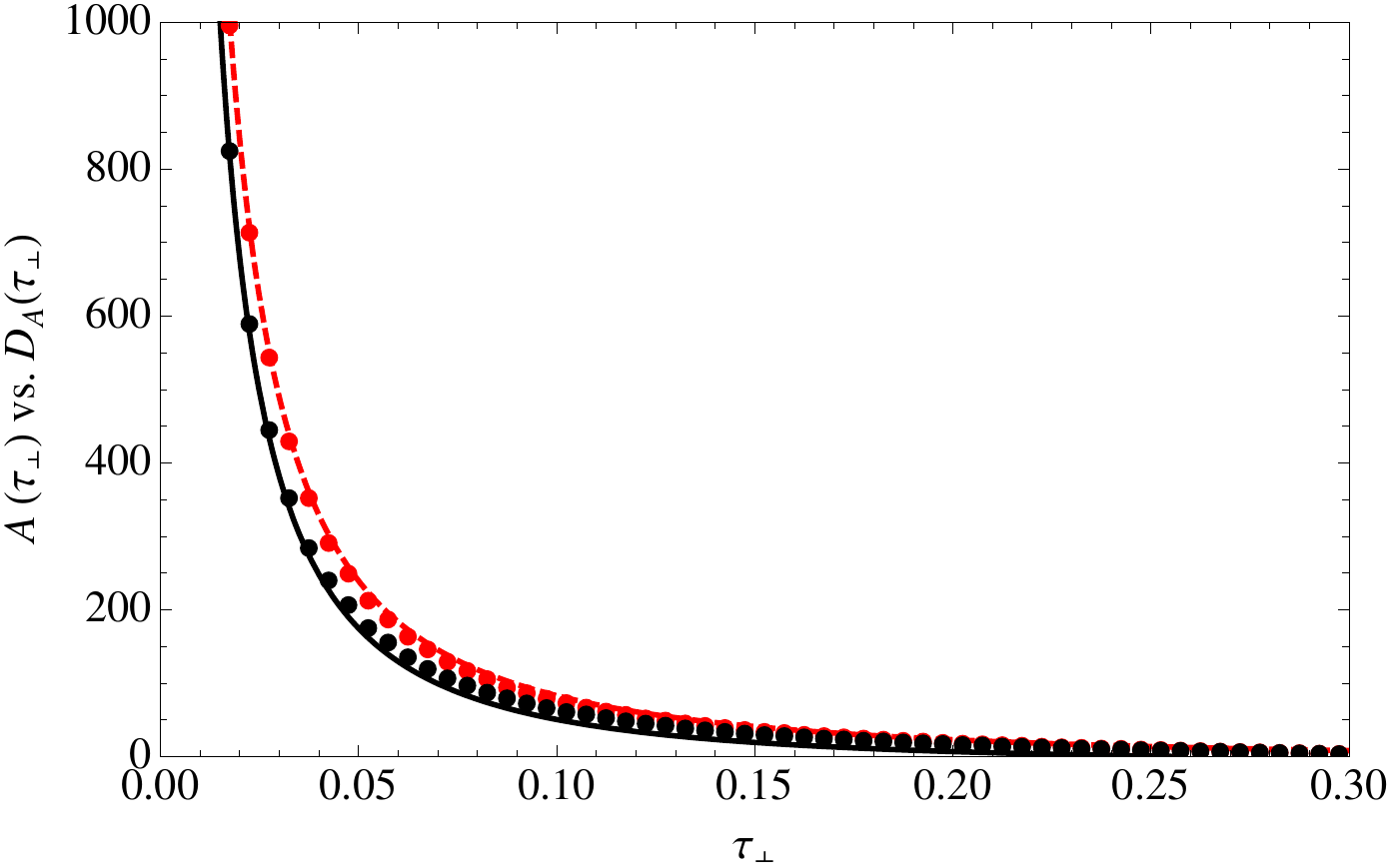}
\includegraphics[width=7.5cm]{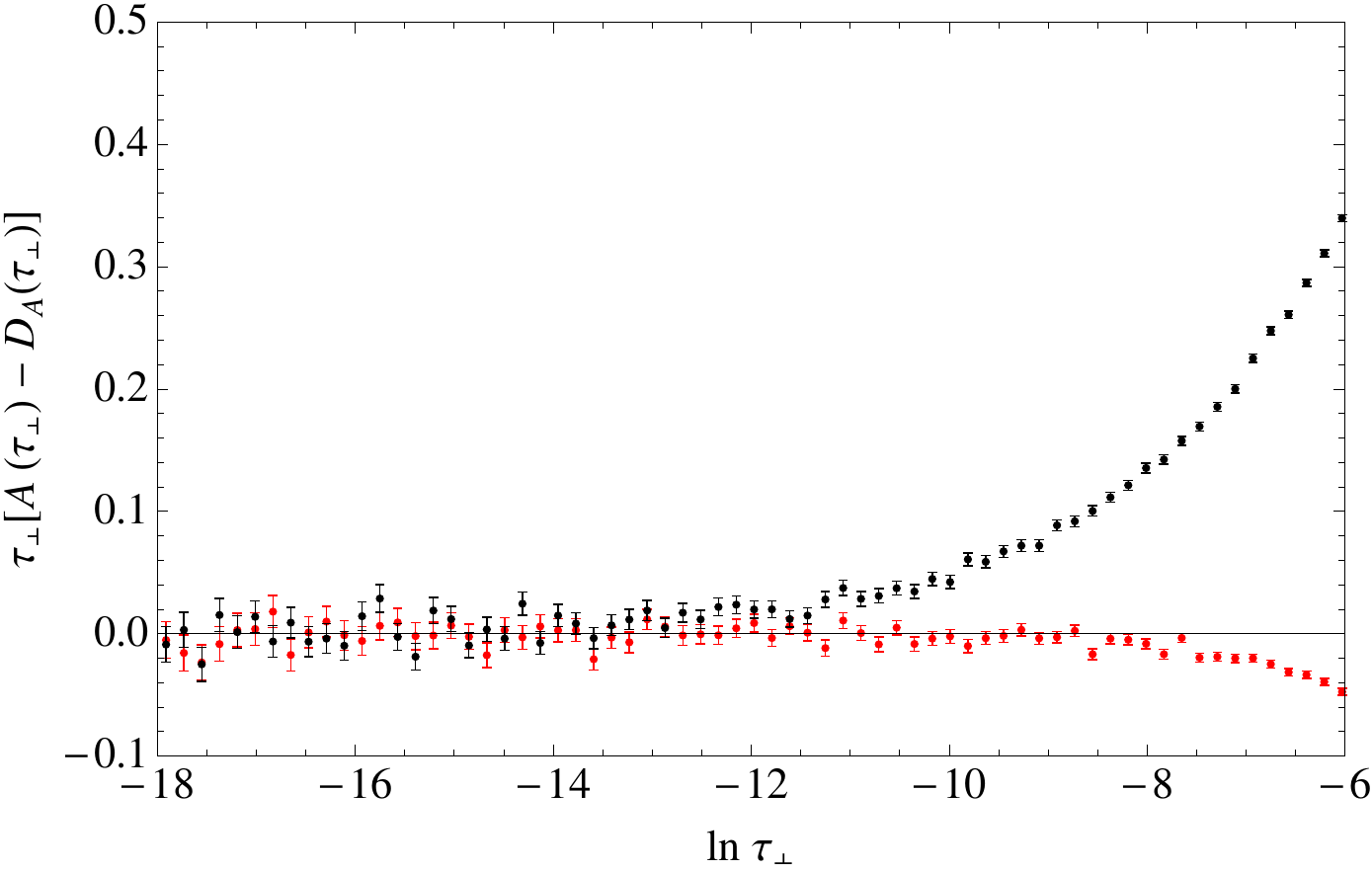}
\caption{Left panel: Comparison of the singular terms $D_A$ in eq.~(\ref{eq:singcoeffsdiffcs}) (solid line) with the one-loop coefficient $A$  from \texttt{EVENT2} (points). Right panel: One-loop term from \texttt{EVENT2} after subtracting the singular terms given by $D_A$. For comparison we show the results both for transverse thrust (black) and ordinary thrust (red).}\label{fig:ascompev2}
\end{figure}

To obtain the fixed-order expansion of the SCET result, one can set the jet scale $\mu_j$, soft scale $\mu_s$, and hard scale $\mu_h$ all equal to $Q$ in eq.~(\ref{eq:dsdttres}). All RG evolution factors become trivial in this limit and one obtains
\begin{align}
\frac{1}{σ_0}\frac{dσ}{d\tT d(\cos\theta)} 
=&\lim_{\eta\to 0} \,\frac{3}{8}\left(1+\cos\theta^2\right)H(Q^2,Q^2) \nonumber \\
&\quad\quad\times \left[\widetilde{j}_{c\perp}\left(\partial_{\eta},Q\right)\right]^2\widetilde{s}_{\perp}\left(\partial_{\eta},Q\right)\tT^{-1+\eta}\frac{e^{-γ_E\eta}}{Γ(\eta)}\left(4\sin^2\theta\right)^{\eta}.
\end{align}
The limit $\eta\to 0$ can be taken after taking the derivatives with respect to it and expanding $\tT^{-1+\eta}$ in terms of distributions.
Using the general expressions for the hard function and the Laplace-transformed soft and jet functions, as given, for instance, in ref.~\cite{Becher:2008cf}, with the one-loop coefficients and anomalous dimensions from the previous sections, we obtain the singular terms in the transverse-thrust distribution. We collect these singular terms in the $D$ coefficients, according to
\begin{equation}\label{eq:singcoeffsdiffcs}
\frac{1}{σ_0}\frac{dσ}{d\tT}=δ(\tT)+\left(\frac{α_s}{2π}\right)D_A(\tT)+\left(\frac{α_s}{2π}\right)^2D_B(\tT)+\cdots,
\end{equation}
where we integrated over $\theta$, to directly compare with the result from \verb|EVENT2|. The $D$ coefficients should reproduce the singularities in $\tT$ of the fixed order result in eq.~(\ref{eq:coeffsdiffcs}). These coefficients contain plus distributions and $\delta$-functions, but away from $\tT=0$ they reduce to regular functions. For $\tT\neq 0$, we find
\begin{equation}\label{eq:DA}
D_A(\tT)=-\frac{C_F}{3}\frac{1}{\tT}\left(-17+48\ln2+12\ln\tT\right),
\end{equation}
and when we compare with the coefficient $A$ from \verb|EVENT2| we find good agreement as $\tT\to0$, as shown in the left panel of figure~\ref{fig:ascompev2}.
To see more clearly that  the singular contributions are indeed reproduced by the effective theory, we subtract the singular contribution in eq.~(\ref{eq:DA}) from the \verb|EVENT2| result, and plot the remainder using logarithmic binning on the $x$-axis. The result of this subtraction is shown in the right panel of figure~\ref{fig:ascompev2} as black points. We can clearly see that the remainder goes to 0 when $\tT\to0$, as it should. For comparison the corresponding result for ordinary thrust is also shown in the figure (red points).

\begin{figure}
\centering
\includegraphics[width=11.9cm]{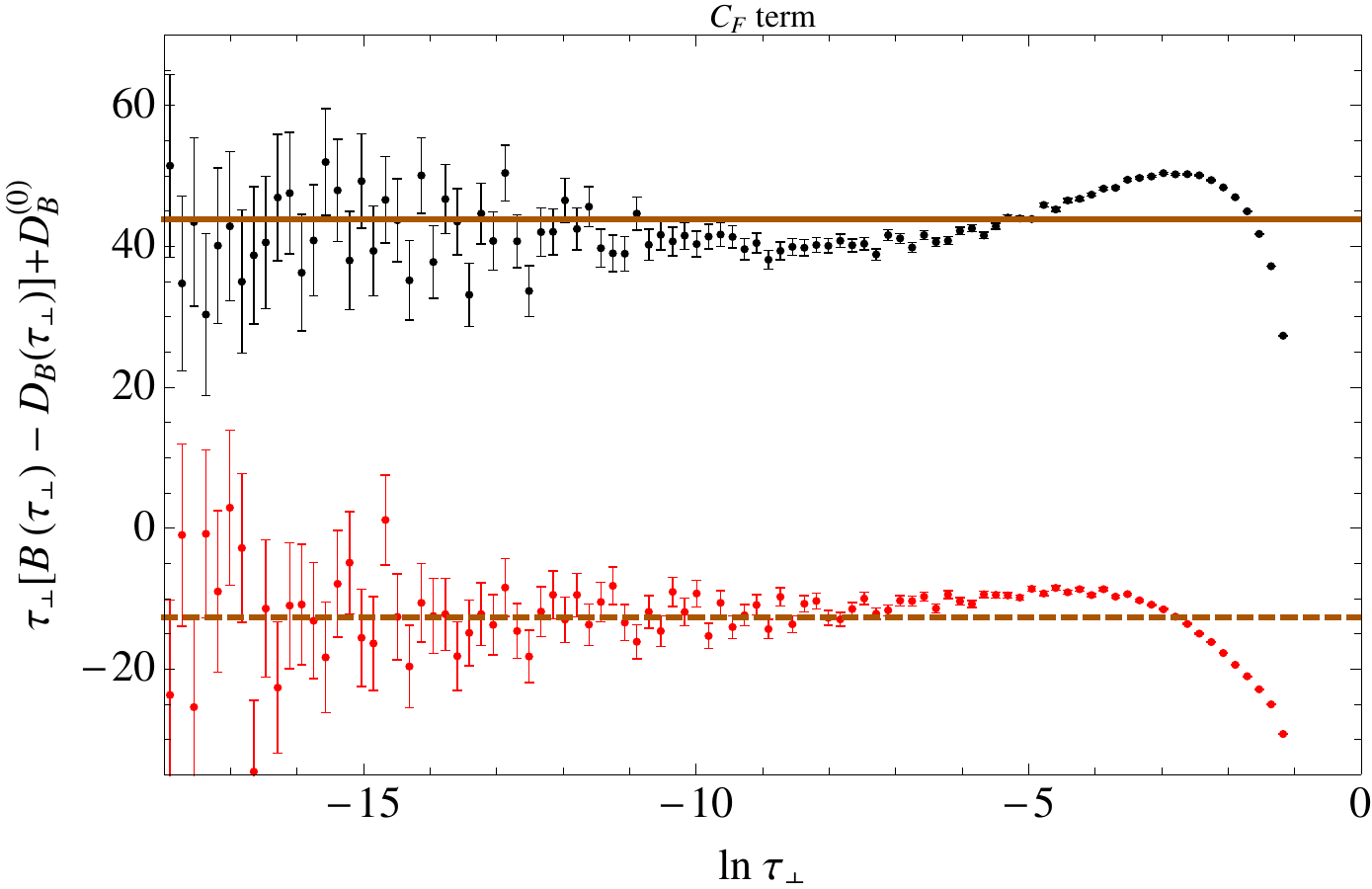}
\caption{$C_F^2$ term in $B$ from \texttt{EVENT2} after subtracting the $D_B^{(i)}$, $i=1,2,3$, singular terms defined in eq.~(\ref{eq:DB}). The horizontal lines corresponds to the $C_F^2$ part of $D_B^{(0)}$. The results  for transverse thrust  are shown in black, the ones for ordinary thrust in red color.}\label{fig:as2CFev2msing}
\end{figure}

The $D_B$ coefficient can be written as
\begin{equation}\label{eq:DB}
D_B=\frac{1}{\tT}\left(D_B^{(0)}+D_B^{(1)}\ln\tT+D_B^{(2)}\ln^2\tT+D_B^{(3)}\ln^3\tT\right)\, ,
\end{equation}
and the one-loop computations in section~\ref{sec:onelin}, together with the RG equations for the ingredients of the factorization formula, determine all coefficients, except for $D_B^{(0)}$, which depends on the two-loop anomalous dimensions of the soft and jet functions. Because of the relation $γ_H=γ_{s\perp}+2γ^{J_c}_{\perp}$, which stems from RG invariance of the cross section, and because the two-loop hard anomalous dimension is known, there is only one unknown coefficient, which we take as $γ_{1s\perp}$. Therefore, we can subtract the singular terms that contain $\ln^i\tT$, with $i=1,2,3$ from the two-loop results from \texttt{EVENT2}, and the remainder, when multiplied by $\tT$, should be constant when $\tT\to0$. By fitting this constant remainder to the expression for $D_B^{(0)}$ in terms of $γ_{1s\perp}$, we can determine the two-loop soft anomalous dimension. To do so, it is useful to separate the three color structures, $C_F^2, C_F T_F n_f$, and $C_FC_A$, which arise at two-loop order. This separation provides an additional check because the part of $γ_{1s\perp}$ proportional to $C_F^2$ must vanish by the non-abelian exponentiation theorem, since the soft function is a matrix element of soft Wilson lines. We show the remainder for the two-loop $C_F^2$ term in figure~\ref{fig:as2CFev2msing}.
From the figure, we can see that it is indeed constant, and nicely agrees with the $C_F^2$ part of $D_B^{(0)}$. For the other color structures, we fit the corresponding $D_B^{(0)}$ terms to obtain $γ_{1s\perp}$.
\begin{figure}
\centering
\includegraphics[width=11.9cm]{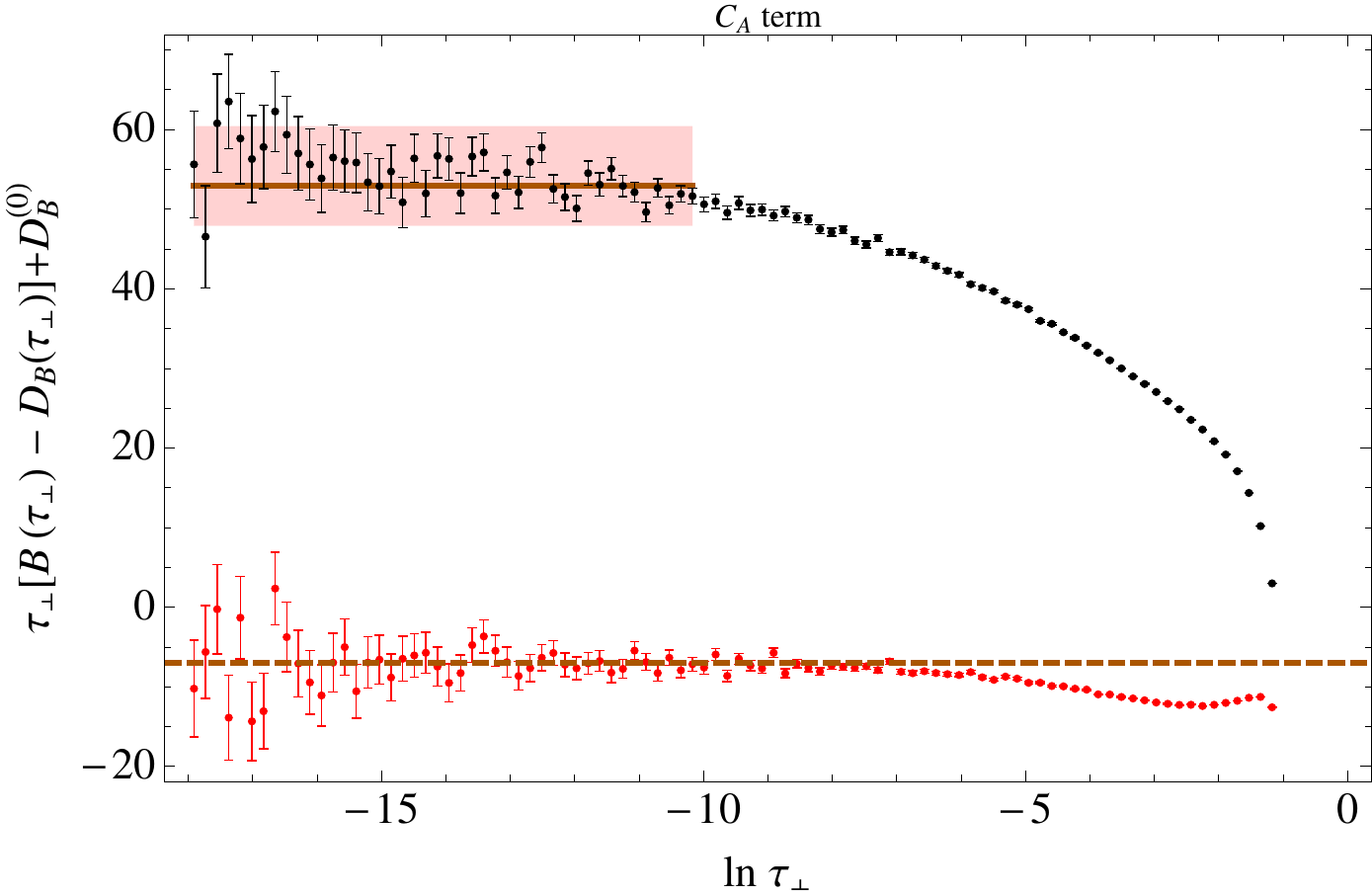}
\caption{$C_F C_Α$ term in  $B$ from \texttt{EVENT2} after subtracting the $D_B^{(i)}$, $i=1,2,3$, singular terms given in eq.~(\ref{eq:DB}). The solid line and the band are the result of the fit for the two-loop anomalous dimension for transverse thrust (see text). The results for ordinary thrust are shown as the red points and dashed line, which corresponds to the known $C_F C_A$ part of $D_B^{(0)}$.}\label{fig:as2CAev2msing}
\end{figure}
\begin{figure}
\centering
\includegraphics[width=11.9cm]{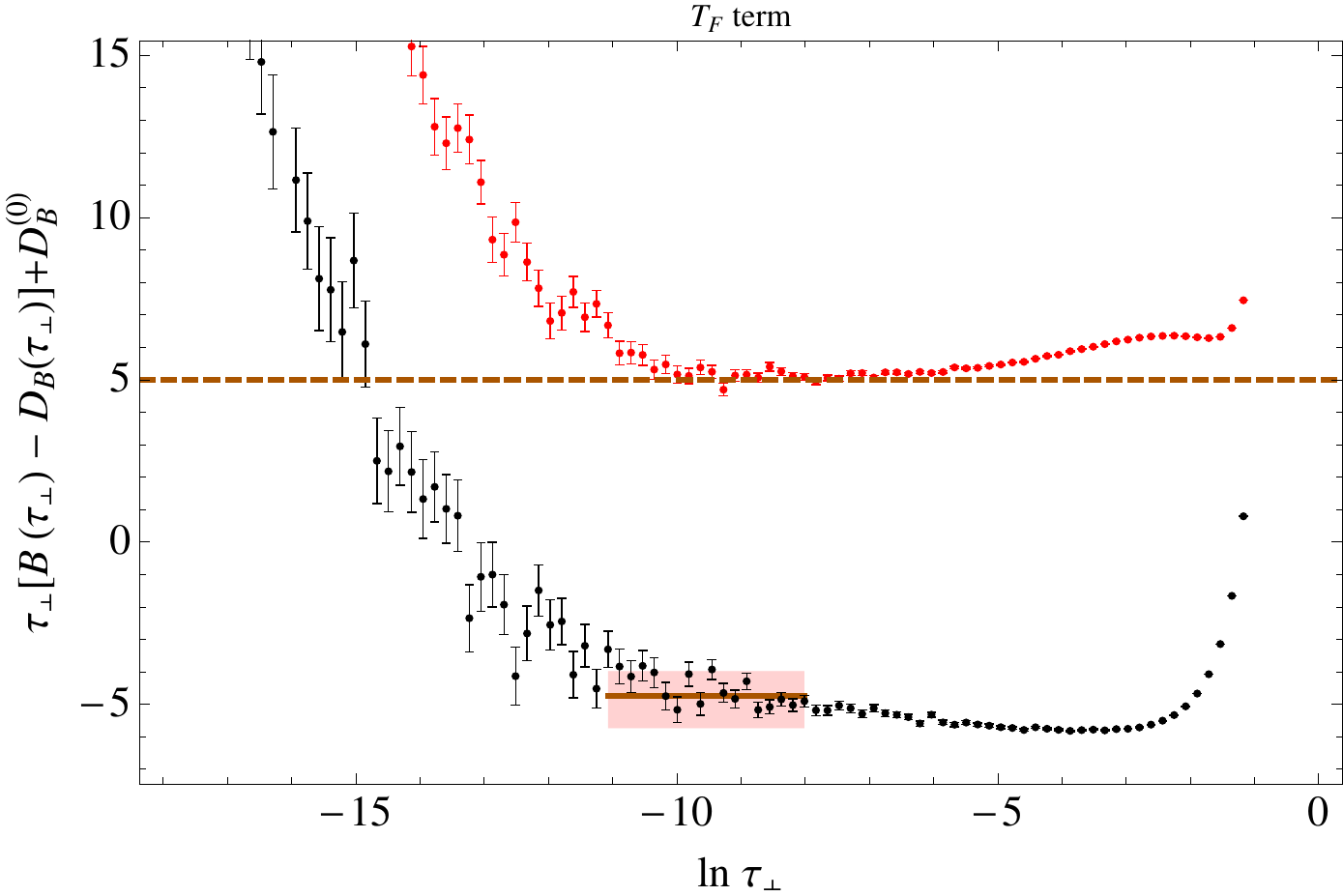}
\caption{$C_F T_F$ term in $B$ from \texttt{EVENT2} after subtracting the $D_B^{(i)}$, $i=1,2,3$, singular terms given in eq.~(\ref{eq:DB}) (black points). The solid line and the band are the result of the fit for the two-loop anomalous dimension (see text). For comparison results for ordinary thrust are shown as red points, together with a dashed line, which corresponds to the known $T_F$ part of $D_B^{(0)}$ for ordinary thrust.}\label{fig:as2TFev2msing}
\end{figure}
The results from these fits can be seen in figures~\ref{fig:as2CAev2msing} and \ref{fig:as2TFev2msing}, where the solid line is the result of a fit to the \texttt{EVENT2} result, and the band is chosen to cover the range spanned by the error bars. The range of $\ln\tT$ where the band and the line are plotted corresponds to the fit range. We can see that the remainder in the $C_F$ and $C_A$ cases nicely goes to a constant for $\tT\to0$. There seems to be some numerical instability below $\ln\tT\sim-11$ in the $T_F$ case, and the remainder grows. It is unclear why this happens\footnote{To obtain our results, we have run \texttt{EVENT2} in quadruple precision with a value of the \texttt{CUTOFF} parameter equal to $10^{-18}$ and the \texttt{NPOW} parameters set to 6. We have also performed runs (up to $\ln\tT=-12$) with \texttt{CUTOFF}$=10^{-12},10^{-15},10^{-16}$, and decreasing \texttt{NPOW} to 4, and found similar behaviour.}, but the problem also arises for the $T_F$ structure in ordinary thrust, see the red points in figure~\ref{fig:as2TFev2msing}. In view of this numerical problem, we perform the fit only for $\ln\tT>-11$ in this case. We obtain
\begin{equation}
\left.γ_{1s\perp}\right|_{C_A}=148^{+30}_{-20}\quad ;\quad\left.γ_{1s\perp}\right|_{T_F}=-18^{+3}_{-2},
\end{equation}
where
\begin{equation}
γ_{1s\perp}=:C_FC_A\left.γ_{1s\perp}\right|_{C_A}+C_FT_Fn_f\left.γ_{1s\perp}\right|_{T_F},
\end{equation}
and we recall that, as discussed in previous sections, the above numbers correspond to the following definition of the anomalous dimension
\begin{equation}
\frac{d}{d\ln\mu}\widetilde{s}_{\perp}\left(\ln\frac{4\kappa Q\sin^2\theta}{μ},μ\right)=-2\left(γ_{\rm cusp}c_S\ln\frac{4\kappa Q\sin^2\theta}{μ}+γ_{s\perp}\right)\widetilde{s}_{\perp}\left(\ln\frac{4\kappa Q\sin^2\theta}{μ},μ\right).
\end{equation}

With the value of the two-loop soft anomalous dimension obtained in the equations above, and the two-loop jet anomalous dimension determined through the relation $\gamma_{1\perp}^{J_c}=(\gamma_{1H}-\gamma_{1s\perp})/2$, we have now determined all ingredients for N$^2$LL resummation accuracy for the transverse-thrust differential cross section in leptonic collisions.

\subsection{Hadronic collisions}

To perform N$^2$LL resummation in the hadron-collider case, we need the two-loop anomalous dimensions of all the ingredients, together with the two-loop anomaly exponent. The general result for the two-loop anomalous dimension of the hard function is known due to factorization constraints, and the two-loop jet function anomalous dimension is known numerically, from our determination in the previous subsection. This leaves the two-loop anomalous dimensions of the beam and soft functions and the anomaly exponent as unknowns. However, the soft and beam functions are evaluated at the same scale, so we only need the anomalous dimension of their product, i.e.\ the anomalous dimension of the remainder function $\widetilde{W}_{JI}$ in eq.~\eqref{eq:SBB}, which is 
\begin{equation}\label{eq:gammaW}
\gamma^{W_{q\bar{q}}}_\perp = \gamma_{s\perp}+2\gamma^{B_q}_{\perp} =\gamma_H-2\gamma^{J_q}_{\perp}\,,
\end{equation}
see eq.~\eqref{eq:RGinv}. The anomaly exponent $F_\perp$ is thus the single unknown ingredient for N$^2$LL accuracy in the quark-jet case. For gluon jets, also the two-loop gluon-jet anomalous dimension $\gamma^{J_g}_{\perp}$ will be needed. We can obtain the two-loop result for $\gamma^{J_g}_{\perp}$ by considering transverse thrust for the process $H \to gg+X$, which involves gluon jet functions instead of the quark jet functions relevant for $\gamma^* \to q\bar{q}+X$. The two-loop soft functions of the two processes are related by Casimir scaling, i.e.\ the one for the gluon case can be obtained by replacing $C_F\to C_A$ in the result for the $q\bar{q}$ final state. Since the hard anomalous dimensions are also known, $\gamma^{J_g}_{\perp}$ follows from RG invariance. Due to Casimir scaling, it is also sufficient to determine $F_\perp$ in an arbitrary channel and because the divergences in the analytic regulator cancel, a computation of the two-loop $1/\alpha$ divergence of either the beam or the soft function will be sufficient. We will now discuss the simplest way to extract the anomaly $F_\perp$.

Since soft matrix elements are simpler than collinear ones, it seems preferable to extract the anomaly from the soft function. However, it involves Wilson lines along four directions, which leads to nontrivial color structure and nontrivial dependence on the scattering angle $\theta$. Both complications can be avoided by considering transverse thrust for $p p \to Z + {\rm jet}$ and $p p \to Z + Z $. The first one, involving a single $Z$-boson, is of interest phenomenologically and has been measured by CMS \cite{Chatrchyan:2013tna}. The factorization theorem for this case has the same structure as the one for the two-jet case, but obviously the hard functions are the ones relevant for $p p \to Z + {\rm jet}$, which were determined to two-loop accuracy in \cite{Becher:2013vva}. The jet function relevant for the $Z$-boson is trivial. While there can be hadronic radiation collinear to the $Z$-boson, this effect is power suppressed in $\tT$ and does not arise in our leading-order factorization theorem. In addition, the soft function involves only three-legs because the $Z$-boson does not carry color charge. The color structure of the soft function is then trivial, see e.g.\ \cite{Becher:2012za}. 

The soft function becomes even simpler in the $p p \to Z + Z $ case, which does not involve hard radiation in the final state, except along the beam directions. We can simplify things even a bit more by considering energetic electrons instead of $Z$-bosons, i.e. by computing transverse thrust in $p p \to \gamma^* \to e^+ \, e^- $, which involves the same hard function we encountered in the lepton-collider case. Neglecting electromagnetic interactions, both jet functions are trivial and the only hadronic contributions arise from the beam and soft functions. In this case, the soft function will be scaleless and vanishes with the standard analytic regulator \cite{Becher:2011dz}, but we can use the form eq.~\eqref{splitreg} to work with a non-vanishing soft function. A simple form of the two-loop soft matrix element can be found in Appendix C of \cite{Becher:2012qc}. Instead of performing the two-loop computation analytically, one can try to extract the anomaly coefficient numerically using a N$^2$LO fixed order code for $p p \to \gamma^* \to e^+ \, e^- $, such as FEWZ \cite{Gavin:2010az} or DYNNLO \cite{Catani:2009sm}. To increase the accuracy, one can run this code using simple model PDFs, since the anomaly does not depend on their form.

It is interesting to ask whether this two-loop computation could be avoided by defining an $e^+ e^- \to  2\, {\rm jets}$ event shape involving the same soft function as transverse thrust in $p p \to e^+\, e^- $. For such an observable, one could run \verb|EVENT2| and extract the anomaly coefficient using the same technique as we used for the jet-function anomalous dimension $\gamma^{J_q}_{\perp}$ in the previous subsection. Naively, it appears that one can achieve this by computing transverse thrust along the electron direction in $e^+ e^- \to  2\, {\rm jets}$, where transverse is defined with respect to the thrust axis. The outgoing jets would play the role of the beams in the hadronic collisions and the incoming electron would define the thrust vector. However, an interesting complication arises: in contrast to the beam axis, the thrust axis is recoiling against soft radiation. This changes the factorization theorem, which then has the same form as the one for jet broadening \cite{Chiu:2011qc,Becher:2011pf}. In the presence of recoil, the soft function is not exactly the same as the one shown in eq.~\eqref{eq:defSoft}, but will have to be computed at a fixed value of the transverse momentum in each hemisphere (where the hemispheres are defined by the thrust vector). The soft transverse momentum is opposite and equal to the transverse momentum of the collinear radiation in the given hemisphere and the event shape will then involve an integral over the transverse momentum. To avoid this problem and make the $e^+ e^- \to  2\, {\rm jets}$ case more similar to the $p p \to e^+\, e^- $ case, one could try to use a recoil-free definition of the jet axis, such as the broadening axis or the winner-take-all axis \cite{Larkoski:2014uqa}. This solves the recoil problem, but the two axes will in general not be back to back, and in order to define them, one will first need to split the event into two hemispheres, for example using the thrust axis. The resulting soft function is thus again more complicated than the original one, and the two-loop anomaly coefficient will likely be a function of the angle between the thrust axis and the broadening axis and extracting this function numerically seems difficult. We conclude that a numerical extraction using $p p \to \gamma^* \to e^+ \, e^- $ looks more promising. We will not pursue the extraction further in the present paper, but plan to come back to this issue in the future when we implement the resummed expression for transverse thrust numerically.

\section{Conclusions and outlook}\label{sec:concl}

In this paper we have analyzed transverse thrust in the dijet limit. Our findings are synthesized in eq.~(\ref{eq:factformgenchann}), the factorized expression of the transverse-thrust cross section for hadronic colliders in Laplace space. This result, which we derived within the framework of SCET, provides the basis for all-order resummations of enhanced perturbative corrections to this observable, beyond the NLL accuracy which has been achieved in the literature.  

The factorization formula for transverse thrust is quite interesting from the point of view of the effective theory and displays several remarkable features, which are worth emphasizing. In particular, it involves collinear modes at more than one invariant mass scale, together with soft modes, and therefore contains the two regimes of SCET, called SCET$_{\mathrm I}$ and SCET$_{\mathrm{II}}$, together in the same problem; this is the first collider-physics example that we are aware of, where this occurs. It also involves a collinear factorization anomaly, beam and jet functions, as well as matrix-valued soft and hard functions, bringing together in one problem several different effective-theory objects that were developed and studied in recent years. Anomaly cancellation involves an intricate interplay between the soft and beam functions, which leads to all-order constraints on the form of the collinear anomaly, similar to the constraints factorization imposes on the structure of soft anomalous dimensions. 

As an instructive starting point of our studies, we analyzed transverse thrust in leptonic collisions, which provide a simplified environment to study transverse event shapes. The resulting factorization formula is much simpler than the hadron-collider result and has the same structure as the one for ordinary thrust. Using our result, we were able to numerically extract a two-loop anomalous dimension that is needed in the hadron-collider case from an analysis of the leptonic result; this provides a nice example of the universality which becomes manifest after separating physics associated with the different relevant energy scales. 

In order to obtain N$^2$LL resummation accuracy for hadronic colliders from our general factorization formula, one needs the two-loop anomalous dimensions as well as the two-loop anomaly exponent. Using RG invariance and the general results for the anomalous dimensions of the hard function, together with our result for the lepton-collider case, we obtain all two-loop anomalous dimensions. The single missing ingredient for N$^2$LL accuracy is thus the two-loop anomaly coefficient, which can be extracted numerically by computing transverse thrust in Drell-Yan production using one of the existing N$^2$LO fixed-order codes.

In order to perform a phenomenological analysis and compare with data, one needs to match the resummed results to a fixed-order computation~\cite{Nagy:2003tz}, and to carefully study power corrections. Our results for leptonic collisions seem to indicate that the matching to fixed-order may be a larger correction than for regular thrust. In the future, we will extract the  missing ingredient for hadronic collisions numerically and implement the N$^2$LL resummed expression for transverse thrust into a numerical code.

It would be interesting to extend the present analysis to other hadronic event shapes. Instead of analytically computing the ingredients of the associated factorization theorems for each given observable, it would be much more efficient to perform a fully numerical evaluation of the corresponding soft, jet, and beam functions; this would provide an effective-theory-based, automated framework for computing such observables, which could be integrated with efforts to construct a SCET-based Monte Carlo event generator~\cite{Alioli:2012fc}. There exists already an automated resummation framework, CAESAR~\cite{Banfi:2004yd}, for hadronic event shapes at NLL accuracy, and for a class of lepton-collider event shapes an extension to N$^2$LL was recently achieved in ref.~\cite{Banfi:2014sua}. Also within SCET, automated N$^2$LL resummation has been performed, so far not for event shapes, but for cross sections with electroweak bosons in the presence of a jet veto \cite{Becher:2014aya}, using the automated fixed-order NLO code MadGraph5\Q{_}aMC@NLO \cite{Alwall:2014hca}.

An interesting application of the event-shape computations is that they could provide an alternative subtraction scheme for dijet and $H/Z + {\rm jet}$ production, along the line of the $q_T$ subtraction scheme by Catani and Grazzini \cite{Catani:2007vq}. The resummed cross section includes the virtual corrections as well as the singularities arising for $\tT \to 0$. It can thus be used as a subtraction in this limit.  Configurations with $\tT> 0 $, on the other hand can be computed using the NLO prediction for $Z + 2 \,{\rm jets}$ (or $3 \,{\rm jets}$ for dijet observables).

Before closing, let us mention that we did not consider Glauber gluons in the derivation of our factorization formula. Our result assumes the standard factorization of the cross section into a perturbative kernel convolved with PDFs and will resum the logarithms encountered in the perturbative computation of the hard-scattering kernel. Recently ref.~\cite{Gaunt:2014ska} showed that the standard diagrammatic arguments (see e.g. refs.~\cite{Bodwin:1984hc,Collins:1985ue,Collins:1988ig,Collins:book}) to show the absence of Glauber contributions fail for observables such as transverse thrust. It would be interesting to analyze the contribution from Glauber gluons within SCET but to date a complete implementation of this mode into the effective theory is not yet available. If Glauber modes are indeed present in transverse thrust, they could mediate spectator-interaction contributions to the cross section, in addition to the terms captured by our result. These would likely involve non-perturbative physics not encoded in the PDFs. A comparison of our factorized results with data may help shed some light on these effects and how they relate to the underlying-event contribution supplied by parton-shower Monte Carlo programs. Once their form is understood, our result can also be used to study these effects quantitatively. To do so, one could use a combination of event shapes that maximizes the sensitivity to underlying-event effects. On the other hand, to, for instance, obtain a determination of the strong coupling from a hadronic event shape analysis, one would choose a combination that is as less sensitive as possible to underlying-event effects. We look forward to investigate these issues in the future.

\begin{acknowledgments}
We thank Silvan Etter for collaboration on the lepton-collider part of the computations. We also thank Guido Bell and Jim Talbert for performing a numerical check of the lepton-collider soft function, Jan Piclum for pointing out typographical mistakes in the draft, and Thomas Gehrmann and Aude Gehrmann-De Ridder for useful discussions. This work is supported by the Swiss National Science Foundation (SNF) under the Sinergia grant number CRSII2\underline{ }141847\underline{ }1.

\end{acknowledgments}

\appendix

\section{Anomalous dimensions}\label{app:anom}
The coefficients of the cusp anomalous dimension that are needed for N$^2$LL accuracy are given by
\begin{eqnarray}
\Gamma_0 & = & 4 \,, \\
\Gamma_1 & = &  \left( \frac{268}{9} 
    - \frac{4\pi^2}{3} \right) C_A - \frac{80}{9}\,T_F n_f \,,\\
\Gamma_2 & = & C_A^2 \left( \frac{490}{3} 
    - \frac{536\pi^2}{27}
    + \frac{44\pi^4}{45} + \frac{88}{3}\,\zeta_3 \right) 
    + C_A T_F n_f  \left( - \frac{1672}{27} + \frac{160\pi^2}{27}
    - \frac{224}{3}\,\zeta_3 \right) \nonumber\\
   &&\mbox{}+ C_F T_F n_f \left( - \frac{220}{3} + 64\zeta_3 \right) 
    - \frac{64}{27}\,T_F^2 n_f^2 \,. 
\end{eqnarray}
The RG evolution functions
\begin{equation}
S(\nu,\mu)=-\int_{\alpha_s(\nu)}^{\alpha_s(\mu)}d\alpha\frac{\gamma_{\rm cusp}}{β(α)}\int_{α_s(ν)}^α\frac{dα'}{β(α')}\, ,
\end{equation}
and
\begin{equation}
A_{\gamma_{\rm cusp}}(\nu,\mu)=-\int_{\alpha_s(\nu)}^{\alpha_s(\mu)}d\alpha\frac{\gamma_{\rm cusp}}{β(α)}\, ,
\end{equation}
resum double and single logarithms and
\begin{equation}
\frac{dα_s(μ)}{d\lnμ}=β(α(μ)) \, ,
\end{equation}
is the QCD beta function. The quantity $A_{\gamma_{J_i}}(\nu,\mu)$ is defined like $A_{\gamma_{\rm cusp}}(\nu,\mu)$ but with $\gamma_{\rm cusp}$ replaced by $\gamma_{J_i}$.

The coefficients of the quark and gluon anomalous dimensions $\gamma^q$ and $\gamma^g$, respectively, read
\begin{eqnarray}
   \gamma_0^q &=& -3 C_F \,, \\
   \gamma_1^q &=& C_F^2 \left( -\frac{3}{2} + 2\pi^2
    - 24\zeta_3 \right)
    + C_F C_A \left( - \frac{961}{54} - \frac{11\pi^2}{6} 
    + 26\zeta_3 \right)\nonumber\\
    &&+ C_F T_F n_f \left( \frac{130}{27} + \frac{2\pi^2}{3} \right)\, ,
    \\
\gamma_0^g &=& - \beta_0 
    = - \frac{11}{3}\,C_A + \frac43\,T_F n_f \,,\\
   \gamma_1^g &=& C_A^2 \left( -\frac{692}{27} + \frac{11\pi^2}{18}
    + 2\zeta_3 \right) 
    + C_A T_F n_f \left( \frac{256}{27} - \frac{2\pi^2}{9} \right)
    + 4 C_F T_F n_f\, .
\end{eqnarray}
The color factors are given by
\begin{align}
C_F&=\frac{N_c^2-1}{2N_c}\,, & C_A&=N_c\,, &T_F&=\frac{1}{2},
\end{align}
where $N_c$ is the number of colors, and $n_f$ the number of light fermions.

\section{One-loop soft, jet, and beam functions}\label{app:oneloop}

\subsection{Jet functions}
In the main text, we gave the result for the quark jet function in light-cone gauge. For convenience we now also give the result of the diagrams in Feynman gauge, where the first three diagrams in figure~\ref{fig:jetf} contribute.
The contribution from the first diagram is
\begin{equation}
D_{1j}(\tau_{c\perp})=α_sC_F\frac{(2-2\varepsilon) Ω_{1-2\varepsilon}π^{\frac{5}{2}}2^{-2+2ε}}{(2π)^{3-2\varepsilon}}\frac{1}{τ_{c\perp}}\left(\frac{τ_{c\perp}Q^2 \sin^2\theta}{\tilde{μ}^2}\right)^{-ε}\frac{Γ(1-ε)}{Γ\left(\frac{3}{2}-ε\right)}\, ,
\end{equation}
the second and third one are identical and give
\begin{equation}
D_{2j}(\tau_{c\perp})=α_sC_F \frac{Ω_{1-2\varepsilon}π^{\frac{5}{2}}2^{2ε}}{(2π)^{3-2\varepsilon}}\frac{1}{τ_{c\perp}}\left(\frac{τ_{c\perp}Q^2 \sin^2\theta}{\tilde{μ}^2}\right)^{-ε}(1-ε)\frac{Γ(-ε)}{Γ\left(\frac{3}{2}-ε\right)}\, ,
\end{equation}
where $\Omega_d = 2\pi^{d/2}/\Gamma(d/2)$ is the $d$-dimensional solid angle. The reader can easily check that the sum of the diagrams gives the same result which is obtained in light-cone gauge in eq.~\eqref{quarkjet}.

In light-cone gauge, the bare gluon jet function is obtained by computing the fifth and sixth diagrams in figure~\ref{fig:jetf}
\begin{equation}
J_{g\perp}^{\rm bare}(τ_{c\perp})=δ(τ_{c\perp})+D_{5j}+D_{6j}\,,
\end{equation}
for which we obtain
\begin{equation}
D_{5j}(\tau_{c\perp})=α_sn_fT_F\frac{Ω_{1-2\varepsilon}π^{\frac{5}{2}}2^{2ε}}{(2π)^{3-2\varepsilon}}\frac{1}{τ_{c\perp}}\left(\frac{τ_{c\perp}Q^2 \sin^2\theta}{\tilde{μ}^2}\right)^{-ε}\frac{Γ(2-ε)}{Γ\left(\frac{5}{2}-ε\right)}\,,
\end{equation}
\begin{equation}
D_{6j}(\tau_{c\perp})=α_sC_A\frac{Ω_{1-2\varepsilon}3π^{2}}{(2π)^{3-2\varepsilon}}\frac{1}{τ_{c\perp}}\left(\frac{τ_{c\perp}Q^2 \sin^2\theta}{\tilde{μ}^2}\right)^{-ε}\frac{(-4+3\varepsilon)\Gamma(-\varepsilon)Γ(2-ε)}{(-3+2\varepsilon)Γ\left(2-2ε\right)}\, .
\end{equation}

\subsection{Beam functions}
The diagrams that contribute to the beam functions at one loop are shown in figure~\ref{fig:beamgraphs}. We denote the contribution of the $i$th diagram in the figure by $D_{iB}$ when it corresponds to the $c_a$ sector, and by $\bar{D}_{iB}$ when it corresponds to the $c_b$ sector. 

The diagrams in the first row of figure~\ref{fig:beamgraphs} contribute to the quark beam function. In Feynman gauge, the first diagram does not need the analytic regulator to be well defined and, therefore, gives the same result in the $c_a$ and $c_b$ sectors. We obtain
\begin{equation}
D_{1B}(z,\tT) =\bar{D}_{1B}(z,\tT)  = \frac{C_F \alpha_s}{\pi}\, C(\varepsilon) (1-\varepsilon) (1-z )\,  \frac{1}{\tT}\left(\frac{\mu}{2Q\tT\sin\theta}\right)^{2\varepsilon}\,,
\end{equation}
where
\begin{equation}
C(\varepsilon) = \frac{4^{ε}e^{\gamma_E  \varepsilon } \left(\psi\left(\frac{\varepsilon
   }{2}+\frac{3}{4}\right)-\psi\left(\frac{\varepsilon
   }{2}+\frac{1}{4}\right)\right) }{\sqrt
   {\pi }\, \Gamma
   \left(\frac{1}{2}-\varepsilon \right)} = \left(1 - \frac{8 G}{\pi}\varepsilon+ \frac{\pi^2}{4} \varepsilon^2   \right) + {\cal O}(\varepsilon^3)\,,
\end{equation}
with $\psi(x)$ the digamma function, and it is understood that $z$ corresponds to $x_a$ ($x_b$) in the $c_a$ ($c_b$) sector. For the sum of the second and third diagrams, which give identical contributions, we obtain
\begin{equation}
D_{(2+3)B}(x_a,\tT) = \frac{C_F \alpha_s}{\pi}  2\,C(\varepsilon) \,x_a\,  (1-x_a )^{-1-\alpha } \,\frac{1}{\tT} \left(\frac{\mu}{2Q\tT\sin\theta}\right)^{2\varepsilon}
\left(\frac{\nu}{\bar{n}_a\cdot P_a}\right)^{\alpha },
\end{equation}
\[
\bar{D}_{(2+3)B}(x_b,\tT) = \frac{C_F \alpha_s}{\pi} 2\, C(\varepsilon) \, x_b \, (1-x_b )^{-1+\alpha} \,\frac{1}{\tT} \left(\frac{\mu}{2Q\tT\sin\theta}\right)^{2\varepsilon} \left(\frac{\nu\, \bar{n}_b\cdot P_b}{4Q^2\tT^2\sin^2\theta}\right)^{\alpha } 
\]
\begin{equation}
\times\left[ 1 +\alpha \,\left(-\frac{8 G}{\pi}+ \varepsilon F\right)\right] \,+\mathcal{O}(\alpha),
\end{equation}
with 
\begin{align}
F &= -\frac{64 G^2}{\pi ^2}-\frac{16 G \ln 2}{\pi }
-\frac{\Phi\left(-\frac{1}{4},3,\frac{1}{4}\right)}{4 \pi }
-\frac{\Phi\left(-\frac{1}{4},3,\frac{1}{2}\right)}{4 \pi }-
\frac{\Phi\left(-\frac{1}{4},3,\frac{3}{4}\right)}{8 \pi }+\frac{9 \pi ^2}{4}+\ln^2 2  \nonumber\\
& \approx 8.20629\,,
\end{align}
where $\Phi(z,s,a)$ is the Lerch transcendent. The fourth diagram in the first row vanishes.

The diagrams in the second row of figure~\ref{fig:beamgraphs} contribute to the gluon beam function, and its computation is analogous to the quark case above. We obtain 
\begin{equation}
D_{5B}(z,\tT) =\bar{D}_{5B}(z,\tT)  = \frac{C_A \alpha_s}{\pi}\, C(\varepsilon) \left(-2 + \frac{2}{z} + 3 z - 2 z^2 \right) \,  \frac{1}{\tT}\left(\frac{\mu}{2Q\tT\sin\theta}\right)^{2\varepsilon}\,,
\end{equation}
\begin{equation}
D_{(6+7)B}(x_a,\tT) = \frac{C_A \alpha_s}{\pi} \,C(\varepsilon) \,x_a\,(1+x_a)\,  (1-x_a )^{-1-\alpha } \,\frac{1}{\tT} \left(\frac{\mu}{2Q\tT\sin\theta}\right)^{2\varepsilon}
\left(\frac{\nu}{\bar{n}\cdot P_a}\right)^{\alpha },
\end{equation}
\begin{multline}
\bar{D}_{(6+7)B}(x_b,\tT) = \frac{C_A \alpha_s}{\pi}  C(\varepsilon) \, x_b\left(1+x_b\right)  (1-x_b )^{-1+\alpha} \frac{1}{\tT}\! \left(\frac{\mu}{2Q\tT\sin\theta}\right)^{2\varepsilon} \!\!\left(\frac{\nu\, \bar{n}_b\cdot P_b}{4Q^2\tT^2\sin\theta^2}\right)^{\alpha }\\
\times  \left[ 1 +\alpha \,\left(-\frac{8 G}{\pi}+ \varepsilon F\right)\right] \,+\mathcal{O}(\alpha),
\end{multline}
and the eighth diagram vanishes.

The diagrams in the third row of figure~\ref{fig:beamgraphs} contribute to the off-diagonal coefficients $I_{q\leftarrow g}$, first diagram in the row, and $I_{g\leftarrow q}$, second diagram in the row. We obtain
\begin{equation}
D_{9B}(z,\tT) = \bar{D}_{9B}(z,\tT) = \frac{T_F \alpha_s}{\pi} C(\varepsilon) \left(1-\frac{2 z  (1-z )}{1-\varepsilon}\right) \frac{1}{\tau}\left(\frac{\mu}{2Q\tau\sin\theta}\right)^{2\varepsilon},
\end{equation}
\begin{equation}
D_{10B}(z,\tT) = \bar{D}_{10B}(z,\tT) = \frac{C_F \alpha_s}{\pi} C(\varepsilon)\frac{1}{z} \left(2-2 z +(1-\varepsilon) z^2\right) \frac{1}{\tau}\left(\frac{\mu}{2Q\tau\sin\theta}\right)^{2\varepsilon}.
\end{equation}

The divergences in the analytic regulator for all the expressions above can be made manifest by expanding
\begin{equation}
(1-x)^{-1-\alpha}=-\frac{1}{\alpha}\delta(1-x)+\left(\frac{1}{1-x}\right)_++\mathcal{O}(\alpha).
\end{equation}

\subsection{Soft functions}
\begin{figure}
\centering
\includegraphics[width=10cm]{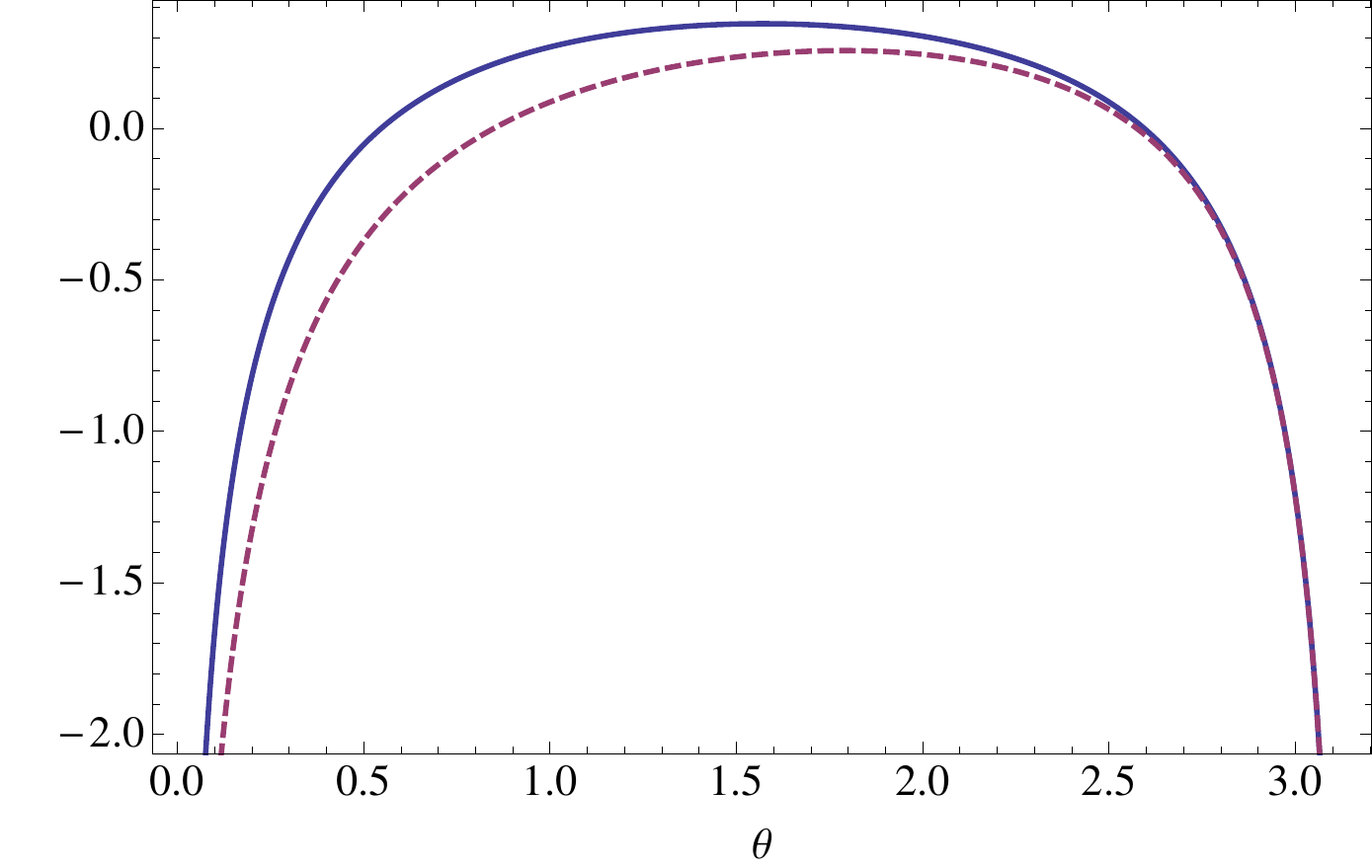}
\caption{Numerical evaluation of the $a^{(1)}(\theta)$ (blue solid line) and $b^{(1)}(\theta)$ (magenta dashed line) terms in eqs.~(\ref{eq:I12expeps}) and (\ref{eq:Iplusres}).}\label{fig:a1theta}
\end{figure}

The $\mathcal{I}$ integrals in eq.~(\ref{eq:Iijints}) are given by
\begin{eqnarray}
\mathcal{I}_{0} & = & 
\int_0^1\!\!\! dx(2-x)^{-\frac{1}{2}-ε}x^{-\frac{1}{2}+ε}\!\!\!\int_{-\infty}^{\infty}\!\!\!\!\!\! dy\frac{1}{\sqrt{1+y^2}}\frac{1}{1+y^2-(y\cos\theta+(1-x)\sin\theta)^2},\\
\mathcal{I}_{\pm} & = & %\frac{n_a\cdot n_1}{2}N\!\!
\int_0^1\!\!\! dx(2-x)^{-\frac{1}{2}-ε}x^{-\frac{1}{2}+ε+α}\!\int_{-\infty}^{\infty}\!\!\!\!\!\! dy\frac{1}{2}\left(y+\sqrt{1+y^2}\right)^{-α}\left(\frac{1}{A}\pm\frac{1}{B}\right),\\
\mathcal{I'}_{\pm} & = & %\frac{n_a\cdot n_1}{2}N\!\!
\int_0^1\!\!\! dx(2-x)^{-\frac{1}{2}-ε}x^{-\frac{1}{2}+ε+α}\!\int_{-\infty}^{\infty}\!\!\!\!\!\! dy\frac{1}{2}\left(-y+\sqrt{1+y^2}\right)^{-α}\left(\frac{1}{A}\pm\frac{1}{B}\right),
\end{eqnarray}
where
\begin{eqnarray}
A & := & \sqrt{1+y^2}\left(-y+\sqrt{1+y^2}\right)\left(\sqrt{1+y^2}-y\cos\theta-(1-x)\sin\theta\right),\\
B & := & \sqrt{1+y^2}\left(y+\sqrt{1+y^2}\right)\left(\sqrt{1+y^2}+y\cos\theta+(1-x)\sin\theta\right),
\end{eqnarray}
and therefore
\begin{eqnarray}
\frac{1}{A}+\frac{1}{B} & = & 2
\frac{1}{\sqrt{1+y^2}}\frac{1+y(1-x)\sin\theta+y^2(1+\cos\theta)}{1+y^2-(y\cos\theta+(1-x)\sin\theta)^2},\\
\frac{1}{A}-\frac{1}{B} & = & 2
\frac{y(1+\cos\theta)+(1-x)\sin\theta}{1+y^2-(y\cos\theta+(1-x)\sin\theta)^2}\, .\label{eq:1oAm1oB}
\end{eqnarray}
The normalization term $N$ reads
\begin{equation}
N:=4^{-1+ε}π^{-3+2ε}Ω_{1-2\varepsilon}\left(Q\ts |\sin\theta| \right)^{-2ε-α}\ts^{-1} \,g_s^2\tilde{μ}^{2ε}\nu^{\alpha},
\end{equation}
where it is understood that we take it at $\alpha=0$ for the $I_{12}$ integral, which does not require the analytic regulator. For convenience, we also define
\begin{equation}
M:=π^{-2+2ε}Ω_{1-2\varepsilon}\left(Q\ts\sin^2\theta\right)^{-2ε}\ts^{-1} \,g_s^2 \tilde{μ}^{2ε}.
\end{equation}
The result for the integrals reads 
\begin{align}
\frac{n_1\cdot n_2}{2}N\mathcal{I}_{0} & =  M\left(\frac{1}{8ε}+ε\, a^{(1)}(\theta)+\mathcal{O}(\varepsilon^2)\right),\label{eq:I12expeps}\\
\frac{n_a\cdot n_1}{2}N\,\mathcal{I}_+ & =  \frac{M}{2}\left(\frac{2\cot\frac{\theta}{2}}{\sin\theta}\right)^{-2ε}\left(\frac{1}{8ε}+ε\, b^{(1)}(\theta)+\mathcal{O}(α,\varepsilon^2)\right)\label{eq:Iplusres},\\
\frac{n_a\cdot n_1}{2}\mathcal{I}_- & =  \int_0^1\!\!\! dx(2-x)^{-1-ε}x^{-1+ε+α}\left\{\frac{\sqrt{2-x}\sqrt{x}}{α}+{\rm sign}(\theta)\frac{π}{2}(1-x)+\mathcal{O}(α)\right\}\nonumber\\
& = \frac{\pi}{2}\left\{\frac{1}{\alpha}\left(1-\frac{8G ε}{\pi}\right)+H+ ε K+{\rm sign}(\theta)\left(\frac{1}{2ε} -\ln 2+ \frac{\pi^2}{12} ε \right) \right\}  +\mathcal{O}(α,ε^2)\,.\label{eq:Iminusres}
\end{align}
For arbitrary values of $\alpha$, one has $\mathcal{I}_+'(\theta) = \mathcal{I}_+(-\theta)$ and $\mathcal{I}_-'(\theta) = -\mathcal{I}_-(-\theta)$. The numerical values of the constants are $H\approx -1.85939$ and $K\approx 8.44015$. The results for the functions $a^{(1)}(\theta)$ and $b^{(1)}(\theta)$ are plotted in figure~\ref{fig:a1theta}, as blue solid and magenta dashed lines, respectively. The functions are symmetric in $\theta$. For positive $\theta$, they are obtained from a numerical evaluation of the expressions 
\begin{align}
a^{(1)}(\theta)= &\int_0^1dx\frac{(\sin\theta)^{-1}}{8\pi(2-x)}\Bigg(\delta(x)\left(\ln2+2\ln\sin\theta\right)^2\nonumber \\
& \quad
+\left[\frac{1}{x}\right]_+\left( 4\ln2 -2\ln(2-x)+4\ln\sin\theta\right)+\left[\frac{\ln x}{x}\right]_+2\Bigg) \nonumber\\
&\quad \times{\rm Im}\left\{\frac{1}{\sqrt{1+y_1^2}}\ln\left(\frac{y_1-\sqrt{1+y_1^2}}{y_1+\sqrt{1+y_1^2}}\right)\right\}, 
\end{align}
\begin{align}
b^{(1)}(\theta) =&\int_0^1dx\frac{(\sin\theta)^{-1}}{8\pi(2-x)}\Bigg(\delta(x)\left(3\ln2+2\ln\cot\frac{\theta}{2}\right)^2 \nonumber\\
&\quad +\left[\frac{1}{x}\right]_+\left(8\ln2-2\ln(2-x)+4\ln\cot\frac{\theta}{2}\right)+\left[\frac{\ln x}{x}\right]_+2\Bigg) \nonumber\\
& \quad\times{\rm Im}\left\{\frac{(1-\cos\theta)\left(1+y_1(1-x)\sin\theta+y_1^2(1+\cos\theta)\right)}{\sqrt{1+y_1^2}}\ln\left(\frac{y_1-\sqrt{1+y_1^2}}{y_1+\sqrt{1+y_1^2}}\right)\right\},
\end{align}
where $y_1$ was given in eq.~(\ref{eq:y1y2}). 

\section{Lepton tensor}\label{app:lepttens}
The lepton tensor, including contributions from photon and $Z$ exchanges, is given by
\begin{eqnarray}
L_{μν}^V & = &
-\frac{e^4}{2Q^2}\left(g_{μν}-2\frac{p_{1μ}p_{2ν}+p_{2μ}p_{1ν}}{Q^2}\right)\left[Q_f^2-\frac{2Q^2v_ev_fQ_f}{Q^2-M_Z^2}+\frac{Q^4(v_e^2+a_e^2)v_f^2}{(Q^2-M_Z^2)^2}\right]\nonumber\\
&=:&-L^V \left(g_{μν}-2\frac{p_{1μ}p_{2ν}+p_{2μ}p_{1ν}}{Q^2}\right),\\
L_{μν}^A & = &
-\frac{e^4}{2Q^2}\left(g_{μν}-2\frac{p_{1μ}p_{2ν}+p_{2μ}p_{1ν}}{Q^2}\right)\frac{Q^4(v_e^2+a_e^2)a_f^2}{(Q^2-M_Z^2)^2}\nonumber\\
&=:&-L^A \left(g_{μν}-2\frac{p_{1μ}p_{2ν}+p_{2μ}p_{1ν}}{Q^2}\right),
\end{eqnarray}
with $p_1$ and $p_2$ the electron and positron momenta,
respectively. The vector and axial charges, $v_f$ and $a_f$, are defined as
\begin{equation}
v_f:=\frac{T^3_f-2Q_fs_W^2}{2s_Wc_W}\quad;\quad a_f:=\frac{T_f^3}{2s_Wc_W},
\end{equation}
with $T_f^3$ the third component of the weak isospin, $Q_f$ the
electric charge of fermion $f$ (with $Q_e=2\,T_e^3=-1$), $c_W:=\cos\theta_W$, $s_W:=\sin\theta_W$, $\theta_W$ the weak mixing
angle, and $M_Z$ the Z-boson mass.


\begin{thebibliography}{99}
\bibitem{Banfi:2004nk}
  A.~Banfi, G.~P.~Salam and G.~Zanderighi,
  %``Resummed event shapes at hadron - hadron colliders,''
  JHEP {\bf 0408} (2004) 062
  [hep-ph/0407287].
  %%CITATION = HEP-PH/0407287;%%
  %73 citations counted in INSPIRE as of 27 May 2014

\bibitem{Banfi:2010xy}
  A.~Banfi, G.~P.~Salam and G.~Zanderighi,
  %``Phenomenology of event shapes at hadron colliders,''
  JHEP {\bf 1006} (2010) 038
  [arXiv:1001.4082 [hep-ph]].
  %%CITATION = ARXIV:1001.4082;%%
  %59 citations counted in INSPIRE as of 27 May 2014

\bibitem{Khachatryan:2011dx}
  V.~Khachatryan {\it et al.}  [CMS Collaboration],
  %``First Measurement of Hadronic Event Shapes in $pp$ Collisions at $\sqrt(s)=7$ TeV,''
  Phys.\ Lett.\ B {\bf 699} (2011) 48
  [arXiv:1102.0068 [hep-ex]].
  %%CITATION = ARXIV:1102.0068;%%
  %43 citations counted in INSPIRE as of 28 Nov 2014

\bibitem{Aad:2012np}
  G.~Aad {\it et al.}  [ATLAS Collaboration],
  %``Measurement of event shapes at large momentum transfer with the ATLAS detector in $pp$ collisions at $\sqrt{s}=7$ TeV,''
  Eur.\ Phys.\ J.\ C {\bf 72} (2012) 2211
  [arXiv:1206.2135 [hep-ex]].
  %%CITATION = ARXIV:1206.2135;%%
  %21 citations counted in INSPIRE as of 28 Nov 2014

\bibitem{Aad:2012fza}
  G.~Aad {\it et al.}  [ATLAS Collaboration],
  %``Measurement of charged-particle event shape variables in $\sqrt{s}=7$ TeV proton-proton interactions with the ATLAS detector,''
  Phys.\ Rev.\ D {\bf 88} (2013) 3,  032004
  [arXiv:1207.6915 [hep-ex]].
  %%CITATION = ARXIV:1207.6915;%%
  %27 citations counted in INSPIRE as of 28 Nov 2014

\bibitem{Khachatryan:2014ika}
  V.~Khachatryan {\it et al.}  [CMS Collaboration],
  %``Study of hadronic event-shape variables in multijet final states in pp collisions at sqrt(s) = 7 TeV,''
  JHEP {\bf 1410} (2014) 87
  [arXiv:1407.2856 [hep-ex]].
  %%CITATION = ARXIV:1407.2856;%%
  %1 citations counted in INSPIRE as of 28 Nov 2014

\bibitem{Aaltonen:2011et}
  T.~Aaltonen {\it et al.}  [CDF Collaboration],
  %``Measurement of Event Shapes in Proton-Antiproton Collisions at Center-of-Mass Energy 1.96 TeV,''
  Phys.\ Rev.\ D {\bf 83} (2011) 112007
  [arXiv:1103.5143 [hep-ex]].
  %%CITATION = ARXIV:1103.5143;%%
  %12 citations counted in INSPIRE as of 28 Nov 2014

\bibitem{Banfi:2004yd}
  A.~Banfi, G.~P.~Salam and G.~Zanderighi,
  %``Principles of general final-state resummation and automated implementation,''
  JHEP {\bf 0503} (2005) 073
  [hep-ph/0407286].
  %%CITATION = HEP-PH/0407286;%%
  %79 citations counted in INSPIRE as of 04 Jun 2014

\bibitem{Bauer:2000yr}
  C.~W.~Bauer, S.~Fleming, D.~Pirjol and I.~W.~Stewart,
  %``An Effective field theory for collinear and soft gluons: Heavy to light decays,''
  Phys.\ Rev.\ D {\bf 63} (2001) 114020
  [hep-ph/0011336].
  %%CITATION = HEP-PH/0011336;%%
  %772 citations counted in INSPIRE as of 04 Jun 2014

\bibitem{Bauer:2001yt}
  C.~W.~Bauer, D.~Pirjol and I.~W.~Stewart,
  %``Soft collinear factorization in effective field theory,''
  Phys.\ Rev.\ D {\bf 65} (2002) 054022
  [hep-ph/0109045].
  %%CITATION = HEP-PH/0109045;%%
  %644 citations counted in INSPIRE as of 04 Jun 2014

\bibitem{Beneke:2002ph}
  M.~Beneke, A.~P.~Chapovsky, M.~Diehl and T.~Feldmann,
  %``Soft collinear effective theory and heavy to light currents beyond leading power,''
  Nucl.\ Phys.\ B {\bf 643} (2002) 431
  [hep-ph/0206152].
  %%CITATION = HEP-PH/0206152;%%
  %341 citations counted in INSPIRE as of 04 Jun 2014
  
  %\cite{Becher:2014oda}
\bibitem{Becher:2014oda} 
  T.~Becher, A.~Broggio and A.~Ferroglia,
  %``Introduction to Soft-Collinear Effective Theory,''
  arXiv:1410.1892 [hep-ph].
  %%CITATION = ARXIV:1410.1892;%%
  %2 citations counted in INSPIRE as of 27 Oct 2014

\bibitem{Becher:2010tm}
  T.~Becher and M.~Neubert,
  %``Drell-Yan production at small q_T, transverse parton distributions and the collinear anomaly,''
  Eur.\ Phys.\ J.\ C {\bf 71} (2011) 1665
  [arXiv:1007.4005 [hep-ph]].
  %%CITATION = ARXIV:1007.4005;%%
  %87 citations counted in INSPIRE as of 17 Jun 2014
 

  \bibitem{Collins:1984kg}
  J.~C.~Collins, D.~E.~Soper and G.~F.~Sterman,
  %``Transverse Momentum Distribution in Drell-Yan Pair and W and Z Boson Production,''
  Nucl.\ Phys.\ B {\bf 250} (1985) 199.
  %%CITATION = NUPHA,B250,199;%%
  %762 citations counted in INSPIRE as of 09 Apr 2015}


\bibitem{Mateu:2013gya}
  V.~Mateu and G.~Rodrigo,
  %``Oriented Event Shapes at N$^{3}$LL $+ O(\alpha_{S}^{2})$,''
  JHEP {\bf 1311} (2013) 030
  [arXiv:1307.3513].
  %%CITATION = ARXIV:1307.3513;%%

\bibitem{MScSilvan}
  S.~Etter,
  MSc thesis. University of Bern (2014).

\bibitem{Stewart:2009yx}
  I.~W.~Stewart, F.~J.~Tackmann and W.~J.~Waalewijn,
  %``Factorization at the LHC: From PDFs to Initial State Jets,''
  Phys.\ Rev.\ D {\bf 81} (2010) 094035
  [arXiv:0910.0467 [hep-ph]].
  %%CITATION = ARXIV:0910.0467;%%
  %78 citations counted in INSPIRE as of 17 Jun 2014

\bibitem{Kelley:2010fn}
  R.~Kelley and M.~D.~Schwartz,
  %``1-loop matching and NNLL resummation for all partonic 2 to 2 processes in QCD,''
  Phys.\ Rev.\ D {\bf 83} (2011) 045022
  [arXiv:1008.2759 [hep-ph]].
  %%CITATION = ARXIV:1008.2759;%%
  %21 citations counted in INSPIRE as of 08 Oct 2014

\bibitem{Broggio:2014hoa}
  A.~Broggio, A.~Ferroglia, B.~D.~Pecjak and Z.~Zhang,
  %``NNLO hard functions in massless QCD,''
  JHEP {\bf 1412} (2014) 005
  [arXiv:1409.5294 [hep-ph]].
  %%CITATION = ARXIV:1409.5294;%%
  %5 citations counted in INSPIRE as of 09 Feb 2015

\bibitem{Becher:2009qa} 
  T.~Becher and M.~Neubert,
  %``On the Structure of Infrared Singularities of Gauge-Theory Amplitudes,''
  JHEP {\bf 0906}, 081 (2009)
  [Erratum-ibid.\  {\bf 1311}, 024 (2013)]
  [arXiv:0903.1126 [hep-ph]].
  %%CITATION = ARXIV:0903.1126;%%
  %130 citations counted in INSPIRE as of 27 Oct 2014  

%\bibitem{Beneke_lecture}
%M.~Beneke,
%\texttt{theor.jinr.ru/\~{}hq2005/Lectures/Beneke/Beneke-Dubna-05.pdf}

\bibitem{Chiu:2007dg}
  J.~-y.~Chiu, F.~Golf, R.~Kelley and A.~V.~Manohar,
  %``Electroweak Corrections in High Energy Processes using Effective Field Theory,''
  Phys.\ Rev.\ D {\bf 77} (2008) 053004
  [arXiv:0712.0396 [hep-ph]].
  %%CITATION = ARXIV:0712.0396;%%
  %41 citations counted in INSPIRE as of 03 Jul 2014

\bibitem{Becher:2011dz}
  T.~Becher and G.~Bell,
  %``Analytic Regularization in Soft-Collinear Effective Theory,''
  Phys.\ Lett.\ B {\bf 713} (2012) 41
  [arXiv:1112.3907 [hep-ph]].
  %%CITATION = ARXIV:1112.3907;%%
  %19 citations counted in INSPIRE as of 20 Jun 2014

\bibitem{Chiu:2011qc}
  J.~-y.~Chiu, A.~Jain, D.~Neill and I.~Z.~Rothstein,
  %``The Rapidity Renormalization Group,''
  Phys.\ Rev.\ Lett.\  {\bf 108} (2012) 151601
  [arXiv:1104.0881 [hep-ph]].
  %%CITATION = ARXIV:1104.0881;%%
  %43 citations counted in INSPIRE as of 03 Jul 2014

\bibitem{Becher:2011pf}
  T.~Becher, G.~Bell and M.~Neubert,
  %``Factorization and Resummation for Jet Broadening,''
  Phys.\ Lett.\ B {\bf 704} (2011) 276
  [arXiv:1104.4108 [hep-ph]].
  %%CITATION = ARXIV:1104.4108;%%
  %18 citations counted in INSPIRE as of 08 Jul 2014

\bibitem{Becher:2012qa}
  T.~Becher and M.~Neubert,
  %``Factorization and NNLL Resummation for Higgs Production with a Jet Veto,''
  JHEP {\bf 1207} (2012) 108
  [arXiv:1205.3806 [hep-ph]].
  %%CITATION = ARXIV:1205.3806;%%
  %41 citations counted in INSPIRE as of 08 Jul 2014

\bibitem{Becher:2013xia} 
  T.~Becher, M.~Neubert and L.~Rothen,
  %``Factorization and $N^{3}LL_{p}$+NNLO predictions for the Higgs cross section with a jet veto,''
  JHEP {\bf 1310}, 125 (2013)
  [arXiv:1307.0025 [hep-ph]].
  %%CITATION = ARXIV:1307.0025;%%
  %32 citations counted in INSPIRE as of 21 Nov 2014

\bibitem{Echevarria:2012js}
  M.~G.~Echevarría, A.~Idilbi and I.~Scimemi,
  %``Soft and Collinear Factorization and Transverse Momentum Dependent Parton Distribution Functions,''
  Phys.\ Lett.\ B {\bf 726} (2013) 795
  [arXiv:1211.1947 [hep-ph]].
  %%CITATION = ARXIV:1211.1947;%%
  %39 citations counted in INSPIRE as of 09 Apr 2015

\bibitem{Catani:1996jh}
  S.~Catani and M.~H.~Seymour,
  %``The Dipole formalism for the calculation of QCD jet cross-sections at next-to-leading order,''
  Phys.\ Lett.\ B {\bf 378} (1996) 287
  [hep-ph/9602277].
  %%CITATION = HEP-PH/9602277;%%
  %355 citations counted in INSPIRE as of 03 Jul 2014

  %\cite{Becher:2009cu}
\bibitem{Becher:2009cu} 
  T.~Becher and M.~Neubert,
  %``Infrared singularities of scattering amplitudes in perturbative QCD,''
  Phys.\ Rev.\ Lett.\  {\bf 102}, 162001 (2009)
  [Erratum-ibid.\  {\bf 111}, no. 19, 199905 (2013)]
  [arXiv:0901.0722 [hep-ph]].
  %%CITATION = ARXIV:0901.0722;%%
  %120 citations counted in INSPIRE as of 21 Nov 2014
  
  %\cite{Gardi:2009qi}
\bibitem{Gardi:2009qi} 
  E.~Gardi and L.~Magnea,
  %``Factorization constraints for soft anomalous dimensions in QCD scattering amplitudes,''
  JHEP {\bf 0903}, 079 (2009)
  [arXiv:0901.1091 [hep-ph]].
  %%CITATION = ARXIV:0901.1091;%%
  %110 citations counted in INSPIRE as of 21 Nov 2014
  
  %\cite{Dixon:2009ur}
\bibitem{Dixon:2009ur} 
  L.~J.~Dixon, E.~Gardi and L.~Magnea,
  %``On soft singularities at three loops and beyond,''
  JHEP {\bf 1002}, 081 (2010)
  [arXiv:0910.3653 [hep-ph]].
  %%CITATION = ARXIV:0910.3653;%%
  %61 citations counted in INSPIRE as of 21 Nov 2014

\bibitem{Becher:2008cf}
  T.~Becher and M.~D.~Schwartz,
  %``A precise determination of $\alpha_s$ from LEP thrust data using effective field theory,''
  JHEP {\bf 0807} (2008) 034
  [arXiv:0803.0342 [hep-ph]].
  %%CITATION = ARXIV:0803.0342;%%
  %126 citations counted in INSPIRE as of 23 Jun 2014

\bibitem{Becher:2006mr}
  T.~Becher, M.~Neubert and B.~D.~Pecjak,
  %``Factorization and Momentum-Space Resummation in Deep-Inelastic Scattering,''
  JHEP {\bf 0701} (2007) 076
  [hep-ph/0607228].
  %%CITATION = HEP-PH/0607228;%%
  %130 citations counted in INSPIRE as of 02 dic 2014

\bibitem{Becher:2012qc}
  T.~Becher and G.~Bell,
  %``NNLL Resummation for Jet Broadening,''
  JHEP {\bf 1211} (2012) 126
  [arXiv:1210.0580 [hep-ph]].
  %%CITATION = ARXIV:1210.0580;%%
  %16 citations counted in INSPIRE as of 02 dic 2014
  
\bibitem{Becher:2010pd} 
  T.~Becher and G.~Bell,
  %``The gluon jet function at two-loop order,''
  Phys.\ Lett.\ B {\bf 695}, 252 (2011)
  [arXiv:1008.1936 [hep-ph]].
  %%CITATION = ARXIV:1008.1936;%%
  %16 citations counted in INSPIRE as of 24 Nov 2014

\bibitem{Collins:1981uw} 
  J.~C.~Collins and D.~E.~Soper,
  %``Parton Distribution and Decay Functions,''
  Nucl.\ Phys.\ B {\bf 194}, 445 (1982).
  %%CITATION = NUPHA,B194,445;%%
  %654 citations counted in INSPIRE as of 12 Mar 2014

\bibitem{Catani:1996vz}
  S.~Catani and M.~H.~Seymour,
  %``A General algorithm for calculating jet cross-sections in NLO QCD,''
  Nucl.\ Phys.\ B {\bf 485} (1997) 291
   [Erratum-ibid.\ B {\bf 510} (1998) 503]
  [hep-ph/9605323].
  %%CITATION = HEP-PH/9605323;%%
  %1082 citations counted in INSPIRE as of 23 Jun 2014

\bibitem{Kelley:2010qs}
  R.~Kelley and M.~D.~Schwartz,
  %``Threshold hadronic event shapes with effective field theory,''
  Phys.\ Rev.\ D {\bf 83} (2011) 033001
  [arXiv:1008.4355 [hep-ph]].
  %%CITATION = ARXIV:1008.4355;%%
  %12 citations counted in INSPIRE as of 19 Jun 2014

%\cite{Glover:2003cm}
\bibitem{Glover:2003cm} 
  E.~W.~N.~Glover and M.~E.~Tejeda-Yeomans,
  %``Two loop QCD helicity amplitudes for massless quark massless gauge boson scattering,''
  JHEP {\bf 0306}, 033 (2003)
  [hep-ph/0304169].
  %%CITATION = HEP-PH/0304169;%%
  %43 citations counted in INSPIRE as of 15 Aug 2014

%\cite{Glover:2004si}
\bibitem{Glover:2004si} 
  E.~W.~N.~Glover,
  %``Two loop QCD helicity amplitudes for massless quark quark scattering,''
  JHEP {\bf 0404}, 021 (2004)
  [hep-ph/0401119].
  %%CITATION = HEP-PH/0401119;%%
  %43 citations counted in INSPIRE as of 15 Aug 2014

%\cite{Bern:2003ck}
\bibitem{Bern:2003ck} 
  Z.~Bern, A.~De Freitas and L.~J.~Dixon,
  %``Two loop helicity amplitudes for quark gluon scattering in QCD and gluino gluon scattering in supersymmetric Yang-Mills theory,''
  JHEP {\bf 0306}, 028 (2003)
  [Erratum-ibid.\  {\bf 1404}, 112 (2014)]
  [hep-ph/0304168].
  %%CITATION = HEP-PH/0304168;%%
  %72 citations counted in INSPIRE as of 15 Aug 2014

%\cite{De_Freitas:2004tk}
\bibitem{DeFreitas:2004tk} 
  A.~De Freitas and Z.~Bern,
  %``Two-loop helicity amplitudes for quark-quark scattering in QCD and gluino-gluino scattering in supersymmetric Yang-Mills theory,''
  JHEP {\bf 0409}, 039 (2004)
  [hep-ph/0409007, hep-ph/0409007].
  %%CITATION = HEP-PH/0409007,;%%
  %29 citations counted in INSPIRE as of 15 Aug 2014

%\cite{Bern:2002tk}
\bibitem{Bern:2002tk} 
  Z.~Bern, A.~De Freitas and L.~J.~Dixon,
  %``Two loop helicity amplitudes for gluon-gluon scattering in QCD and supersymmetric Yang-Mills theory,''
  JHEP {\bf 0203}, 018 (2002)
  [hep-ph/0201161].
  %%CITATION = HEP-PH/0201161;%%
  %144 citations counted in INSPIRE as of 15 Aug 2014
  
  %\cite{Becher:2006nr}
\bibitem{Becher:2006nr} 
  T.~Becher and M.~Neubert,
  %``Threshold resummation in momentum space from effective field theory,''
  Phys.\ Rev.\ Lett.\  {\bf 97}, 082001 (2006)
  [hep-ph/0605050].
  %%CITATION = HEP-PH/0605050;%%
  %92 citations counted in INSPIRE as of 22 Jan 2015

\bibitem{Nagy:2003tz}
  Z.~Nagy,
  %``Next-to-leading order calculation of three jet observables in hadron hadron collision,''
  Phys.\ Rev.\ D {\bf 68} (2003) 094002
  [hep-ph/0307268].
  %%CITATION = HEP-PH/0307268;%%
  %372 citations counted in INSPIRE as of 12 Dec 2014
  
  \bibitem{Chatrchyan:2013tna} 
  S.~Chatrchyan {\it et al.}  [CMS Collaboration],
  %``Event shapes and azimuthal correlations in $Z$ + jets events in $pp$ collisions at $\sqrt{s}=7$ TeV,''
  Phys.\ Lett.\ B {\bf 722}, 238 (2013)
  [arXiv:1301.1646 [hep-ex]].
  %%CITATION = ARXIV:1301.1646;%%
  %28 citations counted in INSPIRE as of 13 Jan 2015
 
  \bibitem{Becher:2013vva} 
  T.~Becher, G.~Bell, C.~Lorentzen and S.~Marti,
  %``Transverse-momentum spectra of electroweak bosons near threshold at NNLO,''
  JHEP {\bf 1402}, 004 (2014)
  [arXiv:1309.3245 [hep-ph]].
  %%CITATION = ARXIV:1309.3245;%%
  %9 citations counted in INSPIRE as of 26 Jan 2015
  
\bibitem{Becher:2012za} 
  T.~Becher, G.~Bell and S.~Marti,
  %``NNLO soft function for electroweak boson production at large transverse momentum,''
  JHEP {\bf 1204}, 034 (2012)
  [arXiv:1201.5572 [hep-ph]].
  %%CITATION = ARXIV:1201.5572;%%
  %16 citations counted in INSPIRE as of 26 Jan 2015
  
\bibitem{Gavin:2010az} 
  R.~Gavin, Y.~Li, F.~Petriello and S.~Quackenbush,
  %``FEWZ 2.0: A code for hadronic Z production at next-to-next-to-leading order,''
  Comput.\ Phys.\ Commun.\  {\bf 182}, 2388 (2011)
  [arXiv:1011.3540 [hep-ph]].
  %%CITATION = ARXIV:1011.3540;%%
  %330 citations counted in INSPIRE as of 02 Feb 2015
  
\bibitem{Catani:2009sm} 
  S.~Catani, L.~Cieri, G.~Ferrera, D.~de Florian and M.~Grazzini,
  %``Vector boson production at hadron colliders: a fully exclusive QCD calculation at NNLO,''
  Phys.\ Rev.\ Lett.\  {\bf 103}, 082001 (2009)
  [arXiv:0903.2120 [hep-ph]].
  %%CITATION = ARXIV:0903.2120;%%
  %297 citations counted in INSPIRE as of 02 Feb 2015 
  
   \bibitem{Larkoski:2014uqa} 
  A.~J.~Larkoski, D.~Neill and J.~Thaler,
  %``Jet Shapes with the Broadening Axis,''
  JHEP {\bf 1404}, 017 (2014)
  [arXiv:1401.2158 [hep-ph]].
  %%CITATION = ARXIV:1401.2158;%%
  %19 citations counted in INSPIRE as of 02 Feb 2015
  
  %\cite{Grazzini:2008tf}
\bibitem{Grazzini:2008tf} 
  M.~Grazzini,
  %``NNLO predictions for the Higgs boson signal in the H ---> WW ---> lnu lnu and H ---> ZZ ---> 4l decay channels,''
  JHEP {\bf 0802}, 043 (2008)
  [arXiv:0801.3232 [hep-ph]].
  %%CITATION = ARXIV:0801.3232;%%
  %153 citations counted in INSPIRE as of 02 Feb 2015
  
\bibitem{Alioli:2012fc}
  S.~Alioli, C.~W.~Bauer, C.~J.~Berggren, A.~Hornig, F.~J.~Tackmann, C.~K.~Vermilion, J.~R.~Walsh and S.~Zuberi,
  %``Combining Higher-Order Resummation with Multiple NLO Calculations and Parton Showers in GENEVA,''
  JHEP {\bf 1309} (2013) 120
  [arXiv:1211.7049 [hep-ph]].
  %%CITATION = ARXIV:1211.7049;%%
  %33 citations counted in INSPIRE as of 09 Apr 2015

\bibitem{Banfi:2014sua}
  A.~Banfi, H.~McAslan, P.~F.~Monni and G.~Zanderighi,
  %``A general method for the resummation of event-shape distributions in e^+e^- annihilation,''
  arXiv:1412.2126 [hep-ph].
  %%CITATION = ARXIV:1412.2126;%%
  
  \bibitem{Becher:2014aya} 
  T.~Becher, R.~Frederix, M.~Neubert and L.~Rothen,
  %``Automated NNLL+NLO Resummation for Jet-Veto Cross Sections,''
  arXiv:1412.8408 [hep-ph].
  %%CITATION = ARXIV:1412.8408;%%
  
  \bibitem{Alwall:2014hca} 
  J.~Alwall, R.~Frederix, S.~Frixione, V.~Hirschi, F.~Maltoni, O.~Mattelaer, H.-S.~Shao and T.~Stelzer {\it et al.},
  %``The automated computation of tree-level and next-to-leading order differential cross sections, and their matching to parton shower simulations,''
  JHEP {\bf 1407}, 079 (2014)
  [arXiv:1405.0301 [hep-ph]].
  %%CITATION = ARXIV:1405.0301;%%
  
  \bibitem{Catani:2007vq} 
  S.~Catani and M.~Grazzini,
  %``An NNLO subtraction formalism in hadron collisions and its application to Higgs boson production at the LHC,''
  Phys.\ Rev.\ Lett.\  {\bf 98}, 222002 (2007)
  [hep-ph/0703012].
  %%CITATION = HEP-PH/0703012;%%

\bibitem{Gaunt:2014ska}
  J.~R.~Gaunt,
  %``Glauber Gluons and Multiple Parton Interactions,''
  JHEP {\bf 1407} (2014) 110
  [arXiv:1405.2080 [hep-ph]].
  %%CITATION = ARXIV:1405.2080;%%
  %6 citations counted in INSPIRE as of 11 Dec 2014

\bibitem{Bodwin:1984hc}
  G.~T.~Bodwin,
  %``Factorization of the Drell-Yan Cross-Section in Perturbation Theory,''
  Phys.\ Rev.\ D {\bf 31} (1985) 2616
   [Erratum-ibid.\ D {\bf 34} (1986) 3932].
  %%CITATION = PHRVA,D31,2616;%%
  %298 citations counted in INSPIRE as of 12 Dec 2014

\bibitem{Collins:1985ue}
  J.~C.~Collins, D.~E.~Soper and G.~F.~Sterman,
  %``Factorization for Short Distance Hadron - Hadron Scattering,''
  Nucl.\ Phys.\ B {\bf 261} (1985) 104.
  %%CITATION = NUPHA,B261,104;%%
  %391 citations counted in INSPIRE as of 12 Dec 2014

\bibitem{Collins:1988ig}
  J.~C.~Collins, D.~E.~Soper and G.~F.~Sterman,
  %``Soft Gluons and Factorization,''
  Nucl.\ Phys.\ B {\bf 308} (1988) 833.
  %%CITATION = NUPHA,B308,833;%%
  %269 citations counted in INSPIRE as of 12 Dec 2014

\bibitem{Collins:book}
J.~Collins,
  ``Foundations of perturbative QCD,''
  (Cambridge monographs on particle physics, nuclear physics and cosmology. 32)

\end{thebibliography}
\end{document}